\documentclass[twocolumn]{aastex701}

\usepackage{placeins}
\usepackage{graphicx}
\usepackage{amsmath}

\usepackage{comment}
\usepackage{color}

\graphicspath{{figures/}}

\newcommand{\ce}{C$^{18}$O}
\newcommand{\ct}{$^{13}$CO}
\newcommand{\cseven}{C$^{17}$O}
\newcommand{\kms}{ km s$^{-1}$}

\newcommand\ntwoh{N$_2$H$^+$}

\definecolor{MidnightBlue}{rgb}{0.0, 0.5, 0.69}
\definecolor{bittersweet}{rgb}{1.0, 0.44, 0.37}
\definecolor{green(ncs)}{rgb}{0.0, 0.62, 0.42}
\definecolor{hotmagenta}{rgb}{1.0, 0.11, 0.81}

\shorttitle{AGE-PRO: I. overview}
\shortauthors{Zhang et al.}

\begin{document}

\title{The ALMA Survey of Gas Evolution of PROtoplanetary Disks (AGE-PRO): \\
I. Program Overview and Summary of First Results}


\author[0000-0002-0661-7517]{Ke Zhang}
\affiliation{Department of Astronomy, University of Wisconsin-Madison, 
475 N Charter St, Madison, WI 53706, USA}
\email{ke.zhang@wisc.edu}

\author[0000-0002-1199-9564]{Laura M. P\'erez}
\affiliation{Departamento de Astronom\'ia, Universidad de Chile, Camino El Observatorio 1515, Las Condes, Santiago, Chile}
\email{}

\author[0000-0001-7962-1683]{Ilaria Pascucci}
\affiliation{Lunar and Planetary Laboratory, the University of Arizona, Tucson, AZ 85721, USA}
\email{}

\author[0000-0001-8764-1780]{Paola Pinilla}
\affiliation{Mullard Space Science Laboratory, University College London, 
Holmbury St Mary, Dorking, Surrey RH5 6NT, UK}
\email{}

\author[0000-0002-2828-1153]{Lucas A. Cieza}
\affiliation{Instituto de Estudios Astrof\'isicos, Universidad Diego Portales, Av. Ejercito 441, Santiago, Chile}
\email{}

\author[0000-0003-2251-0602]{John Carpenter}
\affiliation{Joint ALMA Observatory, Avenida Alonso de C\'ordova 3107, Vitacura, Santiago, Chile}
\email{}

\author[0000-0002-8623-9703]{Leon Trapman}
\affiliation{Department of Astronomy, University of Wisconsin-Madison, 
475 N Charter St, Madison, WI 53706, USA}
\email{}

\author[0000-0003-0777-7392]{Dingshan Deng}
\affiliation{Lunar and Planetary Laboratory, the University of Arizona, Tucson, AZ 85721, USA}
\email{}

\author[0000-0002-7238-2306]{Carolina Agurto-Gangas}
\affiliation{Departamento de Astronom\'ia, Universidad de Chile, Camino El Observatorio 1515, Las Condes, Santiago, Chile}
\email{}

\author[0000-0002-5991-8073]{Anibal Sierra}
\affiliation{Departamento de Astronom\'ia, Universidad de Chile, Camino El Observatorio 1515, Las Condes, Santiago, Chile}
\affiliation{Mullard Space Science Laboratory, University College London, Holmbury St Mary, Dorking, Surrey RH5 6NT, UK}
\email{}

\author[0000-0002-2358-4796]{Nicol\'as T. Kurtovic}
\affiliation{Max Planck Institute for Extraterrestrial Physics, Giessenbachstrasse 1, D-85748 Garching, Germany}
\affiliation{Max-Planck-Institut fur Astronomie (MPIA), Konigstuhl 17, 69117 Heidelberg, Germany}
\email{}

\author[0000-0003-3573-8163]{Dary A. Ru\'iz-Rodr\'iguez}
\affiliation{National Radio Astronomy Observatory, 520 Edgemont Rd., Charlottesville, VA 22903, USA}
\email{}

\author[0000-0002-4147-3846]{Miguel Vioque}
\affiliation{European Southern Observatory, Karl-Schwarzschild-Str. 2, 85748 Garching bei München, Germany}
\affiliation{Joint ALMA Observatory, Alonso de Córdova 3107, Vitacura, Santiago 763-0355, Chile}
\email{}

\author[0000-0002-1575-680X]{James Miley}
\affiliation{Departamento de Física, Universidad de Santiago de Chile, Av. Victor Jara 3659, Santiago, Chile}
\affiliation{Millennium Nucleus on Young Exoplanets and their Moons (YEMS), Chile}
\affiliation{Center for Interdisciplinary Research in Astrophysics and Space Exploration (CIRAS), Universidad de Santiago de Chile, Chile}
\email{}

\author[0000-0002-1103-3225]{Beno\^it Tabone }
\affiliation{Université Paris-Saclay, CNRS, Institut d'Astrophysique Spatiale, 91405 Orsay, France}
\email{}

\author[0000-0003-4907-189X]{Camilo Gonz\'alez-Ruilova}
\affiliation{Instituto de Estudios Astrof\'isicos, Universidad Diego Portales, Av. Ejercito 441, Santiago, Chile}
\email{}

\affiliation{Millennium Nucleus on Young Exoplanets and their Moons (YEMS), Chile}
\affiliation{Center for Interdisciplinary Research in Astrophysics and Space Exploration (CIRAS), Universidad de Santiago de Chile, Chile}
\email{}

\author[0009-0004-8091-5055]{Rossella Anania}
\affiliation{Dipartimento di Fisica, Università degli Studi di Milano, Via Celoria 16, I-20133 Milano, Italy}
\email{}

\author[0000-0003-4853-5736]{Giovanni P. Rosotti}
\affiliation{Dipartimento di Fisica, Università degli Studi di Milano, Via Celoria 16, I-20133 Milano, Italy}
\email{}

\author[0000-0001-9961-8203]{Estephani TorresVillanueva}
\affiliation{Department of Astronomy, University of Wisconsin-Madison, 
475 N Charter St, Madison, WI 53706, USA}
\email{}

\author[0000-0001-5217-537X]{Michiel R. Hogerheijde}
\affiliation{Leiden Observatory, Leiden University, PO Box 9513, 2300 RA Leiden, the Netherlands}
\affiliation{Anton Pannekoek Institute for Astronomy, University of Amsterdam, the Netherlands}
\email{}

\author[0000-0002-6429-9457]{Kamber Schwarz}
\affiliation{Max-Planck-Institut fur Astronomie (MPIA), Konigstuhl 17, 69117 Heidelberg, Germany}
\email{}

\author[0000-0002-6946-6787]{Aleksandra Kuznetsova}
\affiliation{Center for Computational Astrophysics, Flatiron Institute, 162 Fifth Ave., New York, New York, 10025}
\email{}

\begin{abstract}
We present the ALMA Survey of Gas Evolution of PROtoplanetary Disks (AGE-PRO), a Large Program of the Atacama Large Millimeter/submillimeter Array (ALMA). AGE-PRO aims to systematically trace the evolution of gas disk mass and size throughout the lifetime of protoplanetary disks. It uses a carefully selected sample of 30 disks around M3-K6 stars in three nearby star-forming regions: Ophiuchus (0.5-1\,Myr), Lupus (1-3\,Myr), and Upper Sco (2-6\,Myr).
Assuming the three regions had similar initial conditions and evolutionary paths, we find the median gas disk mass appears to decrease with age. Ophiuchus disks have the highest median gas mass (6\,M$_{\rm Jup}$), while the Lupus and Upper Sco disks have significantly lower median masses (0.68 and 0.44\,M$_{\rm Jup}$, respectively). Notably, the gas and dust disk masses appear to evolve on different timescales. This is evidenced by the median gas-to-dust mass ratio, which decreases from 122 in the youngest disks ($<$1\,Myr) to 46 in Lupus disks, and then increases to 120 in the Upper Sco disks. The median gas disk sizes range between 74-110\,au, suggesting that typical gas disks are much smaller than those of well-studied, massive disks. Population synthesis models suggest that magneto-hydrodynamic wind-driven accretion can reproduce median disk properties across all three regions, when assuming compact disks with a declining magnetic field over time. In contrast, turbulent-driven models overestimate gas masses of $>$1\,Myr disks by an order of magnitude. Here we discuss the program's motivation, survey design, sample selection, observation and data calibration processes, and highlight the initial results.

\end{abstract}




\section{Introduction} \label{sec:intro}

Planets form in gas- and dust-rich circumstellar disks around young stars \citep[e.g., see reviews by][]{Williams11, Andrews20_review, Oberg2023ARAA}. Most of these disks last for up to 3-10\,Myr before the gas and dust dissipate, which sets the timescale of giant planet formation \citep[e.g.,][]{Fedele10, Ribas_2014_diskfraction}. 
The structure and evolution of these disks profoundly affect every step of planet formation: from the initial grain growth, the formation of planetesimals, the accretion of planetary atmospheres, to the migration of planets \citep[e.g.,][]{benz14, Drakazkowska_PPVII, Krijt_PPVII}. 

Over the past decade, ALMA has revolutionized our understanding of  \textit{dust} in protoplanetary disks, revealing the prevalence of dust rings/gaps that are intimately linked to planet formation \citep[e.g.,][]{andrews16, zhang16, long18, Andrews20_review, Cieza21_ODISEAIII}, as well as demonstrating that dust disk mass decreases with time by providing statistics of $\sim$1500 nearby disks \citep[for a recent review summarizing surveys of nearby star-forming regions, see][]{Manara_PPVII}.
 
Despite the huge progress on the solid component, we still 
lack a basic understanding of how the dominant mass constituent of disks, the \textit{gas}, evolves with time, and what mechanism drives its global evolution \citep[e.g.,][]{bergin17}. The lack of understanding of the global evolution of gas disks has become a primary bottleneck in the field of planet formation \citep{morbidelli16b_challenges, Mordasini2024RvMG}.

The disk mass and size are the primary determinants of when, where, how many, and what type of planets form for a given protoplanetary disk \citep{benz14,Drakazkowska_PPVII, Mordasini2024RvMG}. All quantitative models of planet formation require an essential input of how the disk gas surface density evolves with time.  

Disk mass is often considered as the most fundamental property of disks.
On the macro-scale of planet formation, the available gas reservoir determines whether a planet ends up as a gas giant, icy giant, or mini-Neptune \citep[e.g.,][]{pollack96,benz14,Savvidou23}. The migration direction and speed of planets also depend on the gas mass distribution. On the micro-scale, the gas density and the gas-to-dust mass ratio regulate the dynamical behaviors of dust grains and larger solid bodies in the disk  \citep[e.g.,][]{Birnstiel23_dustreview}. The speed of dust growth, settling, and drifting all depend on gas density. For example, a locally low gas-to-dust ratio is needed for the formation of kilometer-size planetesimals through the streaming instability in disks \citep[e.g.,][]{johansen14,Stammler19}. 

The disk size also establishes the basic architecture of a planetary system.  For instance,  a very compact disk smaller than 10\,au  will not be able to form Uranus/Neptune analogs, even if the disk contains enough mass to form such planets.
The evolution of the gas disk size provides important constraints to the driving mechanism of the global disk evolution \citep[e.g.,][]{Najita18_evolution,Trapman20_viscous, Trapman21_wind, Toci2021}. In addition to the global disk evolution, the dust disk size is also driven by dust growth and drift in disks. Comparing the difference in gas and dust disk sizes has been suggested as an important test of dust evolution models \citep[e.g.,][]{Rosotti19, Trapman20_drift,Long22_size}.  

The global disk evolution is primarily determined by how angular momentum is transported. Two leading mechanisms proposed are turbulent viscosity and magneto-hydrodynamical (MHD) disk winds \citep[e.g.,][]{pringle81, Blandford82, Lesur_PPVII,Pascucci_PPVII, Pascucci2025wind}. In addition, photoevaporation by UV/X-rays from the central star or surrounding environments may contribute to the dissipation of disks \citep[e.g.,][]{Pascucci_PPVII, Winter2022_external_photoevaporation}.

The general picture is, in a viscous disk, angular momentum is redistributed, most of materials spiral inward, while a smaller fraction moves outward, causing the disk to expand over time. This expansion depends on the viscosity strength (parameter $\alpha$, \citealt{shakura73}). Viscous turbulence, traditionally thought to be driven by magneto-rotational instability (MRI), can be suppressed under certain conditions, making $\alpha$ values in disks unclear \citep[e.g.,][]{turner14, Delage22, Delage23}.

Conversely, wind-driven disks lose angular momentum as gas escapes along magnetic field lines, leading to disk shrinkage and faster mass loss compared to viscous disks \citep[e.g.,][]{Blandford82,bai13,Gressel15,Bethune17,Kadam2025}. The efficiency of this process depends on the disk's magnetic field, which remains largely unknown \citep[e.g.,][]{Harrison2021}.

There are two approaches to testing disk evolution mechanisms: (1) searching for observational signals predicted by the mechanisms, like MHD disk winds or turbulence levels, and (2) examining how average disk properties (such as disk fraction, accretion rate, as well as gas and dust disk masses and sizes) change over time in a population. For the first approach, most of the previous turbulence measurements yielded only upper limits, indicating sub-threshold levels for driving accretion \citep[e.g.,][]{flaherty15, Teague18a,Flaherty20, Rosotti23}. Few disk wind observations had sufficient resolution and sensitivity to convincingly attribute winds to radially extended MHD winds \citep[e.g.,][]{Louvet18, deValon20, Tabone_2020, Pascucci_PPVII}. On the other hand, population data show changes in median disk fraction, accretion rate, and dust mass over ages \citep{Manara_PPVII}. However, none of these directly constrain the gas masses and sizes of protoplanetary disks in a population.

The ALMA survey of Gas Evolution in PROtoplanetary disks (AGE-PRO) is an ALMA Cycle 8 Large Program (2021.1.00128.L, PI Zhang) that aims to systematically trace the evolution of gas disk mass and size throughout the lifetime of protoplanetary disks, which traces the evolution of the bulk disk mass more closely than dust. The two primary goals of AGE-PRO are to: (1) provide a legacy library of high-quality line observations that enable robust measurements of gas masses and sizes of 30 protoplanetary disks across the typical disk lifetime (0.5-6\,Myr); (2) constrain global disk evolution mechanisms by comparing model predictions with mass and size measurements of disks at different ages. The results provide a baseline test for the viscous and MHD disk wind models of protoplanetary disks, serving as an essential context for future in-depth studies of planet formation processes in disks.

This paper aims to present the design of the AGE-PRO program, introduce papers on the first results of AGE-PRO, and provide the essential context on how to interpret the results for the gas and dust evolution in protoplanetary disks. \S\ref{sec:age-pro design} describes the disk sample and the lines targeted to measure gas disk masses and sizes. In \S\ref{sec:obs}, we describe the ALMA observations, self-calibration, and the imaging processes. \S\ref{sec:data_analysis} briefly discusses how the gas masses and sizes were measured from the observations, and \S\ref{sec:data_release} describes the image data and ancillary products available for the community. \S\ref{sec:overview} provides an overview of the AGE-PRO results and introduce the highlights of individual papers of the first batch of AGE-PRO results. \S\ref{sec:discussion} discusses potential limitations of the AGE-PRO survey and how the results have changed our view of gas and dust evolution in protoplanetary disks. And \S\ref{sec:conc} summarizes the key remarks from the AGE-PRO program and discuss possible future directions to confirm trends from the initial AGE-PRO results.

\section{Design of the AGE-PRO Program} \label{sec:age-pro design}

\subsection{Sample Selection}
The AGE-PRO program was designed to measure gas disk masses and sizes at three evolutionary phases: the embedded disk phase, the middle age, and the end of the gas-rich phase. The first phase sets the conditions of early planet formation, as observations have suggested that giant planet formation may be already underway within the first 1\,Myr of disk formation \citep[e.g.,][]{alma15,Harsono18, Cieza21_ODISEAIII,Ohashi_eDisK_2023}. The middle and end phases are crucial to test evolutionary mechanisms, as different mechanisms predict disk properties diverging at these phases \citep[e.g.,][]{Tabone_2020,Trapman20_viscous}.  
The disk sample was selected from three nearby star-forming regions of different ages: Ophiuchus (embedded disks), Lupus (middle age), and Upper Sco (the end of gas disk lifetime). 
All three regions have been characterized by ALMA shallow surveys \citep{ansdell16, barenfeld16, Cieza19_ODISEA_I, Carpenter2025}. 
All three regions are nearby (140-160\,pc), providing high sensitivity and spatial resolution.

\begin{figure*}[!t]
\centering
\vspace{-0.cm}
\includegraphics[width=0.8\textwidth]{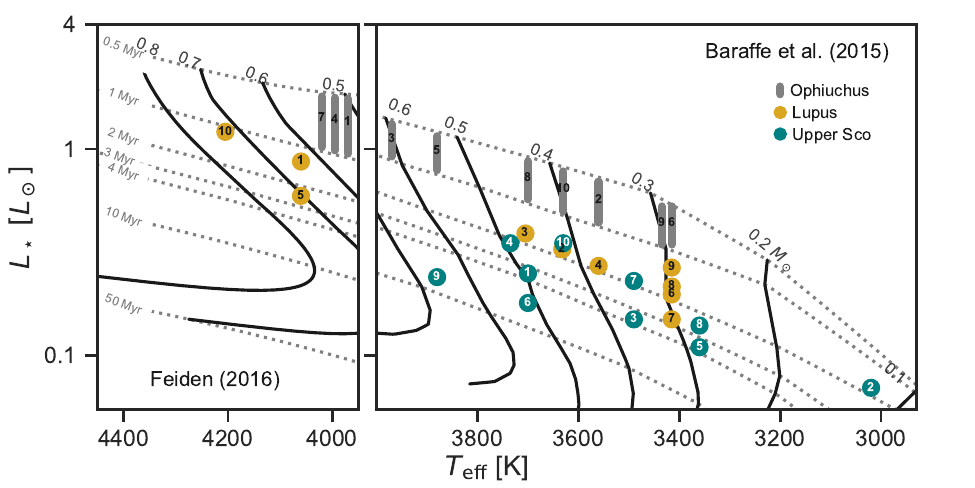}
\vspace{-0.4cm}
\includegraphics[width=0.8\textwidth]{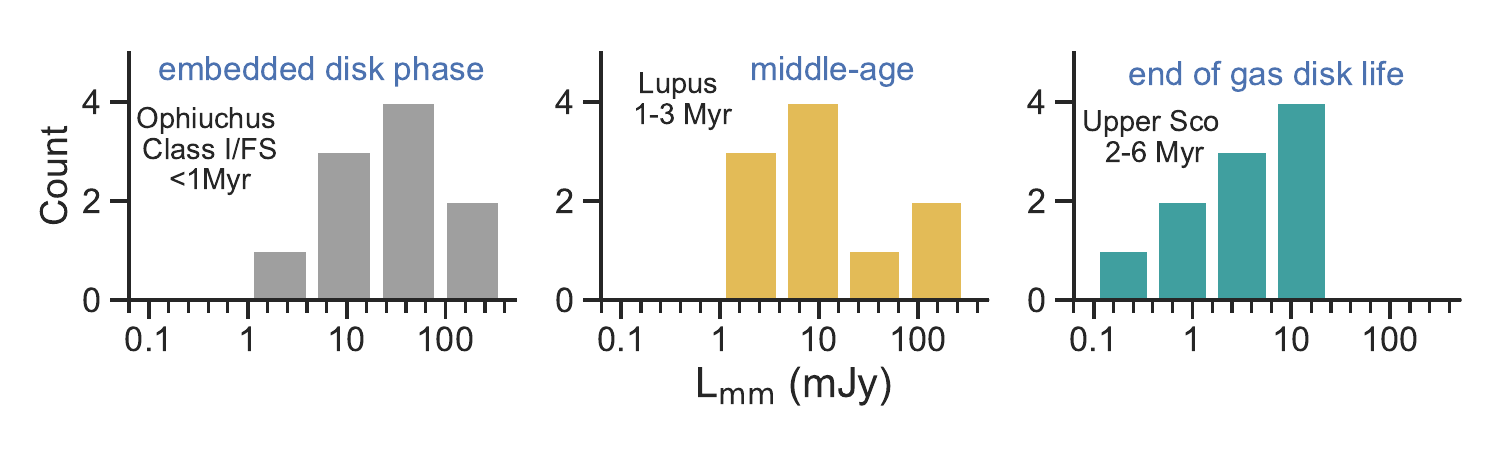}
\vspace{-0.cm}
\caption{ Top: Hertzsprung-Russell diagram of the AGE-PRO sample. The pre-main-sequence evolutionary tracks and isochrones are from \citet{Feiden16, Baraffe15}. Ophiuchus sources do not have well-constrained stellar luminosity due to high extinction. We instead plot their luminosity ranges based on statistical ages of Class I and Flat spectrum sources (0.5-1\,Myr) \citep{Evans09}. Bottom: 1.3mm-continuum luminosity ($L_{\rm mm}$) distributions (using AGE-PRO continuum fluxes and normalized to 140\,pc) of the selected sample in each region. The sample covers the entire $L_{\rm mm}$ range of M3-K6 sources in each star-forming region \citep[see detailed comparisons in ][]{AGEPRO_II_Ophiuchus, AGEPRO_III_Lupus, AGEPRO_IV_UpperSco}. \label{fig:source_selection}
}
\end{figure*}

\begin{deluxetable*}{lccccclcccccc}\tablewidth{\textwidth}
\tabletypesize{\scriptsize}
\tablecaption{AGE-PRO Sample: Host Star Properties \label{table:stars}}
\tablewidth{0pt}
\tablehead{
\colhead{Index} & \colhead{Source Name} & \colhead{RA} & \colhead{Dec}  & \colhead{Dist} & \colhead{Class} & 
\colhead{SpT} & \colhead{$T_{\rm eff}$} & \colhead{$L_\ast$} & \colhead{Av} & \colhead{$M_\ast$} & \colhead{$\log{\dot{M}_\ast}$} & \colhead{Refs.} 
\\
\colhead{} & \colhead{} &  \colhead{(h m s)} & \colhead{(d m s)} & \colhead{(pc)} &  & & 
\colhead{(K)} & \colhead{($L_\odot$)} & &\colhead{($M_\odot$)}  & \colhead{($M_\odot$\, ${\rm yr}^{-1}$)} & \colhead{}
}
\decimalcolnumbers
\startdata
 Oph 1       &  SSTc2dJ162623.6-242439   &  16:26:23.571    &  -24:24:40.102   &  \nodata&  FS    &  K7   &    3970&    \nodata&   16.1&    0.6&     \nodata& 1,2,3\\
 Oph 2       &  SSTc2dJ162703.6-242005   &  16:27:03.580    &  -24:20:06.018   &  \nodata&  FS    &  M1.5 &    3560&    \nodata&   10.1&    0.4&     \nodata& 1,2,3\\
 Oph 3       &  SSTc2dJ162719.2-242844   &  16:27:19.196    &  -24:28:44.499   &  \nodata&  FS    &  K7   &    3970&    \nodata&   31.5&    0.6&     \nodata& 1,2,3\\
 Oph 4       &  SSTc2dJ162728.4-242721   &  16:27:28.436    &  -24:27:21.790   &  \nodata&  FS    &  K6.5 &    3995&    \nodata&   24.5&    0.6&     \nodata& 1,2,3\\
 Oph 5       &  SSTc2dJ162737.2-244237   &  16:27:37.235    &  -24:42:38.543   &  \nodata&  I     &  K9   &    3880&    \nodata&   40.0&    0.5&     \nodata& 1,2,3\\
 Oph 6       &  SSTc2dJ162738.9-244020   &  16:27:38.931    &  -24:40:21.168   &  \nodata&  I     &  M2.5 &    3425&    \nodata&   12.1&    0.3&     \nodata& 1,2,3\\
 Oph 7       &  SSTc2dJ163135.6-240129   &  16:31:35.647    &  -24:01:30.048   &  \nodata&  I     &  K6   &    4020&    \nodata&   23.3&    0.7&     \nodata& 1,2,3\\
 Oph 8       &  SSTc2d J162305.4-230257  &  16:23:05.416    &  -23:02:57.553   &  \nodata&  I     &  M0.5 &    3700&    \nodata&    1.3&    0.4&     \nodata& 1,2,3\\
 Oph 9       &  SSTc2dJ162627.5-244153   &  16:26:27.538    &  -24:41:53.527   &  \nodata&  FS    &  M2.5 &    3425&    \nodata&    5.0&    0.3&     \nodata& 1,2,3\\
 Oph 10      &  SSTc2dJ162718.4-243915   &  16:27:18.371    &  -24:39:15.325    &  \nodata&  I     &  M1   &    3630&    \nodata&   44.0&    0.4&     \nodata& 1,2,3\\
 Lupus 1     &  Sz65                     &  15:39:27.753    &  -34:46:17.577   &  153.0&  II    &  K7   &    4060&   0.87&    0.6&   0.68&    -9.5& 4,5,6\\
 Lupus 2     &  Sz71                     &  15:46:44.709    &  -34:30:36.054   &  154.1&  II    &  M1.5 &    3632&   0.33&    0.5&   0.42&    -9.0& 4,5,6\\
 Lupus 3     &  J16124373-3815031        &  16:12:43.736    &  -38:15:03.471   &  158.7&  II    &  M1   &    3705&   0.39&    0.8&   0.48&    -9.1& 4,5,6\\
 Lupus 4     &  Sz72                     &  15:47:50.608    &  -35:28:35.779   &  155.5&  II    &  M2   &    3560&   0.27&    0.8&   0.39&    -8.6& 4,5,6\\
 Lupus 5     &  Sz77                     &  15:51:46.941    &  -35:56:44.531   &  154.8&  II    &  K7   &    4060&   0.59&    0.0&   0.73&    -8.7& 4,5,6\\
 Lupus 6     &  J16085324-3914401        &  16:08:53.227    &  -39:14:40.553   &  161.5&  II    &  M3   &    3415&   0.20&    1.9&   0.31&   -10.0& 4,5,6\\
 Lupus 7     &  Sz131                    &  16:00:49.414    &  -41:30:04.263   &  159.1&  II    &  M3   &    3415&   0.15&    1.3&   0.31&    -9.2& 4,5,6\\
 Lupus 8     &  Sz66                     &  15:39:28.264    &  -34:46:18.450   &  154.4&  II    &  M3   &    3415&   0.22&    1.0&   0.30&    -8.5& 4,5,6\\
 Lupus 9     &  Sz95                     &  16:07:52.293    &  -38:58:06.446   &  159.2&  II    &  M3   &    3415&   0.27&    0.8&   0.30&    -9.4& 4,5,6\\
 Lupus 10    &  V1094Sco                 &  16:08:36.160    &  -39:23:02.879   &  154.8&  II    &  K6   &    4205&   1.21&    1.7&   0.82&    -7.9& 4,5,6\\
 UppSco 1 &  J16120668-3010270        &  16:12:06.664    &  -30:10:27.617   &  131.9&  II    &  M0.5 &    3700&   0.25&    0.1&   0.51&    -9.4& 7-12\\
 UppSco 2 &  J16054540-2023088        &  16:05:45.379    &  -20:23:09.330   &  137.6&  II    &  M4.5 &    3020&   0.07&    0.3&   0.13&    -9.4& 7-12\\
 UppSco 3 &  J16020757-2257467        &  16:02:07.556    &  -22:57:47.424   &  139.6&  II    &  M2   &    3490&   0.15&    0.5&   0.37&   -11.0& 7-12\\
 UppSco 4 &  J16111742-1918285        &  16:11:17.406    &  -19:18:29.231   &  136.9&  II    &  M0.25&    3735&   0.35&    0.9&   0.50&     \nodata& 7-12\\
 UppSco 5 &  J16145026-2332397        &  16:14:50.249    &  -23:32:40.238   &  144.0&  II    &  M3   &    3360&   0.11&    1.4&   0.29&     \nodata& 7-12\\
 UppSco 6 &  J16163345-2521505        &  16:16:33.429    &  -25:21:51.163   &  158.4&  II    &  M0.5 &    3700&   0.18&    1.1&   0.52&   -10.9& 7-12\\
 UppSco 7 &  J16202863-2442087        &  16:20:28.622    &  -24:42:09.174   &  152.7&  II    &  M2   &    3490&   0.23&    1.7&   0.34&     \nodata& 7-12\\
 UppSco 8 &  J16221532-2511349        &  16:22:15.324    &  -25:11:35.672   &  139.0&  II    &  M3   &    3360&   0.14&    1.9&   0.29&     \nodata& 7-12\\
 UppSco 9 &  J16082324-1930009        &  16:08:23.247    &  -19:30:00.980   &  137.0&  II    &  M0   &    3880&   0.24&    0.9&   0.56&    -9.1& 7-12\\
 UppSco 10&  J16090075-1908526        &  16:09:00.739    &  -19:08:53.284   &  136.9&  II    &  M0   &    3630&   0.35&    1.2&   0.53&    -8.8& 7-12\\
 \enddata
\tablecomments{Col.~(1) AGE-PRO index (2) Source name,
(3) RA, (4) Dec, (5) Distance, Col.~(6) Disk class, (7) Spectral type, (8) Stellar effective temperature, (9) Stellar luminosity, (10) V band extinction magnitude, (11) Stellar mass, (12) Mass accretion rate, (13) References.}
\tablerefs{All distances are adopted from \citet{BailerJones2021}. In Col.~(13), the references for Class, Spectral type, \{$T_{\rm eff}$, $L_\ast$\}, A$_v$, and accretion rates:  1 = \citet{Evans09}, 2 = \citet{Cieza19_ODISEA_I}, 3 = \citet{AGEPRO_II_Ophiuchus},  4 = \citet{Alcala14_Lupus}, 5 = \citet{Alcala17_Lupus}, 6 = \citet{AGEPRO_III_Lupus}, 7 = \citet{Luhman2022_uppstars}, 8 = \citet{Manara2020}, 9 = Carpenter et al. 2024, submitted, 10 = \citet{Manara_PPVII}, 11 = \citet{Fang2023}, 12 = \citet{AGEPRO_IV_UpperSco}.}
\end{deluxetable*}

The AGE-PRO sample was selected with four main criteria: (1) sources with known stellar spectral type between M3-K6 \footnote{Upper Sco 2 was classified as M2 in \citet{Preibisch2001} which we adopted in the AGE-PRO sample selection stage. But more recently \citet{Manara2020} reclassified the source as M4.5 with X-shooter spectra. }, roughly corresponding to a stellar mass (M$_\star$) of 0.3-0.8\,M$_\odot$. The majority of stellar spectral types were adopted from literature and the rest from Ruiz-Rodriguez in prep. The literature stellar luminosities were updated with Gaia DR3 distances (see the list of references in Table~\ref{table:stars}).  The focus on a narrow space of stellar mass is because disk properties are expected to depend on stellar mass (e.g., dust disk masses are known to be correlated with M$_\star$; \citealt{Pascucci16,Andrews18a}). The stellar properties of the AGE-PRO sample are listed in Table~\ref{table:stars}. (2) sources without known companions or that are in wide-separation binaries ($>$600\,AU), as close binaries may evolve differently due to tidal interactions \citep[e.g.,][]{Cuello2023}. (3) For the Ophiuchus region,
Class I/Flat spectrum sources were chosen to ensure the sub-sample is $<$1\,Myr old \citep{Evans09}; 
for Lupus and Upper Sco 
regions, Class II sources were selected to exclude debris disks. (4) sources were selected from previous detections of mm continuum and CO line emission. These four criteria led to 10-34 viable candidates for each region.  
To ensure that our disks are representative of the selected M$_\star$ range, ten disks were selected from each region that covers the spread of continuum luminosities in the region (\citealt{ansdell16, barenfeld16, Cieza19_ODISEA_I, Carpenter2025}).
We 
excluded known edge-on disks ($>$70 deg) as they have smaller emitting areas and thus have fainter line emission. 
Figure~\ref{fig:source_selection} shows the stellar properties of the AGE-PRO sample in the Hertzsprung-Russell diagram and their continuum luminosity distributions.

To evaluate the representativeness of the AGE-PRO sample in each star-forming region, \citet{AGEPRO_III_Lupus} and \citet{AGEPRO_IV_UpperSco} compare the distributions of mass accretion rates and millimeter fluxes of the AGE-PRO sample with the full M3-K6 Class II disk populations in the Lupus and Upper Sco regions. Two-sample Kolmogorov-Smirnov tests showed that the AGE-PRO samples are indistinguishable from the populations of Class II disks around $\sim$M3-K6 sources in these two star-forming regions. For the Ophiuchus region, the spectral types are not well constrained for the population of embedded sources due to their high extinctions.

One potential concern is the selection of sources with existing CO detections. Previous shallow ALMA surveys found the CO detection rate of Class II disks around M3-K6 stars of 79\% (19/24) for the Lupus region and 88\% (30/34) for the Upper Sco region \footnote{The evolutionary stages of Upper Sco sources are typically characterized in the classification scheme of full/transitional/evolved/evolved transitional/debris disks rather than the Class I/II/III scheme \citep[e.g.,][]{Luhman2022_uppstars}. We considered full or transitional disk types as Class II disks. If all disk types except for debris disks are considered as Class II, the CO detection rate in the Upper Sco disks around M3-K6 stars is 80\% (31/39). } \citep{ansdell16,ansdell18,barenfeld16, Carpenter2025}. It might seem surprising that the CO detection rate in the younger Lupus region was slightly lower, but the Lupus survey was less sensitive than the Upper Sco surveys, and the Lupus detections were from $^{12}$CO\,(2-1) or \ct\,(3-2), which are expected to be fainter than the $^{12}$CO\,(3-2) line used in Upper Sco surveys. Therefore, the differences in detection rates should be not considered as intrinsic differences among the two regions. Given the high detection rates of CO in both regions, we consider the AGE-PRO sample a reasonably representative sample of mm continuum flux and accretion rates of the M3-K6 Class II disks that survived at the Lupus and Upper Sco ages. It remains to be investigated if this is true for the Ophiuchus sources.

AGE-PRO selected the approach of using three star-forming regions to sample different ages, under the assumption that these regions have similar initial conditions and evolutionary paths. Please see more discussion on the validation of these assumptions in \S\ref{sec:assumption_on_similarity}. 
Another option would be to measure gas properties of disks at different evolutionary stages within the same star-forming region. However, this approach is less practical due to the small number of available disks in each age bin, particularly for a given stellar mass range, and the large uncertainty of individual ages. Nearby star-forming regions typically have $\sim$100-200 Class 0/I/II disks in each region, most of which (60-70\%) are around stars later than M3 \citep[e.g.,][]{ansdell16,barenfeld16,Pascucci16}. The disks around later spectral types are much fainter, making it difficult to study their gas emission. For the M3-K6 type AGE-PRO targeted, there are typically only $\sim$20-30 disks in each region. Furthermore, although embedded sources are relatively easy to distinguish from Class II disks, the individual ages of single Class II sources are highly uncertain and model-dependent. The AGE-PRO’s approach uses different regions as a more robust separation of ages.

\subsection{Ages of the AGE-PRO sample} \label{sec:age_discussion}

Since the primary goal of the AGE-PRO program is to study the evolution of gas disks, constraining the ages of the sample is crucial. However, estimating the absolute ages of very young stars ($<$10\,Myr, all pre-main sequence PMS)  is challenging due to the scarcity of direct age indicators and limited knowledge of star formation processes (see review by \citealt{Soderblom2014_PPVI}).  Typically, the ages of young PMSs are estimated by comparing observations with theoretical evolutionary tracks in either Hertzsprung-Russell Diagrams (HRDs) or color-magnitude diagrams (CMDs). If the stellar parameters and extinctions of individual stars are sufficiently constrained, ages can be determined fitting evolutionary tracks to each star, though these individual ages are often considered less reliable. Median ages are then adopted for the ages of clusters or star-forming regions. Another technique, often used when the extinctions and spectral types of individual stars are uncertain, is to derive a single age for an entire cluster by fitting all stars with one theoretical isochrone, assuming all clustered stars to be coeval.

Although the absolute ages derived from these methods are highly model-dependent, relative ages can be more reliably established for low-mass stars ($\leq$1\,M$_\odot$). More luminous stars of the same effective temperature are generally younger than less luminous ones, allowing for comparisons of relative ages between different star-forming regions or clusters. The AGE-PRO program carefully selected samples from three star-forming regions with known offsets on the HRDs of their PMS stars. This selection was driven by the need to compare regions with different evolutionary stages.

Ophiuchus is typically considered one of the youngest star-forming regions, as it has a larger fraction of Class I/FS sources than other nearby star-forming regions, such as Taurus, Lupus, and Chamaeleon \citep{Evans09}. The Lupus region is generally considered 1-3\,Myr old \citep{Comeron2008,Alcala14_Lupus,Galli2020}. The Upper Sco star-forming region has an estimated age of 5-10\,Myr \citep{deGeus1989,Preibisch2002,Pecaut2012,Sullivan2021}. Recent studies with Gaia have further divided the Upper Sco region into several clusters based on peaks and dips in 5-dimensional phase space (spatial and velocity). The isochronal ages of Upper Sco clusters were estimated between $\sim$4-14\,Myr \citep{Ratzenbock2023a,Ratzenbock2023b}. These age estimations were based on isochronal fit to hundreds of PMS stars in each cluster, the majority of which are disk-less.

Given the age uncertainties, we have taken the following approach for the AGE-PRO sample. Due to the high extinctions toward the Ophiuchus region, the stellar luminosities are highly uncertain. For our Ophiuchus disk sample (Class I/FS sources), we adopted an age of 0.5-1\,Myr, based on the statistical ages of different classes from the Spitzer C2D program \citep{Evans09}. The stellar masses of the Ophiuchus sample were derived by comparing the stellar spectral types with evolutionary tracks, assuming an age between 0.5-1\,Myr. We also compared the bolometric temperature (T$_{\rm bol}$) of our sample with that of the whole embedded disk population in the Ophiuchus region \citep{Evans09}. Our sample has T$_{\rm bol}$ between 300-700\,K, which is on the larger half of the T$_{\rm bol}$ distribution of the embedded Ophiuchus sources, supporting that our sample is among the relatively older part of the embedded disks of Ophiuchus. 

To ensure a consistent comparison of relative ages between the Lupus and Upper Sco samples, we collected effective temperatures and stellar luminosities from the literature (Table~\ref{table:stars}), and estimated their ages and stellar masses by comparing them to evolutionary tracks. We use the \texttt{Python} package \texttt{ysoisochrone}\footnote{\url{https://github.com/DingshanDeng/ysoisochrone}} \citep{Deng_2025_ysoisochrone} to estimate stellar masses and ages from stellar evolutionary tracks. The \texttt{ysoisochrone} package uses tracks of \citet{Feiden16} for sources with T$_{\rm eff}$$>$3900\,K and that of \citet{Baraffe15} for sources with T$_{\rm eff}$$\le$3900\,K, following the approach of \citet{Pascucci16}.  We utilized the Bayesian approach to derive stellar masses and ages and the corresponding uncertainties (see details and individual ages in \citet{AGEPRO_III_Lupus, AGEPRO_IV_UpperSco}. 

In Appendix Figure~\ref{fig:age_distribution}, we show the estimated ages of individual sources. The ages of our Lupus sample are between 1-3\,Myr with a median of 2\,Myr, consistent with the typical 1-3\,Myr estimations for the Lupus region. For Upper Sco, our sample shows two groups: five sources between 2-3\,Myr with a median of 2.34\,Myr, comparable to the Lupus sample (2 Myr); another five Upper Sco sources appear to be consistently older, with ages between 4-6 Myr and a median of 4.6\,Myr. 

Although the absolute ages reported by different methodologies differ, we note that in the work of \citet{Ratzenbock2023b} based on fitting CMDs of astrometric clusters, the AGE-PRO Upp Sco sample is also divided into two age groups, the ages of 3 sources overlaps with the Lupus 1-4 cluster age, and 7 sources are 2-3\,Myr older (see Appendix Figure~\ref{fig:age_comparison}). Hence, while every age determination method is uncertain, two independent methodologies for age derivation agree that 3-5 sources of the AGE-PRO Upp Sco sample have ages comparable to the AGE-PRO Lupus sample, with the other 5-7 Upp Sco sources being older. 

In short, for the discussion of the AGE-PRO results, we adopt the following age ranges: Ophiuchus disk sample (0.5-1\,Myr), Lupus disk sample (1-3\,Myr), and Upper Sco disk sample (2-6\,Myr).

\subsection{Stellar masses of the AGE-PRO sample} \label{sec:mstar_discussion}

\begin{figure}[!t]
\centering
\includegraphics[width=0.46\textwidth]{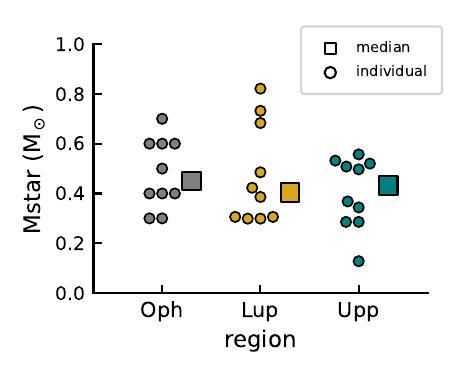}
\vspace{-0.5cm}
\caption{Stellar mass distributions in the AGE-PRO samples of the Ophiuchus, Lupus, and Upper Sco star-forming regions. The squares show the median stellar mass in each region, suggesting similar median stellar masses.  \label{fig:mstar}
}
\end{figure}

In Figure~\ref{fig:mstar} we compare the distribution of stellar masses across the three star-forming regions. The stellar masses are derived from the comparison with evolutionary tracks as described above. The bolometric temperature of our Ophiuchus sample indicates these are more evolved embedded disks, and therefore ,their current stellar masses should be close to their final masses. The median stellar masses of the three regions 0.45, 0.40, and 0.43\,M$_\odot$ for the Ophiuchus, Lupus, and Upper Sco region, respectively. The uncertainties of stellar masses from our estimations are typically 0.1-0.2\,M$_\odot$, based on the uncertainties of the stellar effective temperature and luminosity  \citep[see details in][]{AGEPRO_III_Lupus,AGEPRO_IV_UpperSco}. However, these uncertainties do not include uncertainties of evolutionary tracks which are known to be model-dependent \citep{Soderblom2014_PPVI}. Comparisons between stellar masses derived by evolutionary tracks with dynamical masses measured in binary systems suggest typical 30-80\% uncertainties \citep{Simon19}. Given these uncertainties on stellar masses, we consider the stellar masses in our sample are similar across the three regions and therefore suitable for comparison. 

\subsection[Line target]{Line targets}

\subsubsection{Lines to measure gas disk masses} \label{sec:lines for gas masses}

The gas masses of protoplanetary disks have been challenging to measure \citep{bergin17, Miotello23_PPVII}. 
The dominant gas constituent in disks is H$_2$, but its emission is undetectable in most regions of a disk due to its lack of a permanent dipole moment and low disk temperature. The best proxy of H$_2$ mass is HD, which has only been detected in three protoplanetary disks by \textit{Herschel} \citep{bergin13,mcclure16, Kama20} and is not observable with current facilities. Millimeter continuum fluxes have been the most widely used tracer to constrain disk masses. However, the conversion from continuum fluxes to gas disk masses suffers from high uncertainties of dust opacity and gas-to-dust mass ratios. For gas line tracers, CO is the most widely detected molecule in disks and is considered the most robust molecule to trace gas distribution, because of its low condensation temperature and stable chemistry \citep[e.g.,][]{Molyarova17}. 

But the conversion from CO gas mass to the total gas disk mass suffers from the high uncertainty of CO-to-H$_2$ ratio ($x_{\rm CO}$) (\citealt{favre13,schwarz16,bergin17,Miotello23_PPVII}). 
The uncertainty of $x_{\rm CO}$ likely increases over time due to chemical processes converting CO into more complex carbon carriers and dust growth carries CO ice from the disk atmosphere to the mid-plane \citep[e.g,][]{schwarz16,Bosman18,Krijt18, Krijt20}. This expectation has been seen in some recent observations. 
 \citet{Zhang20_evolution, Bergner_20evolution} compared CO isotopologue line fluxes with those predicted by thermo-chemical models and found that C/H and O/H elemental ratios of Class I/FS disks is consistent with the Interstellar Medium (ISM) ratio ($\sim$10$^{-4}$). In contrast, they found that the C/H and O/H abundance of Class II disks needs to be reduced to $0.1-0.01\times$ ISM level. This is consistent with the timescales of the chemical processing of CO and the growth from ISM-sized dust particles to pebbles being $\sim$1\,Myr \citep[e.g.,][]{schwarz18, Bosman18,Krijt20}. However, other works that include conversion of CO into CO$_2$ ice do not require significant reductions in the C/H and O/H elemental abundances with respect to ISM values \citep{Ruaud2022,Deng2023,Pascucci2023,Ruaud2024}.

Fortunately, recent studies have provided a pathway forward to constrain $x_{\rm CO}$ and subsequently the gas disk mass. 
  \ntwoh~line flux is sensitive to changes of $x_{\rm CO}$, because when gas-phase CO is around, \ntwoh~is hard to form, as CO gas competes with N$_2$ for available H$_3^{+}$ that is needed for \ntwoh~formation. Furthermore, even if some \ntwoh~forms, it is rapidly destroyed by proton transfer to gas-phase CO.
 Therefore, the line flux ratios of CO isotopologue and \ntwoh~provide constraints of $x_{\rm CO}$ and thus largely improve the precision of gas mass estimates (\citealt{Anderson19, Anderson22,Trapman22_mass,Sturm2023}). 
 
\citet{Trapman22_mass} did benchmark test by comparing gas masses constrained by the \ce~and~\ntwoh~fluxes in the three disks with independent gas masses from HD (1-0) line fluxes, and found consistent mass results between the two approaches.  One complication is that \ntwoh~is an ion, so its abundance also depends on the ionization rate of the disk, particularly on the cosmic-ray ionization rate ($\zeta_{\rm CR}$), which is also highly uncertain \citep{cleeves13a, cleeves15, Aikawa2021_MAPS}. However, \citet{Trapman22_mass} found that the constrained gas masses only changed a factor of 2-3 when $\zeta_{\rm CR}$ changes two orders of magnitude (see also \citealt{Sturm2023}). 

Given these observations and theoretical expectations, the AGE-PRO program adopted different approaches to measure gas masses for sources younger and older than 1\,Myr. For Ophiuchus sources ($<$1\,Myr), we used \cseven~(2-1) and assume ISM level $x_{\rm CO}$ of 10$^{-4}$ to measure gas disk masses. Based on the dust masses and an ISM level of CO abundance, radiative transfer models predict that \ce~(2-1) is optically thick and therefore the more rare \cseven~line  (typical \ce/\cseven~abundance ratio of 3.6, \citealt{wilson99}) is used for mass measurements \citep{miotello16}. For Class II disks, $x_{\rm CO}$ needs to be constrained for accurate gas measurements, and therefore \ct, \ce\,(2-1), and \ntwoh\,(3-2) lines were used to constrain $x_{\rm CO}$ and the gas disk masses.

\subsubsection{Lines to measure gas disk sizes} 
There is no consensus on the definition of disk sizes, as in most of the theoretical models of disk surface density it does not promptly drop to zero. Therefore, empirical measurements of disk sizes often adopt a radius within which the cumulative flux accounts for for a fixed fraction (typically 68\% or 90\%) of the total flux \citep[e.g.,][]{Tripathi2017, ansdell18, Sanchis21,Long22_size}.

For Class II sources, AGE-PRO measured the gas radius as the radius enclosing 90\% of the total integrated $^{12}$CO~(2-1) emission. This radius is not affected by the uncertainty of $x_{\rm CO}$ as $^{12}$CO is highly optically thick \citep{Trapman20_viscous}. For Class I/FS sources, the disk outer edge is often contaminated by the surrounding envelope/cloud emission and therefore the empirical flux fraction based on size is no longer a good measure \citep[e.g.,][]{Flores2023}. Our approach is to use cloud-free channels to measure the transition radius from Keplerian disk ($v\propto r^{-0.5}$) to  a rotational motion with conserved angular momenta in the infalling materials ($v\propto r^{-1}$) \citep[e.g.,][]{Ohashi2014, Aso2017}. The results will be published in Agurto-Gangas et al.\, 2025b in prep.

\section{Observations}\label{sec:obs}
The ALMA observations of the AGE-PRO program were conducted with the 12-meter array from 2021 December to 2022 August, with baseline lengths between 15 to 2917\,m.  Observational logs of each execution, including observational dates, array configurations, baseline length range, number of antennas, phase and flux calibrators, are shown in \citet{AGEPRO_II_Ophiuchus,AGEPRO_III_Lupus,AGEPRO_IV_UpperSco}, for sources in Ophiuchus, Lupus, and Upper Sco, respectively. 

These observations were carried out at ALMA band 6 and 7 \citep{ALMA_B6_receiver,ALMA_B7_receiver}. Table~\ref{tab:corr} lists the correlator set-ups used for the AGE-PRO program. For Ophiuchus sources, two Band 6 setups were used: one for $^{12}$CO/\ct/\ce~$J$=2-1 lines, and the other one for \cseven~$J$=2-1 line. The average on-source time for Ophiuchus targets was $\sim$30\,minutes for the $^{12}$CO setup and $\sim$60\,minutes for the \cseven~setup. For Lupus and Upper Sco sources, one Band 6 setup covered the CO/\ct/\ce/ $J$=2-1 lines and one Band 7 setup for the \ntwoh~$J$=3-2 line. The average on-source time of the Band 6 CO setup was $\sim$30 and 60\,minutes for the Lupus and Upper Sco targets, respectively. The Band 7 \ntwoh~setup had 60 and 120\, minutes of on-source time for the Lupus and Upper Sco targets, respectively. In addition to the key science lines, these spectral setups also covered various molecular lines for serendipitous discoveries, including H$_2$CO, DCN, DCO$^+$, N$_2$D$^{+}$, and CH$_3$CN. One or two continuum bands with 1875\,MHz bandwidth were included in all correlator setups for self-calibration. Because Upper Sco disks are generally the faintest in continuum emission among the AGE-PRO sample, we used a slightly different spectral setup in Band 6 for Upper Sco than for Lupus sources, doubling the continuum bandwidth to obtain higher sensitivity, which lowers the spectral resolution for some non-key science lines.

\begin{deluxetable*}{llccccccc}
\tablecaption{AGE-PRO Correlator Set-ups \label{tab:corr}}
\tablehead{
\colhead{Set-up}   &\colhead{Center Freq.}  &\colhead{Line Targets }   & \colhead{Vel. Res.}    & \colhead{Bandwidth}\\
 \colhead{}   & \colhead{[GHz]} &  \colhead{} &\colhead{[km~s$^{-1}$]}    & \colhead{[MHz]} 
}
\startdata
Oph/Lupus-B6-CO &217.238530  &DCN J=3--2 &0.097  &58.59\\
 &218.222192  &H$_2$CO $3_{03}-2_{02}$  &0.097  &58.59\\
&219.560358  &C$^{18}$O J=$2-1$  &0.193  &58.59\\
&220.398684  &$^{13}$CO J=$2-1$  &0.192  &58.59\\
&220.679320  &CH$_3$CN  12$_1$--11$_1$   &0.192  &58.59\\
&220.742990  &CH$_3$CN   12$_4$--11$_4$    &0.192  &58.59\\
&230.538000  &CO J=2--1  &0.092  &58.59\\
&231.321828  &N$_2$D$^+$ J=3--2  &0.091  &58.59\\
&234.000000  &Continuum band  &1.446  &1875\\
\hline
UppSco-B6-CO  &218.000000  &Continuum 1  &1.552  &1875\\
&219.560358  &C$^{18}$O J=$2-1$  &0.096  &58.59\\
&220.398684  &$^{13}$CO J=$2-1$  &0.096  &58.59\\
&230.538000  &CO J=2--1  &0.092  &58.59\\
&231.321828  &N$_2$D$^+$ J=3--2  &0.091  &58.59\\
&234.000000  &Continuum 2  &1.446  &1875\\
\hline
Oph-B6-C17O  &220.500000  &Continuum band  &1.535  &1875\\
&223.883569  &SO$_2$ 6$_{4,2}$-7$_{3,5}$  &0.096  &58.59\\
&224.714373  &C$^{17}$O J=$2-1$  &0.094  &58.59\\
&234.683390  &CH$_3$OH  4$_{2,3}$--5$_{1,4}$  &0.180  &58.59\\
&235.151720  &SO$_2$ 4$_{22}$-3$_{13}$  &0.180  &58.59\\
&236.452293  &SO 1$_2$--2$_1$  &0.179  &58.59\\
&237.983380  &$^{13}$CH$_3$OH  5$_{1,4}$--4$_{1,3}$  &0.089  &58.59\\
&239.137925  &CH$_3$CN  13$_0$--12$_0$  &0.088  &58.59\\
\hline
Lupus/UppSco-B7-N2Hp    &279.511760  &\ntwoh~J=3-2  &0.076  &117.2\\
&278.200000  &Continuum 1  &1.217  &1875\\
&288.143858 &DCO$^+$   &0.147  &58.59\\
&289.209066  &C$^{34}$S J=6--5   &0.146  &58.59\\
&289.644907  &DCN J=4-3  &0.146  &58.59\\
&290.850000 &Continuum 2  &1.164  &1875\\
\hline
\enddata
\end{deluxetable*}

\begin{deluxetable}{lcccc}
\tablecaption{Archival ALMA Data Sets Used by AGE-PRO\label{tab:archival_data}}
\tablehead{
\colhead{Source}   &\colhead{Lines}  &\colhead{Program}   & \colhead{PI}    & \colhead{Ref}\\
}
\startdata
Oph 7  & $^{12}$CO, \ct, \ce &  2019.A.00034.S  & Tobin &1\\
Lupus 1 & $^{12}$CO, \ct, \ce & 2017.1.00569.S & Yen & 2\\
Lupus 1, 6 & \ntwoh & 2019.1.01135.S & Anderson & 3\\
UppSco 9, 10 & \ntwoh & 2015.1.01199.S & Anderson & 4\\
\enddata
\tablenotetext{}{Ref: 1. \citet{Flores2023}, 2. \citet{Miley24}, 3. \citet{Anderson22}, 4. \citet{Anderson19}.}
\end{deluxetable}

\subsection{Pipeline calibration and self-calibration}

All execution blocks (EBs) were initially calibrated using the Common Astronomy Software Application (CASA; \citealt{CASA2022}) pipeline version 6.2. Several archival datasets were used in the AGE-PRO program (see Table~\ref{tab:archival_data}). These archival data were calibrated using the CASA pipeline version recommended for each data set. We ran the \texttt{scriptforPI.py} script to restore the calibrated data under the same CASA version as the calibration pipeline script. 

After the standard pipeline calibration, we conducted self-calibration using CASA version 6.4 to improve the signal-to-noise (SNR) of the data. Self-calibration of the continuum visibilities was performed for sources with continuum SNR larger than 20. Our methodology was based on the self-cal procedures of the DSHARP and MAPS large programs \citep{Andrews18b, Czekala21_MAPS}, with modifications tailored to the AGE-PRO program. Unlike the targets of the DSHARP and MAPS large programs which were bright in the millimeter continuum, many of our sources only have intermediate and faint continuum emission ($<$10\,mJy at 1.3\,mm). The modifications were therefore tailored for self-calibrations of fainter sources. We describe the detailed processes below. 

For self-calibration, we prepared continuum-only visibility data, by flagging potential line emission in each spectral window and averaging each spectral window into smaller number of channels with a maximum channel width of 125 MHz.  All known line locations were flagged, regardless of detection or not.  The systemic velocities of our sources are between 2.0-7.0 km s$^{-1}$ in the kinematic local standard of rest frame (LSRK). We flagged line channels with velocities between $\sim$-20--20 km s$^{-1}$ centered on either the known systemic velocity of each source/cloud or centered on 0\,km s$^{-1}$. 
 
 For astrometric alignment, we imaged EB individually and measured the continuum peak emission of each image by fitting a Gaussian shape. We then aligned the different EBs (with CASA tasks \texttt{phaseshift} and \texttt{fixplanets}) to a common phase center.

After the astrometric alignment, we checked the relative flux scaling among EBs. We generated deprojected visibilities for each EB, using the average inclination and position angle of each source from the continuum Gaussian fitting described above. For each source, we plotted the azimuthally averaged visibility amplitude vs. baseline length, and compare the profiles among all EBs of the source. We found that most EBs had fluxes consistent within 10\%. If flux scales among EBs differ more than 10\%, we re-scaled the flux to the EB that was the closest in time with an amplitude flux calibrator measurement from the ALMA calibrator monitoring \footnote{\url{https://almascience.eso.org/alma-data/calibrator-catalogue}}. 

Following flux scaling, we undertook the self-cal processes. The Band 7 setups have only compact configuration EBs, while our Band 6 setups have both compact and extended configurations. For setups with both configurations, the compact configuration data were self-scaled first and then the self-calibrated compact configuration data were concatenated with pre-self-cal extend-configuration data. The combined data were then self-calibrated together. 

For the self-calibration, we first chose a reference antenna, based on the ranking of antennas on the data calibration weblog. We confirmed that the chosen antenna was among the central antennas of the array configuration with the CASA task \texttt{plotants}.

The phase solutions were calculated by combining all spectral windows to improve the SNR, using CASA task \texttt{gaincal} with keyword \texttt{combine=`spws'}. For solution intervals longer than the scan length we also combined scans, using \texttt{combine=`spws,scans'} in the CASA \texttt{gaincal} task. The first round of gaincal solutions were computed with \texttt{solint=`inf'} and further iterations use decreasing time intervals, typically between 300 to 60 seconds. 

We applied the calibration table to the continuum visibilities in each iteration with the \texttt{applycal} task with \texttt{applymode=`calonly'} to ensure no visibilities were flagged if a solution was not found.

At every iteration, the self-calibrated visibilities were imaged with the \texttt{tclean} task with multiscale \texttt{CLEAN} and \texttt{scales=[0, 5, 15, 45]} pixels, where the pixel size was chosen to correspond to approximately one seventh of the beam FWHM. We adopted a Briggs robust parameter between 0.5 and 1.0, depdending on the source SNR, and we defined an elliptical mask around the source (and any companion if required).
 The aspect ratio and position angle of the clean mask were selected using the inclination and position angle of each disk. The resulting peak intensity to noise ratio on the image were used to evaluate whether another iteration is needed. If the peak intensity increased and the signal-to-noise improved by a factor larger than a few percent, we proceeded with the next iteration choosing a shorter solution time interval, usually by a factor of 1.5-2.0 smaller.

As a final step, one amplitude self-calibration was attempted, usually with a solution interval equal to the scan length (attained with \texttt{solint=`inf'} and \texttt{combine=`spws'} but not scans). The amplitude self-calibration solutions were applied only if the criteria outlined above was met.

The astrometric alignment, flux-scale alignment, and resulting calibration tables obtained from the self-calibration process were applied to the original visibilities that have not had any spectral averaging or line flagging. 

After applying the self-calibration, we then used the task \texttt{uvcontsub} to subtract the continuum with all spectral windows together, providing a continuum-subtracted measurement set for each source.

\subsection{Imaging Processes}

Prior to imaging, we generated measurement sets for each targeted line, using the \texttt{split} task with \texttt{timebin=`30s'}. The choice of a 30-second binning interval balances accuracy of the data with storage space. The line visibilities from multiple EBs were then regridded to a constant LSRK velocity grid using the \texttt{cvel2} task, correcting for Doppler shifts throughout the observation period. We adopted a velocity channel width of 0.1\,\kms~for the $^{12}$CO (2-1) line and 0.2\,\kms~for all other lines to ensure detailed spectral resolution while managing data volume and computation time.

For the imaging process, we utilized the \texttt{tclean} task in CASA with the multiscale clean option, employing scales of [0, 4, 12, 24, 48] pixels for Band 6 images and [0, 4] for Band 7 images to effectively capture structures at different sizes. Band 6 images were produced with a resolution of 1600$\times$1600 pixels and a pixel size of 0\farcs025, typical for achieving a CLEAN beam of 0\farcs35. For Band 7 images, we used 864$\times$864 pixels and a pixel size of 0\farcs04 to accommodate a larger beam size of 0\farcs7. 

To enhance the fidelity of CLEAN component locations, a binary image plane mask was employed in CLEAN processes. Given that most Lupus and Upper Sco sources exhibit line emissions following a Keplerian rotation pattern, we constructed ``cleaning'' Keplerian masks for these sources. All parameters for these Keplerian masks, such as stellar mass, disk inclination angle, and position angle measured from continuum visibility fitting (see Table~\ref{table:stars} and \citealt{AGEPRO_X_dust_disks}) can be found in \citet{AGEPRO_III_Lupus, AGEPRO_IV_UpperSco}. Masks were constructed using the \texttt{keplerian\_mask} function\footnote{https://github.com/richteague/keplerian\_mask}. A conservative maximum radius (R$_{\rm max}$) was used for each disk, based on eye evaluation to encompass the observed extent of $^{12}$CO line emission. For imaging the $^{12}$CO line of Lupus and Upper sources, we use uvrange =`$>$50klambda' to remove cloud contamination at scale$>$4\arcsec~(except for Lupus 10, which has a large disk). The choice of 4\arcsec~is based on the size of $^{12}$CO images in all Class II sources except for Lupus 10. 

For the Ophiuchus sources, where line emissions are often contaminated by the envelope and/or outflows, we utilized the \texttt{automask} option in \texttt{tclean}, as detailed in \citet{AGEPRO_II_Ophiuchus}. No baseline filtering was used for Ophiuchus sources.

A robust parameter of 0.5 was adopted for $^{12}$CO line images to achieve high spatial resolution, while value of 1 was used for all other lines to favor sensitivity. We performed cleaning down to the 1$\sigma$ noise level for all images, justifying the choice of a deep clean in the subsequent discussion on the usage of JvM correction technique \citep{JvM95}. See more discussions in \S3.3.

For Band 6 line images, we applied tapering to visibilities to generate a circular beam that is useful to compute radial intensity profiles. We calculated the tapering profiles needed to deliver a circularized CLEAN beam, and applied the tapering profiles to \texttt{tclean} with the \texttt{uvtaper} keyword. For most of the observations, the resulted beam is within 10\% difference in major and minor axis. \texttt{imsmooth} was used to make the final image with a circular beam. The Band 7 line images are mostly unresolved and therefore we did not circularize the beam.

\begin{figure*}[htbp]
    \centering
    \includegraphics[width=\textwidth]{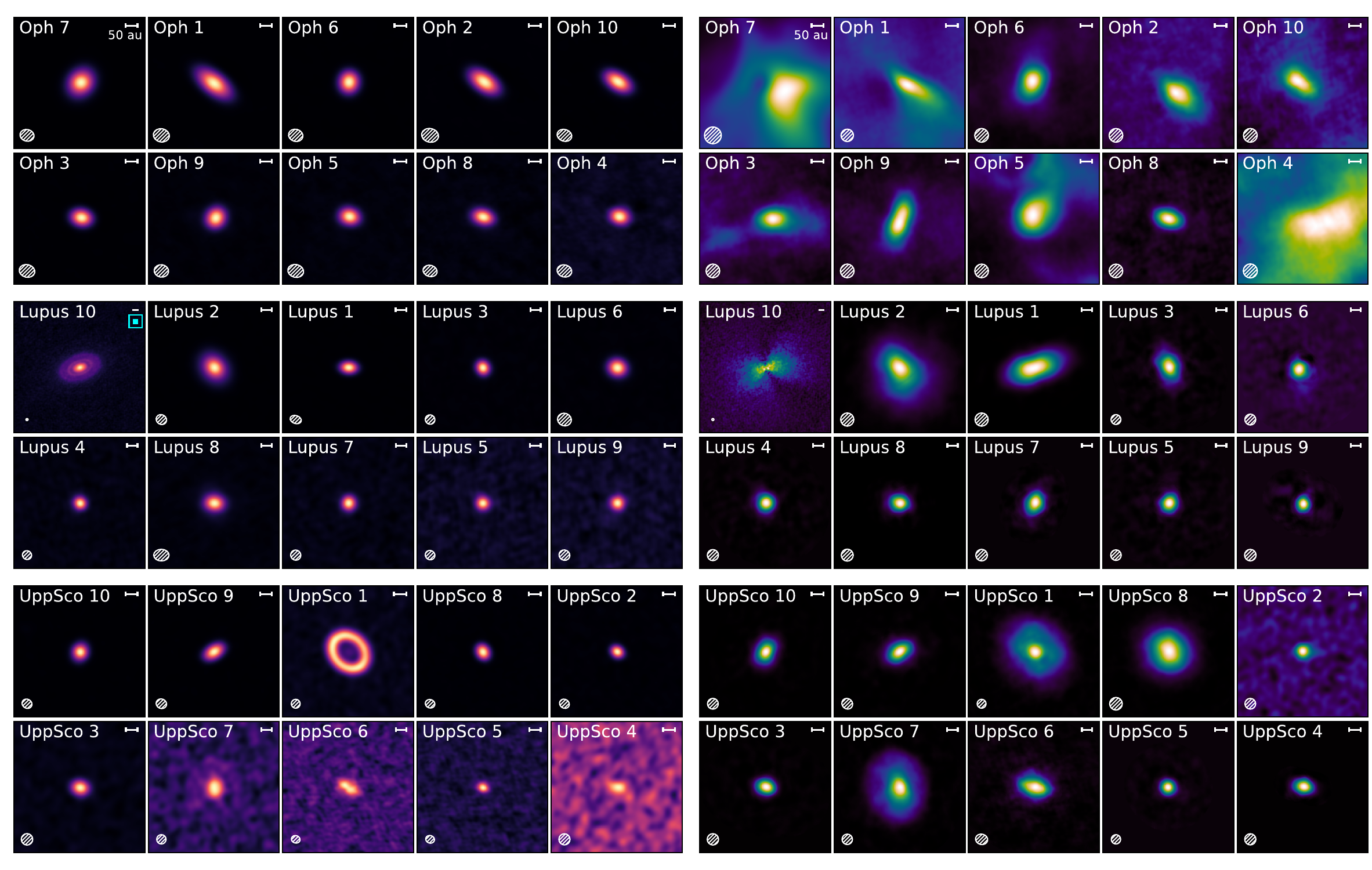}
   \caption{1.3\,mm (Left panels) and $^{12}$CO (2-1) moment zero images (Right panels) of the AGE-PRO sample. The orders are from the brightest 1.3\,mm continuum source to the faintest one within each star-forming region. Images are 2\arcsec$\times$2\arcsec and colors are peak normalized and shown with a linear color stretch. The exception is Lupus 10 (V1094 Sco) which is shown in a 5\arcsec$\times$5\arcsec image and its continuum image is shown with an asinh stretch to accentuate fainter details. The difference in image size between Lupus 10 and the other sources is visualized in cyan in its continuum image. The beam sizes are shown in a circle on the lower-left of each panel and a 50\,au size bar is shown in the upper right corner of each panel.
    \label{fig:cont_co_gallery}}
\end{figure*}

\subsection{The choice of non-JvM correction\label{subsec:JvM}}

The line fluxes were integrated over a restored image that contains the clean components convolved with the CLEAN beam plus the residuals. In this way, the measured flux may suffer from systematic errors on the flux scale, due to the so-called Jorsater \& van Moorsel effect \citep{JvM95, Czekala21_MAPS}. This scaling error is due to the discrepancy of the beam areas between the CLEAN beam used to restore the CLEAN components and the dirty beam in the residual map. This scaling error is usually more significant in data taken with multiple array configurations, where the area of the dirty beam can be substantially larger than that of the CLEAN beam, leading to an overestimation of the total flux measured in the final image (restored components + residual map). To correct for this, the residual map can be scaled to the unit of Janky per CLEAN beam using the ratio between areas of the CLEAN and the dirty beam, the so-called JvM correction. Although JvM correction normally fixes the flux error, it might introduce other issues in the final image, such as underestimating the surface brightness if the CLEAN process is not sufficiently deep, and potentially overestimating the signal-to-noise ratio of the final image \citep{Casassus22}.

To adopt a robust procedure for AGE-PRO data, we conducted a series of tests to assess the impact of the JvM correction on the total flux and radial profiles of line images. We varied the CLEAN threshold between the 1-4$\sigma$ noise level, with/without applying the JvM corrections. We found that CLEAN to 1$\sigma$ and without JvM correction gives the most accurate results for both the total flux and the radial profiles (see more details and discussions of the comparison in Appendix C). As a result, we adopted this approach for all AGE-PRO images.

\begin{figure*}[!t]
\centering
\vspace{-0.cm}
\includegraphics[width=0.82\textwidth]{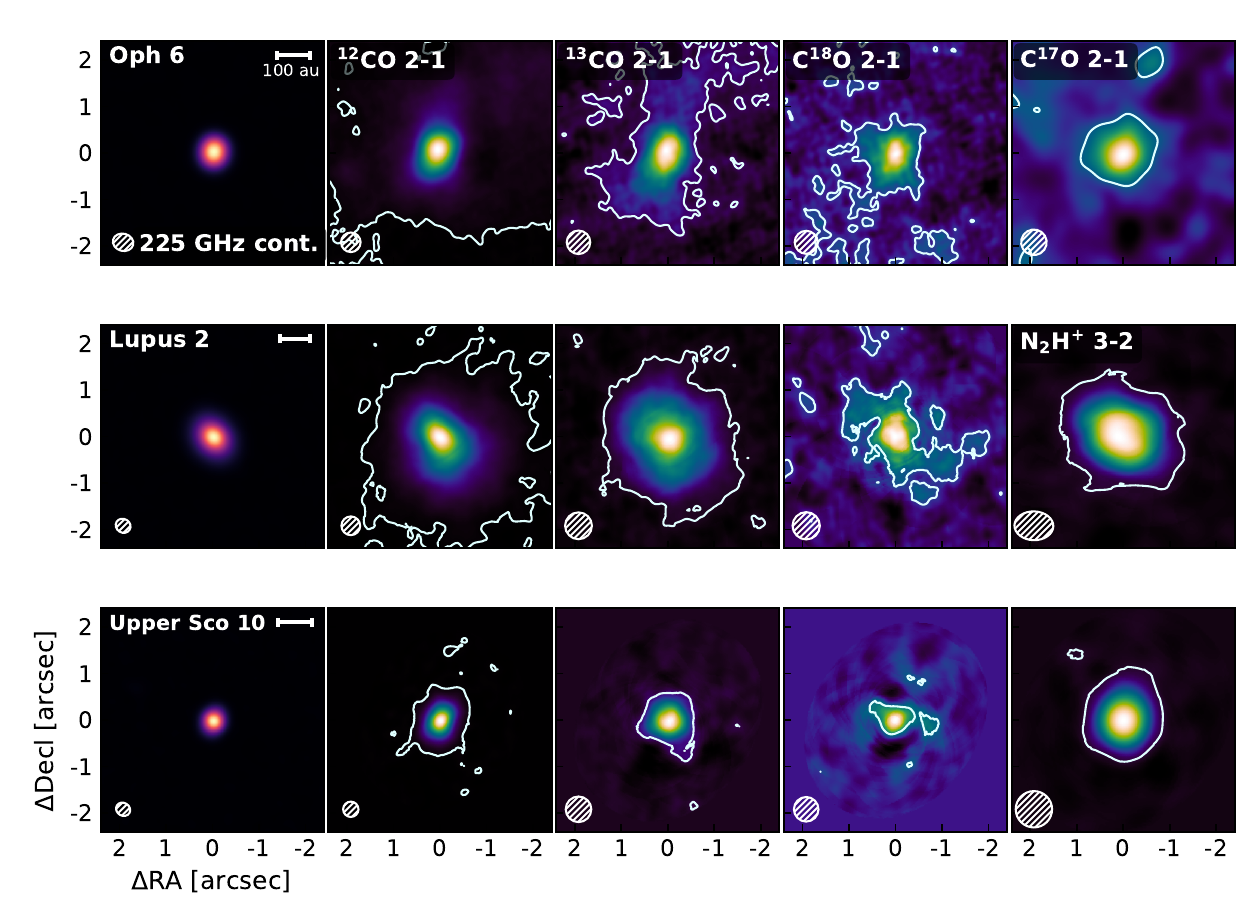}

\vspace{-0.1cm}
\caption{ Example of the 1.3\,mm continuum and moment zero images of key science lines observed in the AGE-PRO program. The full gallery of line images are shown in \citet{AGEPRO_II_Ophiuchus, AGEPRO_III_Lupus, AGEPRO_IV_UpperSco}. The white contours indicate the 3\,$\sigma$ level. }\label{fig:example_gallery}

\vspace{-0.cm}
\end{figure*}

\begin{figure*}[!t]
\centering
\vspace{-0.cm}
\includegraphics[width=0.8\textwidth]{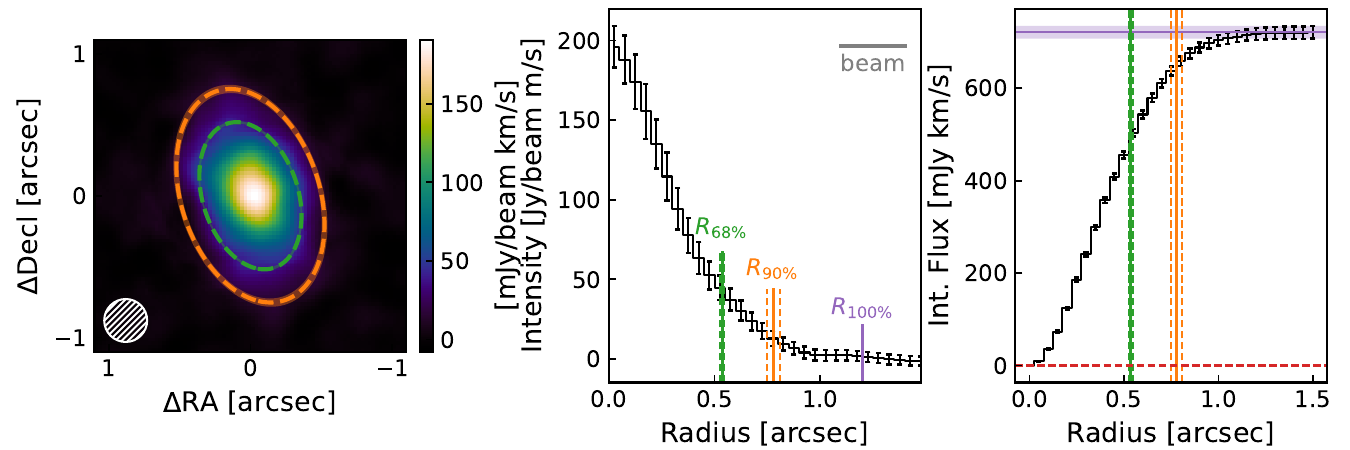}
\vspace{-0.0cm}
\caption{Example of how the curve-of-growth method is used for measuring integrated fluxes and disk sizes. Shown from left to right are the $^{12}$CO (2-1) integrated intensity map of Lupus 3, its deprojected azimuthally averaged radial profile, and the integrated flux curve-of-growth. In each panel, the radii that enclose 68\% and 90\% of the total $^{12}$CO flux are shown in green and orange, respectively. In the middle and right panel, 100\% of the total flux, defined as the first peak in the curve-of-growth, is shown in purple. The two vertical dashlines on both sides of 68\% and 90\% radii are the 1$\sigma$ uncertainty of those radii. \label{fig:curve-of-growth}}
\end{figure*}

\begin{figure*}[!t]
\centering
\vspace{-0.cm}
\includegraphics[width=0.82\textwidth]{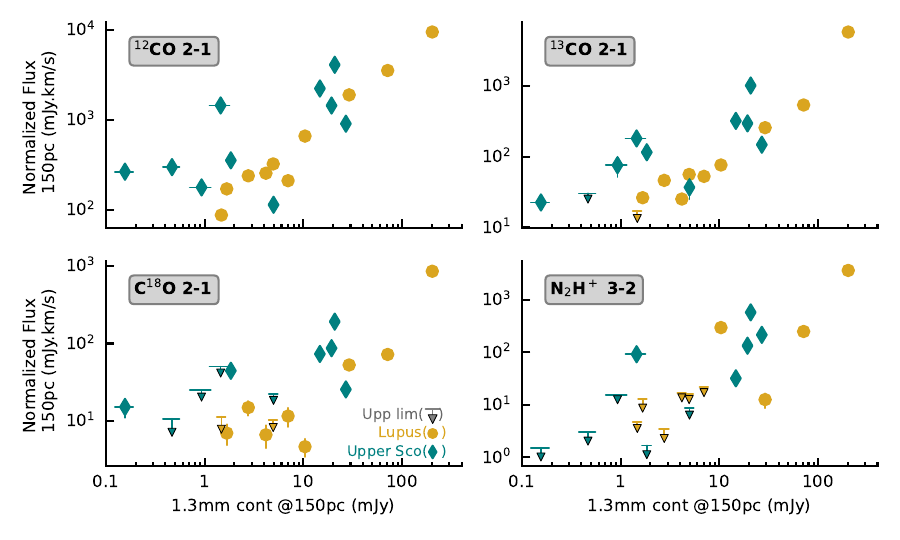}
\vspace{-0.3cm}
\caption{Integrated line fluxes of the Lupus and Upper Sco disks. \label{fig:line_flux}
}
\vspace{-0.cm}
\end{figure*}

\section{Analysis}\label{sec:data_analysis}
Figure~\ref{fig:cont_co_gallery} shows a gallery of 1.3\,mm continuum and $^{12}$CO (2-1) moment zero images of the 30 AGE-PRO sources.
In Figure~\ref{fig:example_gallery}, we show examples of continuum and key science lines of three disks, one from each star-forming region. Here we briefly describe our general approach to measure line fluxes, radii, and the gas and dust disk masses. More in-depth descriptions of the analysis processes can be found in individual AGE-PRO papers \citep{AGEPRO_II_Ophiuchus, AGEPRO_III_Lupus,AGEPRO_IV_UpperSco, AGEPRO_V_gasmasses, AGEPRO_X_dust_disks,AGEPRO_XI_gas_disk_sizes}.

\subsection{Handling cloud contamination}
The majority of the Lupus and Upper Sco sources do not show cloud contamination in their line emissions, except for $^{12}$CO line emission of several sources.
Lupus 2, 6, 8, 10, and Upper Sco 6 show signs of significant cloud contamination in their $^{12}$CO line images \citep{AGEPRO_III_Lupus, AGEPRO_IV_UpperSco}. In most of these cases, the cloud contamination is seen as negative flux of $^{12}$CO around the systemic velocity. In several cases, $^{12}$CO cloud emission can also be seen at a wide velocity range in the channel maps. Thus, to generate integrated line spectra and moment zero maps of disk emission we apply a new Keplerian mask to channel maps, using the \texttt{gofish} and \texttt{bettermoments} package. These Keplerian masks are similar to the ``cleaning'' Keplerian masks used in imaging (with the same stellar mass, disk inclination, and position angle), but these have slightly smaller R$_{\rm max}$ based on eye inspection of final channel maps. The Keplerian mask parameters used to generate moment zero maps are published in AGE-PRO XI \citep{AGEPRO_XI_gas_disk_sizes}. Using a Keplerian mask in most cases masks out the cloud emission as the cloud does not follow Keplerian rotation. Finally, after applying a Keplerian mask, we also visually inspected the channel and moment zero maps. In a few cases where significant emission unrelated to the disk is still seen, we masked out everything outside a maximum radius from the disk center. The masked regions are shown in the Appendix of \citet{AGEPRO_XI_gas_disk_sizes}. In a few cases, the integrated $^{12}$CO line spectra show a blue/red-shifted side asymmetry, probably due to foreground cloud absorption. Given the extinction of the Lupus and Upper Sco are small (A$_{\rm V}<$2), it is unlikely that both blue and red sides suffer from severe absorption and therefore we measure the CO gas radius from the uncontaminated/less contaminated half.  For cases with blue/red-shifted side asymmetry, the $^{12}$CO line fluxes are estimated from the velocity side with less/no cloud contamination and multiplied by 2 to estimate the total disk flux. More details are reported in \citet{AGEPRO_III_Lupus, AGEPRO_IV_UpperSco}.

Most of the molecular line emissions of Ophiuchus sources are contaminated by envelope and/or cloud emission, especially in $^{12}$CO and \ct~lines. The \ce~ and \cseven~lines appear to be mostly Keplerian and for the gas mass estimations of Ophiuchus sources we obtain these line fluxes after Keplerian masking.

\subsection{Measuring line fluxes and radii}

The line fluxes of the Lupus and Upper Sco disks were measured using a curve-of-growth method, where the integrated flux is measured using apertures of increasing size, until the integrated flux no longer increases with aperture sizes or reaches the first peak (see an example in Figure~\ref{fig:curve-of-growth}). The orientation and aspect ratio of the apertures are set to match the orientation of the disk, making them approximately equivalent to circular apertures in the disk's coordinate frame. The integrated line fluxes of $^{12}$CO/\ct/\ce~(2-1) and \ntwoh~(3-2) of the Lupus and Upper Sco disks are shown in Figure~\ref{fig:line_flux}. 

\begin{figure*}[!t]
\centering
\vspace{-0.cm}
\includegraphics[width=0.82\textwidth]{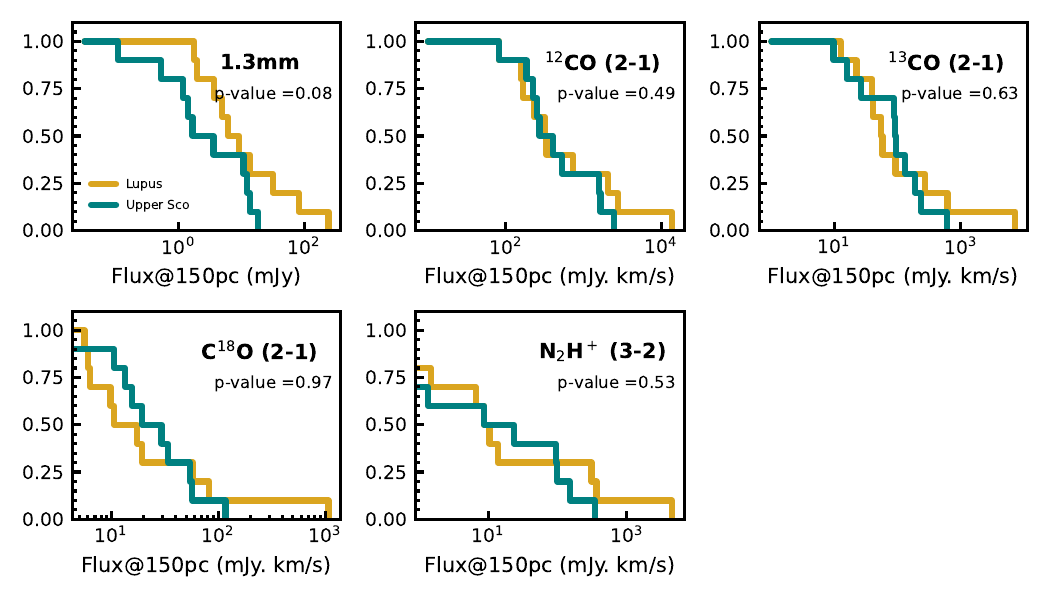}
\vspace{-0.2cm}
\caption{ The cumulative distributions of continuum and line fluxes of the Lupus and Upper Sco sample. The $p$-values are from logrank tests for the two cumulative distributions, including upper limits.  The $p$-values suggest that the continuum distributions of the two regions might be from two different populations and the line flux distributions between the two regions cannot be statistically distinguished as two populations.  \label{fig:cumulative_flux}
}
\vspace{-0.cm}
\end{figure*}

Figure~\ref{fig:cumulative_flux} compares the cumulative distributions of the 1.3\,mm continuum flux and line fluxes of $^{12}$CO/\ct/\ce~(2-1) and \ntwoh~(3-2). We also calculated the p-value of the logrank test for the two samples, including the upper limits of fluxes. The low $p$-value ($p=0.08$) of the 1.3\,mm continuum comparison suggests that the continuum distributions of Lupus and Upper Sco samples might be from different parent populations, but still not statistically significant. The $p$-values of the comparison tests for the line flux distributions are significantly higher, suggesting that the line flux distributions of the Lupus and Upper samples are similar and cannot be statistically distinguished.

\begin{figure*}[!t]
\centering
\vspace{-0.cm}
\includegraphics[width=0.9\textwidth]{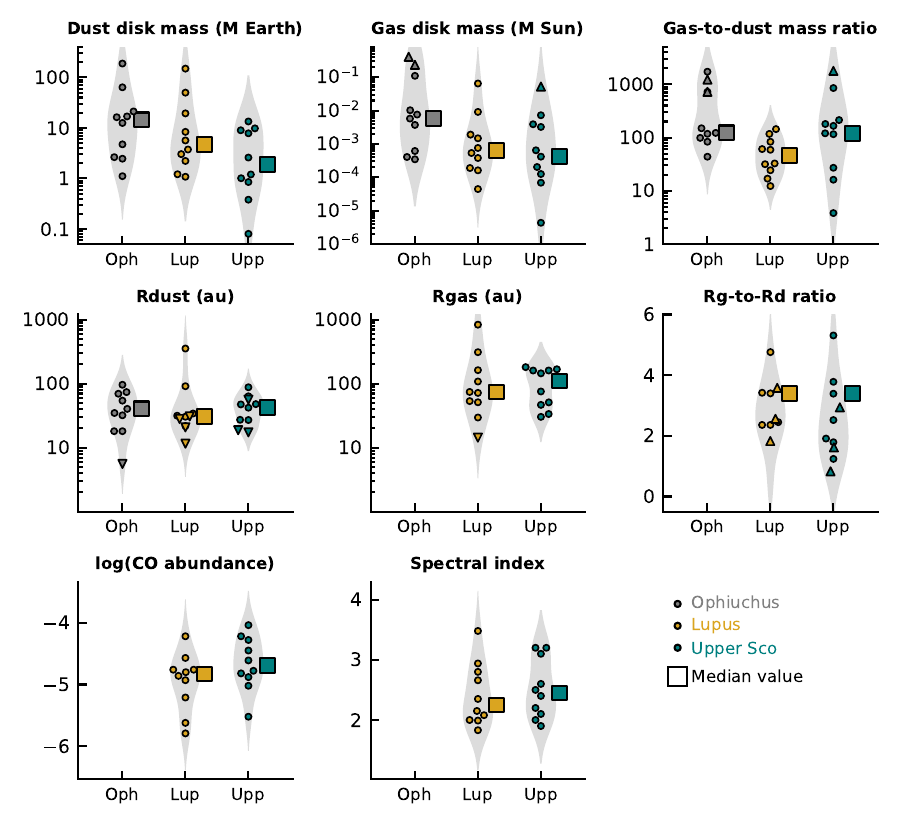}
\vspace{-0.1cm}
\caption{ Comparison of disk properties in the Ophiuchus, Lupus, and Upper Sco samples. The circles are individual values and the squares are median values listed in Table~\ref{tab:median}. The gas disk sizes, R$_{\rm CO, 90\%}$ are measured as radius that encloses 90\,\% of the $^{12}$CO (2-1). Due to the severe cloud/envelope contamination in the $^{12}$CO emission of the Ophiuchus disks, we only measure R$_{\rm CO, 90\%}$ for the Lupus and Upper Sco disks \citep{AGEPRO_XI_gas_disk_sizes}. The CO abundances of Ophiuchus disks are assumed to be at the ISM level \citep{Zhang20_evolution, AGEPRO_V_gasmasses} rather than constrained by \ntwoh, and therefore, are not plotted. The spectral index of Ophiuchus disks is not constrained due to the lack of Band 7 observations.} \label{fig:disk_properties}
\vspace{-0.cm}
\end{figure*}

The gas disk sizes in Lupus and Upper Sco were measured as the radius that encloses 90\,\% of the $^{12}$CO flux in the integrated intensity (moment zero) map.
As can be seen in Figure~\ref{fig:cont_co_gallery}, $^{12}$CO emission is only marginally resolved for a large fraction of the Lupus and Upper Sco disks. As a result, gas disk sizes measured directly from the integrated intensity maps will be dominated by the size of the beam of the observations. To minimize this effect, AGE-PRO XI \citep{AGEPRO_XI_gas_disk_sizes} fit the $^{12}$CO integrated intensity maps with a 2D Nuker profile that was convolved with the clean beam of the observations, and measures gas disk sizes from the unconvolved best fit Nuker profiles using the curve-of-growth method described previously. We adopt the 90\% flux CO sizes based on best-fitting model profile as the AGE-PRO recommended gas disk sizes \citep{AGEPRO_XI_gas_disk_sizes} and use these in all following discussions.

For the dust disk sizes, AGE-PRO X \citep{AGEPRO_X_dust_disks} employs visibility fitting to derive the surface brightness distribution of continuum emission and measure the 68\%, 90\%, and 95\% flux radius for the 30 disks.

The final disk geometries established by AGE-PRO (inclinations and position angles) can be found in AGE-PRO X \citep{AGEPRO_X_dust_disks}. 

\subsection{Gas and dust disk masses}
The gas disk masses are estimated by comparing line fluxes with a large grid of thermo-chemical models. Please see AGE-PRO V \citep{AGEPRO_V_gasmasses} for details. Here we only briefly outline the lines used for gas mass estimation. For Ophiuchus sources, the cloud contamination may affect the gas mass estimations, which is discussed in detail in AGE-PRO V \citep{AGEPRO_V_gasmasses}. We minimize the effect of cloud contamination by using the fluxes of the most optically thin \cseven~line measured within a tight Keplerian mask.  For the majority of Ophiuchus disks, we only use \cseven~line fluxes for gas mass estimations. In cases of Oph 2, 6, and 9, their \ce~emission also appears to be reasonably clear of cloud contamination, we compare \cseven~and \ce~line fluxes with predictions of thermochemical models to constrain gas disk masses  \citep{AGEPRO_V_gasmasses}. We assume the element C/H and O/H ratios are ISM level (correspondingly the ISM level of bulk CO gas abundance) in embedded disks, as suggested by previous study \citep{Zhang20_evolution}.

For the Lupus and Upper Sco sources, we estimate the gas disk masses by comparing the \ct, \ce, and \ntwoh~line fluxes, along with the 1.3 mm continuum fluxes, to a large grid of thermo-chemical models that compute the 2D temperature structure of the gas disk \citep{AGEPRO_V_gasmasses}. As discussed in \S\ref{sec:lines for gas masses}, the \ntwoh-to-CO isotopologue flux ratios constrain the CO abundance ($x_{\rm CO}$), which, in turn, together with the line fluxes of CO isotopologues, provides constraints on the gas disk mass.

For dust disk masses, we employ the standard approach in \citet{Hildebrand1983, Andrews2013, Manara_PPVII}, assuming a dust opacity of $\kappa_\nu=2.3~(\nu/230\,{\rm GHz})\,$cm$^2$ g$^{-1}$ and an average dust temperature of 20\,K, which has been empirically demonstrated to be a good disk average value and widely used in dust mass estimations of Class II disks \citep[e.g.,][]{Tazzari2021b,Manara_PPVII}. 

\subsection{Measuring spectral index of continuum emission}

Spectral index of millimeter fluxes at different wavelengths ($\alpha_{\rm mm}$) provides constraints to the dust size distributions in disks \citep[e.g.,][]{Natta2007, testi14}. For the Lupus and Upper Sco sources,
AGE-PRO observations covered two continuum wavelengths 1.05 (Band 7 setup) and 1.3\,mm (Band 6 setup), which provides additional constraints on dust growth in these sources. 
We measured the spectral index from the total flux of these twenty disks, using the formula: 
\begin{equation}
    \alpha_{\rm{mm}} \, = \,  \frac{ \log_{10} F_{\text{B7}} - \log_{10} F_{\text{B6}}}{\log_{10} \nu_{\text{B7}} - \log_{10} \nu_{\text{B6}}} \text{,}
\label{eq:spectral_index}
\end{equation}
\noindent where values of $\alpha_{mm}\approx2$ are expected for disks that are optically thick at the observed wavelengths, while higher values are typically associated with more optically thin emission.

\section{Data Release}\label{sec:data_release}
The self-calibrated visibility data, image cubes, and radial profiles of spectral lines are available for download from the AGE-PRO website  
\url{https://agepro.das.uchile.cl}. 
There, we also provide a spreadsheet with measured properties of the whole AGE-PRO sample, including integrated line fluxes, stellar properties, gas and dust disk masses and sizes, disk geometries, and mm-spectral indexes.

\section{Overview of the AGE-PRO first results}\label{sec:overview}

The main goals of the AGE-PRO program are to measure the gas disk masses and sizes in three star-forming regions at different ages.
In Figure~\ref{fig:disk_properties}, we show the individual and median disk properties in the three regions, including dust and gas disk masses and sizes, gas-to-dust mass and size ratios, CO abundances, and mm-continuum spectral indexes. The properties of individual disks are derived in AGE-PRO II-V and X-XI, and therefore these papers should be cited as the primary references for those properties \citep{AGEPRO_II_Ophiuchus, AGEPRO_III_Lupus, AGEPRO_IV_UpperSco, AGEPRO_V_gasmasses, AGEPRO_XI_gas_disk_sizes, AGEPRO_X_dust_disks}. 

We calculate the median values for each region using the Kaplan–Meier estimator in the \texttt{lifelines} python package, to include upper and lower limits of individual data points. The median values, listed in Table~\ref{tab:median}, do not take into consideration survival bias, which is inherent in our selected sample as we are probing the disks that survived evolution and eventual dispersal. This bias is properly accounted for in population synthesis models of \citet{AGEPRO_VII_population}. 

As discussed in \S\ref{sec:age_discussion}, the Upper Sco sample might be clustered around two age groups. Therefore, we also test separating our results into four age groups and find the general trends still hold (see Appendix~\ref{app:age_groups}, Figure \ref{fig:4group_disk_properties}). As the ages of individual sources are generally considered less reliable than the median age of a star-forming region, below we focus on trends identified across the three regions. 

\subsection{Gas Evolution and Comparison with models}

\textbf{Gas disk mass:} In each star-forming region, the inferred gas disk masses have a wide spread of $\sim$2 dex, as shown in Figure~\ref{fig:disk_properties}. The median gas mass of the Ophiuchus disk sample is one order of magnitude higher than that of the Lupus and Upper Sco regions. There is no significant difference between the masses of the Lupus and Upper Sco surviving disks. This suggests that most of the gas disk mass evolution happens at early times (between $<$1\,Myr to 1-3\,Myr). In our sample,  there are seven, four, and five disks with gas masses higher than 1\, M$_{\rm Jup}$ in the Ophiuchus, Lupus, and Upper Sco sub-samples, respectively.  
Please see individual gas masses and uncertainties in AGE-PRO III and V \citet{ AGEPRO_III_Lupus,AGEPRO_V_gasmasses}. 

\vspace{0.3cm}

\textbf{Gas disk size:} The median gas disk size obtained from the $^{12}$CO increases from Lupus ($\sim$74\,au) to Upper Sco ($\sim$110\,au) \citep[AGE-PRO XI][]{AGEPRO_XI_gas_disk_sizes}.  We note that Upper Sco disks are subject to mild external photoevaporation that may truncate gas disk sizes, as shown in \citep{AGEPRO_VIII_ext_photoevap}.

\vspace{0.3cm}

\textbf{Comparison with population synthesis models:} 
The gas disk masses and gas disk sizes obtained from AGE-PRO are compared to population synthesis models of turbulent-accretion models, including internal photoevaporation vs. MHD wind-driven accretion models \citep[AGE-PRO VII][]{AGEPRO_VII_population}. 
The turbulence-driven models fail to simultaneously reproduce the median accretion rate and the disk mass. The MHD-driven accretion models reproduce better median disk properties. These MHD models require initially compact disks (5-10\,au) and a magnetic field that declines with time.

\vspace{0.3cm}

\textbf{Gas-to-dust mass ratio}: Although both the median gas and dust masses decrease with age, the decrease rate of gas and dust appear to vary at different times. As a result, the gas-to-dust mass ratio (g2d) has a binomial behavior: from ISM values ($\sim$122) in Ophiuchus, to lower values in Lupus ($\sim$46), to ISM values in the surviving Upper Sco disks ($\sim$120).  The lower g2d ratio in Lupus compared to that of Ophiuchus is due to the median gas mass decreasing faster than the dust mass, while for the surviving Upper Sco sources, the increase of g2d ratio is mainly driven by the decrease of dust mass. 

\vspace{0.3cm}

\textbf{CO abundance}: In the AGE-PRO Class II sample, CO abundances with respect to H$_2$ are inferred from comparing \ntwoh\ and CO isotopologue fluxes to grids of DALI thermochemical models \citep[AGE-PRO V][]{AGEPRO_V_gasmasses}. 
We find the majority of our Class II disks have inferred CO abundances around 10$^{-5}$, suggesting an order of magnitude CO depletion with respect to the ISM ratio.

\subsection{Dust evolution and Comparison with models}

\textbf{Dust disk mass}: the dust disk mass as inferred from the mm-fluxes (assuming optically thin emission and the same temperature and dust opacity) decreases from the youngest to the oldest region, in agreement with previous works \citep[e.g.,][]{ansdell16, Pascucci16,barenfeld16,Tobin20,Manara_PPVII}.

\vspace{0.3cm}

\textbf{Dust disk size}: the dust median sizes of the surviving disks, measured from  visibility fitting of the 1.3\,mm continuum emission, are similar across the three regions \citep[AGE-PRO X][]{AGEPRO_X_dust_disks}. The apparent lack of size evolution differs from the findings of \citet{Hendler20} which, using shallower surveys but a much larger sample of disks, reported more compact dust disks in Upper Sco compared to Lupus. However, our median dust disk sizes of Lupus and Upper Sco regions are consistent with their values within 3\,$\sigma$ \citep[AGE-PRO X][]{AGEPRO_X_dust_disks}.  Sensitive and larger sample observations are needed to improve the statistics on the evolution of dust disk size.

\vspace{0.3cm}

\textbf{Dust Substructures}: the continuum visibility fitting indicates that dust substructures are present in half of the AGE-PRO sample, with a tentative increase in the detection of large cavity-type structures in older disks \citep[AGE-PRO X][]{AGEPRO_X_dust_disks}. Substructures are detected in the young sources as well, suggesting these can form in older embedded disks. This is consistent with the ALMA Large Program eDISK, which found more substructures in their Class I disks compared to Class 0 ones \citep{Ohashi_eDisK_2023}. Similarly, the CAMPOS survey reported continuum substructures in  6 Class I/FS disks  \citep{Hsieh_2024_CAMPOS}.

\vspace{0.3cm}

\textbf{Comparison with dust evolution models}: The decrease in mm-fluxes (or dust disk mass) with age can be explained when dust radial drift has been partially halted in pressure bumps \citep[AGE-PRO VI][]{AGEPRO_VI_dustevolution}.

\begin{deluxetable*}{lcccccccc}
\tablecaption{Median disk properties of the AGE-PRO sample \label{tab:median}}
\tablehead{
\colhead{}&\colhead{Mgas}&\colhead{Mdust}&\colhead{g2d}&\colhead{R$_{\rm d}$}&\colhead{R$_{\rm co}$}&\colhead{R$_{\rm co}$/R$_{\rm d}$}&\colhead{$x_{\rm co}$} & \colhead{$\alpha_{\rm mm}$}\\
\colhead{}&\colhead{(10$^{-4}$M$_\odot$)}&\colhead{(M$_\oplus$)}&\colhead{}&\colhead{(au)}&\colhead{(au)}&\colhead{}&\colhead{}&\colhead{}}
\startdata
Ophiuchus&58&14.4& 122&   40.3&\nodata&\nodata&\nodata&\nodata\\
Lupus&6.5&4.7&  46&   30.4&   74.1&    3.4&  -4.83&   2.25\\
Upper Sco&4.2&1.9& 120&   42.0&  109.9&    3.4&  -4.69&   2.45\\
\enddata
\end{deluxetable*}

 \begin{deluxetable*}{lll}
\tablecaption{First results from AGE-PRO\label{tab:papers}}
\tablehead{
\colhead{AGE-PRO}&\colhead{Reference}   &\colhead{Title} 
}
\startdata
I   &\citet{AGEPRO_I_overview}    &Program Overview and Summary of First Results \\ 
II  & \citet{AGEPRO_II_Ophiuchus} &Dust and Gas Disk Properties in the Ophiuchus Star-forming Region\\
III & \citet{AGEPRO_III_Lupus}    &Dust and Gas Disk Properties in the Lupus Star-forming Region\\
IV  & \citet{AGEPRO_IV_UpperSco}  &Dust and Gas Disk Properties in the Upper Sco Star-forming Region\\
V   & \citet{AGEPRO_V_gasmasses}  & Protoplanetary Gas Disk Masses\\
VI  & \citet{AGEPRO_VI_dustevolution} & Comparison of Dust Evolution Models to AGE-PRO Observations \\
VII & \citet{AGEPRO_VII_population}   & Testing Accretion Mechanisms from Disk Population Synthesis \\
VIII & \citet{AGEPRO_VIII_ext_photoevap}   & Impact of External Photoevaporation on Disk Masses and Radii \\
IX & \citet{AGEPRO_IX_Upp1} & Hints of Planet Formation Signatures in a Large-Cavity Disk in Upper Sco\\
X & \citet{AGEPRO_X_dust_disks} & Dust Substructures, Disk Geometries, and Dust Disk Radii\\
XI & \citet{AGEPRO_XI_gas_disk_sizes} & Beam-corrected Gas Disk Sizes for the AGE-PRO sample\\
XII & \citet{AGEPRO_XII_mm_var_USco7} & Extreme mm Variability in an Upper Sco Class II Disk\\
\enddata
\end{deluxetable*}

\subsection{Highlights of individual AGE-PRO papers}
This section provides an overview of results from the first batch of twelve AGE-PRO papers, which are listed in Table~\ref{tab:papers}. We briefly highlight the key findings and unique contributions of each individual paper, guiding readers toward the original studies for detailed discussions.

\subsubsection{Disk properties in individual star-forming regions}
AGE-PRO II \citep{AGEPRO_II_Ophiuchus}, III \citep{AGEPRO_III_Lupus}, and IV \citep{AGEPRO_IV_UpperSco} offer detailed observational results and discuss the statistical properties of AGE-PRO disks in the Ophiuchus, Lupus, and Upper Sco star-forming regions, respectively. These papers provide comprehensive analyses of the disks in each region, including statistical tests on the representativeness of the AGE-PRO sample, as well as detailed information on line emission morphology and flux distributions. Below, we highlight a key result from each paper. 

Based on the morphology of 1.3\,mm continuum images, AGE-PRO II \citep{AGEPRO_II_Ophiuchus} find 4 edge-on disks ($\geq$ 70 deg), and 3 highly inclined disks ($\geq$ 60 deg) in our Ophiuchus disk sample. For highly extincted objects, edge-on disks can potentially be misclassified as Class I/FS.  \citet{AGEPRO_II_Ophiuchus} provide new spectral type estimations and consider extinction for individual sources. The results confirm that 8 out of 10 objects are Class I/FS objects, while Oph 8 and 9 might belong to a more evolved stage, but not necessarily can be classified as Class II objects.

AGE-PRO III \citep{AGEPRO_III_Lupus} compares the gas masses of the Lupus disk sample estimated from the literature \citep{miotello17}, theoretical models by \citet{Ruaud2022}, and gas masses from AGE-PRO \citep[AGE-PRO\,V,][]{AGEPRO_V_gasmasses}. A good agreement in gas mass estimates, for large and massive disks, is found between the AGE-PRO estimates and those from the \citet{Ruaud2022} model grids, with both being significantly higher than the estimates from \citet{miotello17}. For smaller disks, the two estimates show more variations but remain within the uncertainties.

AGE-PRO IV \citep{AGEPRO_IV_UpperSco} finds moderate correlations between the line fluxes of CO isotopologues and 1.3\,mm continuum flux, suggesting a similar CO gas-to-dust mass ratio across the Upper Sco sample, with a larger scatter compared to the younger Lupus sample. The strong correlation between C$^{18}$O and N$_2$H$^+$ fluxes is nearly identical for Lupus and Upper Sco, suggesting that $x_{\rm{CO}}$ is constant over the Class II stage of disk evolution. Further, the solid mass budget has diminished in older disks, while dust disk sizes and mm spectral indexes are, on average, comparable or even larger than that in the younger regions. This may be explained by the presence of dust traps impeding radial drift in these older disks.

\subsubsection{Measurements of disk properties}

AGE-PRO V \citep{AGEPRO_V_gasmasses} describes how gas disk masses of the 30 AGE-PRO sources are constrained. A large grid of thermochemical models is employed to simultaneously fit the observed 1.3\,mm continuum, CO isotopologue, and \ntwoh\ line fluxes, as well as the 90\% radii of $^{12}$CO line emission. The results showed that the median gas mass of the three regions appears to decrease by one order of magnitude from Ophiuchus to Lupus sample, but the median gas disk masses of the Lupus and Upper Sco samples are comparable. 

AGE-PRO X \citep{AGEPRO_X_dust_disks} performs visibility fitting to the 1.3\,mm continuum of these 30 disks. Substructures are identified in 15 out of the 30 disks, with 5 disks having large ($>$15\,au) inner disk cavities. They discuss the evolution of dust disk radii and substructures among AGE-PRO disks across the three star-forming regions. A tentative trend is found, where the occurrence rate of disks with a large central cavity increases with age. In addition, \citet{AGEPRO_X_dust_disks} reports the disk geometries (inclination angle and position angle) and dust disk sizes (R$_{\rm dust,68\%}$ and R$_{\rm dust,90\%}$, R$_{\rm dust,95\%}$) used by AGE-PRO.

AGE-PRO XI \citep{AGEPRO_XI_gas_disk_sizes} measures the gas disk sizes for the 20 disks in our Lupus and Upper Sco samples, by fitting $^{12}$CO moment zero maps with beam-convolved models. The results show that median gas disk sizes are located at 74\,au in Lupus and 110\,au in Upper Sco. Combined with dust disk sizes measured in \citet{AGEPRO_X_dust_disks}, we also find the gas-to-dust size ratios are between 1-5.5. Contrary to models of dust evolution that predict an increasing size ratio with time \citep[e.g.,][]{Birnstiel2010,Trapman19},  the AGE-PRO results show that the younger disks in Lupus have similar median ratios compared to the older disks in Upper Sco.

\subsubsection{Theoretical interpretation of the AGE-PRO results}
AGE-PRO VI \citep{AGEPRO_VI_dustevolution} compares the mm continuum fluxes, dust disk sizes, and spectral indexes observed in the AGE-PRO sample with dust evolution simulations. They find that the AGE-PRO results are consistent with models with weak or strong pressure traps. The evolution of the spectral index in the AGE-PRO sample is also suggestive of an unresolved population of dust traps.

AGE-PRO VII \citep{AGEPRO_VII_population} develops a population synthesis approach to test two scenarios for disk evolution: MHD disk-wind driven and turbulence-driven with internal photoevaporation winds. A systematic method was used to fit the distribution of four disk properties of Class II disks: disk fractions, mass accretion rates, gas disk masses and sizes. For Class I/FS sources, only the gas masses are used for constraining models due to their large uncertainties in accretion rates and gas disk sizes.  The results show that the MHD wind-driven accretion can reproduce the bulk disk properties from Ophiuchus to Upper Sco, assuming initial compact disks with a magnetic field that declines with time. In contrast, the best-fitting models of turbulent-driven disks with internal photoevaporation wind overpredict the median disk masses in the Lupus and Upper Sco regions by one order of magnitude.

\subsubsection{Interesting individual disks}

AGE-PRO IX \citep{AGEPRO_IX_Upp1} reported dust and gas observational hints of planet formation in the disk around 2MASS-J16120668-301027 (Upper Sco 1 in AGE-PRO). They detect a tentative compact emission at the dust gap center and a kinematic feature in $^{12}$CO line emission. 

AGE-PRO XII \citep{AGEPRO_XII_mm_var_USco7} reports a flare detected in the continuum emission of the  2MASS J16202 863-2442087 (Upper Sco 7 in AGE-PRO) system, which decreases by a factor of 8 in less than an hour, and by a factor of 13 within 8 days. In contrast, the continuum emission of other four Upper Sco sources observed within the same execution blocks were consistent within 5\% among different execution blocks, suggesting the variation of Upper Sco 7 was real.  \citet{AGEPRO_XII_mm_var_USco7} discusses various possibilities of the flare and suggests that synchrotron emission was a probable cause of the continuum flare.

\section{Discussion}\label{sec:discussion}

\subsection{Similarity of initial conditions and external environments across star-forming regions}\label{sec:assumption_on_similarity}
Surveys like AGE-PRO aim to use samples of different ages to constrain the general evolutionary pathway of individual sources. The validation of this population approach relies on two fundamental assumptions: (1) the initial and environmental conditions of different star-forming regions are sufficiently similar; (2) the selected sample can correctly reflect the general characteristics of the target populations.  Here we discuss evidence and potential issues of these assumptions and provide the context for the interpretation of the AGE-PRO results.

\subsubsection{Similarity of initial and environmental conditions}

\textit{How similar are the initial and environmental conditions across nearby low-mass star-forming regions?} Surveys showed Class II disks in the nearby 1-3\,Myr-old low-mass star-forming regions (Taurus, Lupus, Chameleon) have similar cumulative distributions in their continuum mm luminosities and accretion rates \citep[e.g.,][]{Manara_PPVII}. These similarities support that these star-forming regions probably had similar initial conditions and are going through a general evolutionary pathway. 

However, the similarity between the younger and older regions remains unclear. 
The Class II population of the Ophiuchus region appears to have a slightly lower average mm luminosities compared with that of 1-3\,Myr-old regions \citep{Williams19}. 
The younger population of the Ophiuchus disks (Class 0/I/FS) have systemically lower dust disk masses than those in Orion and Perseus \citep{Tobin20, Tychoniec20}. However, the difference could be attributed to different initial conditions, stellar mass distribution, and/or age differences. For example, Orion has more high mass stars and a higher fraction of Class 0 sources compared to that of Ophiuchus \citep{Enoch09}. 

Our knowledge of older star-forming regions ($>$3\,Myr) is also limited. Upper Sco has been the only reference point of old star-forming regions. Although the $\sigma$-Orionis region is also considered relatively old (3-5\,Myr), a large fraction of its disk population is under a strong external FUV field \citep[$>$1000\,G$_0$,][]{ansdell17}, which causes strong external photoevaporation and significantly affects disk evolution \citep{Mauco23}. In the Upper Sco region the average external FUV field is more modest ($\sim$10\,G$_0$), but is still significantly higher than that of Taurus and Lupus (a few G$_0$). The modest FUV strength can still affect the size of disks and the disk mass to a lesser degrees. See detailed discussions on the effect of external UV on Upper Sco disks in  AGE-PRO VIII \citep{AGEPRO_VIII_ext_photoevap}. In the future, studies of more star-forming regions are needed to understand the similarities and differences among different regions.

\subsubsection{Sample selection biases}

In Section~\ref{sec:age-pro design}, we discussed the selection criteria of the AGE-PRO sample. The sample was selected mainly based on spectral type, disk class type, and mm-luminosity. For the Lupus and Upper Sco regions, K-S tests of mm-luminosity and accretion rate between the AGE-PRO sample and the full population of Class II disks around M3-K6 stars showed that these are statistically indistinguishable \citep{AGEPRO_III_Lupus, AGEPRO_IV_UpperSco}. The representativeness of the Ophiuchus sample is more challenging to quantify, as the spectral types and mass accretion rates of the full sample of embedded disks are unknown due to high extinctions. 
One particular surprise of the Ophiuchus sample is that the majority of these disks are highly-inclined ($\gtrsim$\,60\,deg, \citealt{AGEPRO_II_Ophiuchus,AGEPRO_X_dust_disks}), at a much higher proportion than the chance of random selection of inclination. Since most of the AGE-PRO Ophiuchus disks were not spatially resolved in previous surveys, only the new AGE-PRO observations revealed that these disks are highly-inclined. One possible explanation is that we selected Class I/FS sources in the Ophiuchus region from previous classifications that were based on their shallow SED slope at near-IR wavelengths. But highly-inclined disks are known to produce shallow SED near-IR slope. Please see \citet{AGEPRO_II_Ophiuchus} for in-depth discussion on other indicators of the youth of the Ophiuchus sample.

An inevitable bias for surveys to study properties in old disks is the survivor bias, because observations can only be done for the disk population that have survived up to the given age. Almost all mm disk surveys selected samples from sources with IR excesses and/or accretion signatures \citep[e.g.,][]{ansdell16, Pascucci16, barenfeld16,Cieza19_ODISEA_I}. The disk fraction with these features is known to decrease with age, from $\sim$80-90\% in $<$1\,Myr-old regions to 5-10\% in 5-10\,Myr-old regions, with an exponentially decrease timescale of 2-3\,Myr \citep[e.g.,][]{Haisch01,Fedele10,Ribas_2014_diskfraction}. As a result, the median disk properties are the median properties of the survivors. Therefore, the evolution of the median properties cannot be directly interpreted as the general evolutionary pathway of an individual disk. AGE-PRO VII \citep{AGEPRO_VII_population} combines disk fractions, mass accretion rates, and the AGE-PRO results of disk masses and sizes, in a population analysis approach, using these constraints together to test both turbulent viscous and MHD evolution models.

\subsection{Decoupled evolution in gas and dust disk masses \label{sec:g2d_evolution}}

It is of great importance to understand the relative timescales of the dust and gas mass evolution in disks, as the relative ratio of gas and dust contents would fundamentally determine the types of planets that can form inside a given disk \citep[e.g.,][]{benz14,Lee2016}.

Protoplanetary disks are believed to start with an ISM gas-to-dust ratio of $\sim$100 \citep{Bohlin78} and eventually evolve to nearly gas-free debris disks \citep{Hughes18_debrisdisks}. In the classic picture of in-situ planetesimal formation, the gas-to-dust mass ratio in disks is usually assumed to monotonically decrease over time \citep{kennedy08}. When radial drift of dust grains is considered, the local gas-to-dust mass ratio is determined by the competition between gas and dust evolution timescales. For example, in a viscously evolving disk with an $\alpha\le10^{-3}$, the timescales for dust growth and drift are significantly shorter than the timescale of gas depletion. Consequently, models predict that the outer disk regions will develop a high gas-to-dust mass ratio ($\gg$100) after $\sim$1\,Myr \citep[e.g.,][]{Takeuchi05,Birnstiel2010}.

The AGE-PRO program is the first systematic study to observationally trace gas disk mass evolution in protoplanetary disks. The new constraints on gas disk masses provide evidence that the evolution of dust and gas disk masses probably does not operate on the same timescale. 
The different evolution timescale between gas and dust masses leads to a surprising swing in the gas-to-dust mass ratio, with a median ratio first decreasing and then increasing. These differences confirm that dust masses cannot be used to obtain reliable gas disk masses. AGE-PRO VII \citep{AGEPRO_VII_population} compares two gas evolution mechanisms: viscosity and MHD-disk driven, and discusses how AGE-PRO results constrain the timescales of these mechanisms.

\subsection{The typical gas and dust disk sizes}
The sizes of gas and dust disks provide important insight into the disk evolution. As shown in Table~\ref{tab:median}, the median dust disk sizes are remarkably similar among the three samples, from 30-42\,au. 
The possible lack of evolution in the dust disk sizes suggests that some level of pressure traps must be present to keep some mm-sized particles in the outer region of the disks. On the other hand, the decreases in dust masses over time suggest that pressure traps in the outer disk regions cannot completely block the inward pebble drift. Please see \citet{AGEPRO_VI_dustevolution} for an in-depth discussion on the level of pressure traps needed to explain the dust disk size and mm continuum fluxes. 

The median gas disk sizes of the Lupus and Upper Sco samples appear to be around 74\,au and 110\,au, respectively. This is significantly smaller compared to the CO gas disk sizes of 100-1000\,au seen in well-studied massive disks \citep{Long22_size,law_maps_radial}. Therefore, to better estimate the representative sizes of protoplanetary disks, a larger disk sample would be needed in the future.  

The apparent lack of gas disk size evolution might seem to disprove the theory of viscous evolution, as viscously evolving disks are expected to expand over time \citep{Trapman20_viscous}. However, both internal and external photoevaporation can cause gas disk sizes to shrink over time \citep{AGEPRO_VII_population, AGEPRO_VIII_ext_photoevap}. Therefore, gas disk size alone is not decisive evidence against viscous evolution. Nonetheless, it remains a critical input for population synthesis models to constrain disk evolution mechanisms \citep{AGEPRO_VII_population}.

\subsection{CO gas abundance evolution}
CO is one of the major carbon and oxygen carriers, and the most widely detected molecule in protoplanetary disks. Therefore, it is of great importance to understand the abundance evolution of CO in disks. 

For embedded disks (younger than 1\,Myr), \citet{Zhang20_evolution} and \citet{Bergner_20evolution} found that CO abundance is initially around the ISM ratio and then decreases on a timescale of 1\,Myr, although these previous studies were based on a small sample of Class I/FS sources. Using the Ophiuchus sample from AGE-PRO, we confirm that the CO abundance at the embedded stage is indeed close to the ISM level. This is supported by our finding of high gas disk masses derived from \cseven~(2-1) line fluxes and a median gas-to-dust mass ratio close to the ISM ratio of 100.

For disks older than 1\,Myr, previous studies reported a wide range of CO abundances, from the ISM ratio to two orders of magnitude depletion \citep[e.g.,][]{favre13,williams14,schwarz16,kama16,mcclure16,zhang17, Cleeves18, Anderson19,Calahan21_twhya,Trapman22_mass, Schwarz2021_MAPS,Cauley2021,Ruaud2022,Deng2023,Pascucci2023}.

In the AGE-PRO study, we see a wide range of CO gas abundances (\(x_{\rm CO}\)) between \(10^{-6}\) and \(10^{-4}\) in the Lupus and Upper Sco disks \citep{AGEPRO_V_gasmasses}. However, the majority of our disks have $x_{\rm CO}$ around \(10^{-5}\). These results support significant CO depletion in 1-6 Myr disks, but few extreme cases like \(x_{\rm CO} \sim 10^{-6}\) observed in the TW Hya disk. Notably, we do not observe significant variation in \(x_{\rm CO}\) between the Lupus and Upper Sco sources, suggesting that CO depletion generally ceases to progress efficiently after the first 1-3 Myr.

There are two general mechanisms proposed to explain the depletion of CO gas abundance. The first one involves chemical processes: once CO freezes onto dust grains, it can be converted into H$_2$CO, CO$_2$, and CH$_3$OH via hydrogenation on grain surfaces. However, this process requires an ISM level of cosmic ionization rate of \(10^{-17}\,\text{s}^{-1}\) or higher to reduce \(x_{\rm CO}\) to \(10^{-5}\) within 1\,Myr \citep[e.g.,][]{schwarz18, Bosman18}. The second theory suggests that dust growth sequesters CO gas from the atmosphere to the cold mid-plane, locking CO as ice in the mid-plane \citep[e.g.,][]{xu17,Krijt18}. CO depletion due to dust growth alone is relatively inefficient, only reducing CO by a factor of 2-3 over 1\,Myr. However, the combination of chemical processes and dust growth can be highly efficient, depleting CO abundance by a factor of 100 within 3\,Myr under ISM level cosmic-ionization rate \citep{Krijt20}.

We propose two hypotheses to explain the CO abundance evolution pattern observed in AGE-PRO. The first is that the cosmic-ray ionization rate is \(10^{-17}\,\text{s}^{-1}\) during the embedded stage but decreases during the Class II stage. The second hypothesis is that, once most of the dust grains have grown to a large size and settled to the mid-plane after 1 Myr, CO depletion in the disk atmosphere is no longer an effective process.

\section{Conclusion and Future directions}\label{sec:conc}
The AGE-PRO program measures the gas and dust disk masses and sizes for 30 disks around M3-K6 stars in three star-forming regions, including the Ophiuchus sample ($<$1\,Myr), the Lupus sample (1-3\,Myr), and the Upper Sco sample (2-6\,Myr). The results provide the first systematic sample to study gas evolution in protoplanetary disks. Our main findings are: 

\begin{enumerate}
    \item In each star-forming region, the gas disk masses show a wide spread of 2-3 orders of magnitude. The median gas mass of the youngest Ophiuchus disk sample is one order of magnitude higher than that of the older disks in the Lupus and Upper Sco samples. The median gas masses of the Lupus and Upper Sco samples only differ by a factor of 1.5, despite that the median dust mass of the Upper Sco sample is 2.5 times lower than that of Lupus. 

    \item The gas masses measured from AGE-PRO enable the first comparison of gas and dust mass evolution in disks. The gas and dust appear to evolve on different timescales, and the gas-to-dust mass ratio does not monotonically decrease over time. The median gas-to-dust mass ratio starts at 122 (close to the ISM ratio of 100), decreases to a median value of 46, and then increases to 120. These results suggest that dust growth and drift may proceed faster than the gas evolution after 1\,Myr. 

    \item The median dust disk sizes measured from 1.3\,mm-continuum emission appear similar across the three regions (30-42\,au). The median mm-continuum spectral index of the Lupus region is slightly lower than that of Upper Sco.  The consistent continuum disk size and increasing spectral index align with predictions of dust evolution with pressure traps in disks. 

    \item  Substructures are observed in millimeter-continuum emission of disks across all ages. There is a tentative trend that the occurrence rate of disks with a large central cavity increases with age.

    \item External photoevaporation can significantly reduce gas disk sizes, even under moderate far-ultraviolet (G$_0$$\sim$10) radiation fields. While it also decreases gas disk masses, the effect is less pronounced than for disk sizes. 

    \item By combining the current constraints of disk fractions, accretion rates, and gas disk masses and sizes, population synthesis models show that the MHD wind-driven accretion can reproduce the median disk properties from Ophiuchus to Upper Sco, assuming compact disks with a magnetic field that declines with time. Turbulent-driven disks can match the accretion rates and gas disk sizes, but predict larger median gas disk masses in the Lupus and Upper Sco regions by one order of magnitude.

\end{enumerate}

The AGE-PRO program represents the first systematic ALMA initiative to study gas evolution in protoplanetary disks. Its findings highlight significant differences in the evolutionary timescales of gas and dust. To advance our understanding of these processes, future observations need to target a broader range of disk ages, stellar masses, and environmental conditions. High-resolution observations of gas and dust emissions will be essential for comparing their spatial distributions and identifying substructures within disks. On the theoretical front, developing population synthesis models with advanced physics will be crucial for accurately unraveling the interplay between dust and gas evolution. This integrated approach will ultimately enhance our ability to predict planet formation pathways and timescales.

\newpage

\begin{acknowledgments}
This paper makes use of the following ALMA data: ADS/JAO.ALMA\#2021.1.00128.L,   \\
ADS/JAO.ALMA\#2019.A.00034.S, \\
ADS/JAO.ALMA\#2017.1.00569.S, \\
ADS/JAO.ALMA\#2019.1.01135.S, \\
ADS/JAO.ALMA\#2015.1.01199.S. \\
ALMA is a partnership of ESO (representing its member states), NSF (USA) and NINS (Japan), together with NRC (Canada), MOST and ASIAA (Taiwan), and KASI (Republic of Korea), in cooperation with the Republic of Chile. The Joint ALMA Observatory is operated by ESO, AUI/NRAO and NAOJ. The National Radio Astronomy Observatory is a facility of the National Science Foundation operated under cooperative agreement by Associated Universities, Inc.

We thank the anonymous referee whose comments helped us improve the manuscript.
We thank Alena Rottensteiner and Phil Armitage for the useful discussions that contributed to this work.

K.Z. and L.T. acknowledge the support of the NSF AAG grant \#2205617. 
G.R. acknowledges funding from the Fondazione Cariplo, grant no. 2022-1217, and the European Research Council (ERC) under the European Union’s Horizon Europe Research \& Innovation Programme under grant agreement no. 101039651 (DiscEvol). Views and opinions expressed are however those of the author(s) only, and do not necessarily reflect those of the European Union or the European Research Council Executive Agency. Neither the European Union nor the granting authority can be held responsible for them.
P.P. and A.S. acknowledge the support from the UK Research and Innovation (UKRI) under the UK government’s Horizon Europe funding guarantee from ERC (under grant agreement No 101076489).
A.S. also acknowledges support from FONDECYT de Postdoctorado 2022 $\#$3220495.
B.T. acknowledges support from the Programme National “Physique et Chimie du Milieu Interstellaire” (PCMI) of CNRS/INSU with INC/INP and co-funded by CNES.
I.P. and D.D. acknowledge support from Collaborative NSF Astronomy \& Astrophysics Research grant (ID: 2205870).
C.A.G. and L.P. acknowledge support from FONDECYT de Postdoctorado 2021 \#3210520.
L.P. also acknowledges support from ANID BASAL project FB210003.
L.A.C and C.G.R. acknowledge support from the Millennium Nucleus on Young Exoplanets and their Moons (YEMS), ANID - Center Code NCN2024\_001 and 
L.A.C. also acknowledges support from the FONDECYT grant \#1241056.
N.T.K. acknowledges support provided by the Alexander von Humboldt Foundation in the framework of the Sofja Kovalevskaja Award endowed by the Federal Ministry of Education and Research.
K.S. acknowledges support from the European Research Council under the Horizon 2020 Framework Program via the ERC Advanced Grant Origins 83 24 28. E.T. acknowledges support from the National Science Foundation Graduate Research Fellowship program. 
\end{acknowledgments}

\vspace{5mm}
\facilities{ALMA}

\software{astropy \citep{Astropy18},  CASA \citep{CASA2022}, GoFish \citep{GoFish}, \texttt{MATPLOTLIB} \citep{Hunter2007}, \texttt{lifelines} \citep{lifelines_davidson_pilon_2024}
          }

\newpage
\appendix

\section{SEDs}

\begin{figure*}[htbp]
\centering
\vspace{-0.cm}
\includegraphics[width=0.9\textwidth]{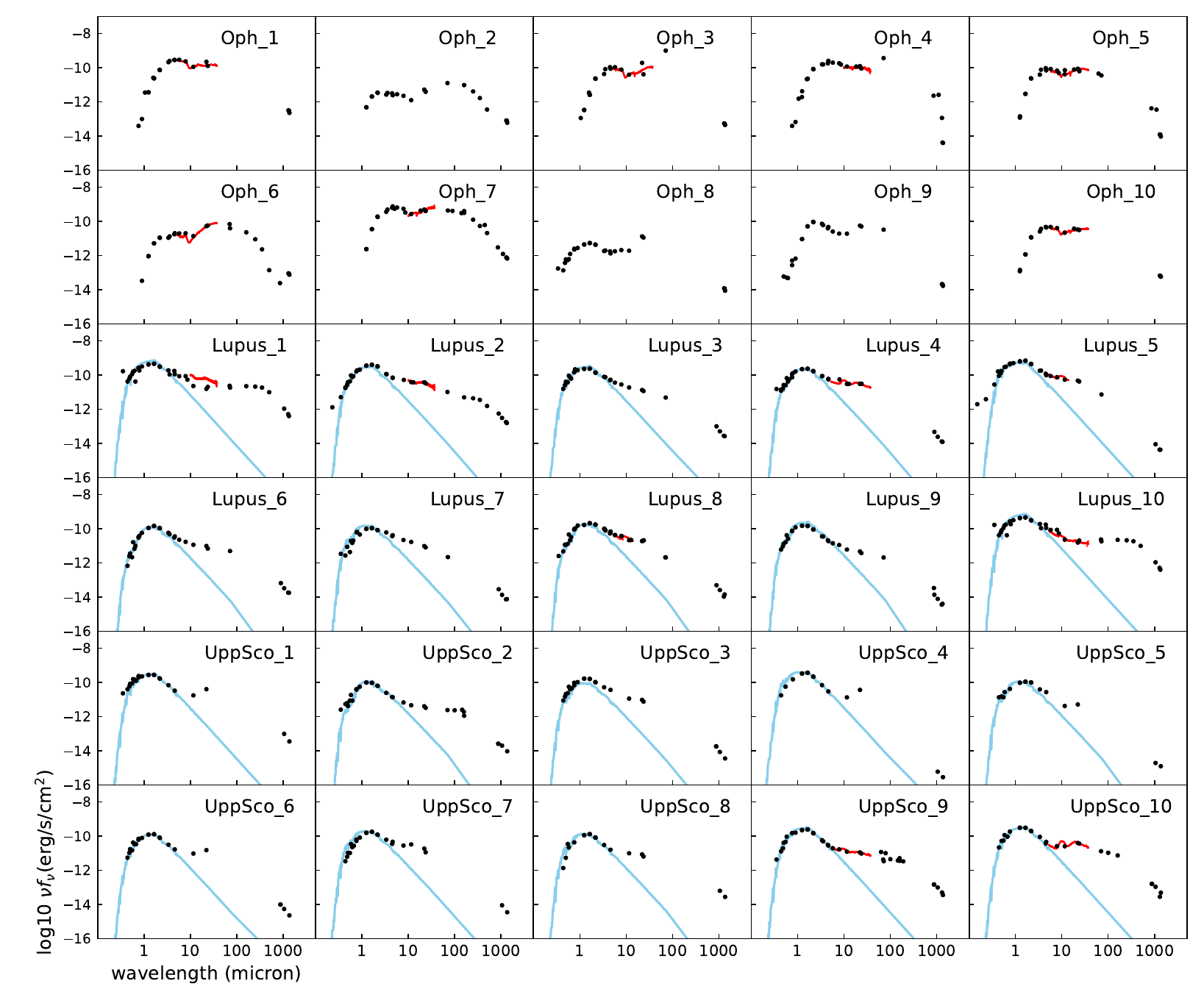}
\vspace{-0.2cm}
\caption{ Spectral Energy Distributions (SED) of AGE-PRO sources. Blue curves show the PHOENIX stellar photosphere models using the stellar parameters and extinction values listed in Table~\ref{table:stars}. The red curves are archival Spitzer IRS spectra.  These SEDs and detailed references are shown in \citet{AGEPRO_II_Ophiuchus,AGEPRO_III_Lupus, AGEPRO_IV_UpperSco}.\label{fig:seds}} 
\vspace{-0.cm}
\end{figure*}

Figure~\ref{fig:seds} shows the spectral energy distributions (SED) of the 30 disks included in the AGE-PRO sample.

\section{Ages of the Lupus and Upper Sco sources in AGE-PRO \label{app:ages}}

\begin{figure*}[htbp]
\centering
\vspace{-0.cm}
\includegraphics[width=0.5\textwidth]{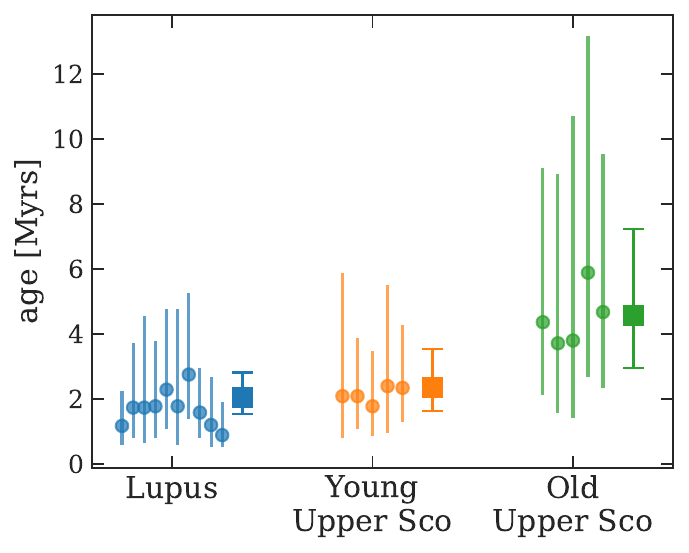}
\vspace{-0.2cm}
\caption{Isochronal ages of the Lupus and Upper Sco samples of AGE-PRO. The circles are individual ages and the squares represent the median ages of each group. The ages were derived based on stellar properties listed in Table~\ref{table:stars} and PMS evolutionary tracks of \citet{Feiden16, Baraffe15}. The individual ages and their uncertainties can be found in \citet{AGEPRO_III_Lupus, AGEPRO_IV_UpperSco}.  \label{fig:age_distribution} 
}
\vspace{-0.cm}
\end{figure*}

\begin{figure*}[htbp]
    \centering
    \includegraphics[width=0.9\textwidth]{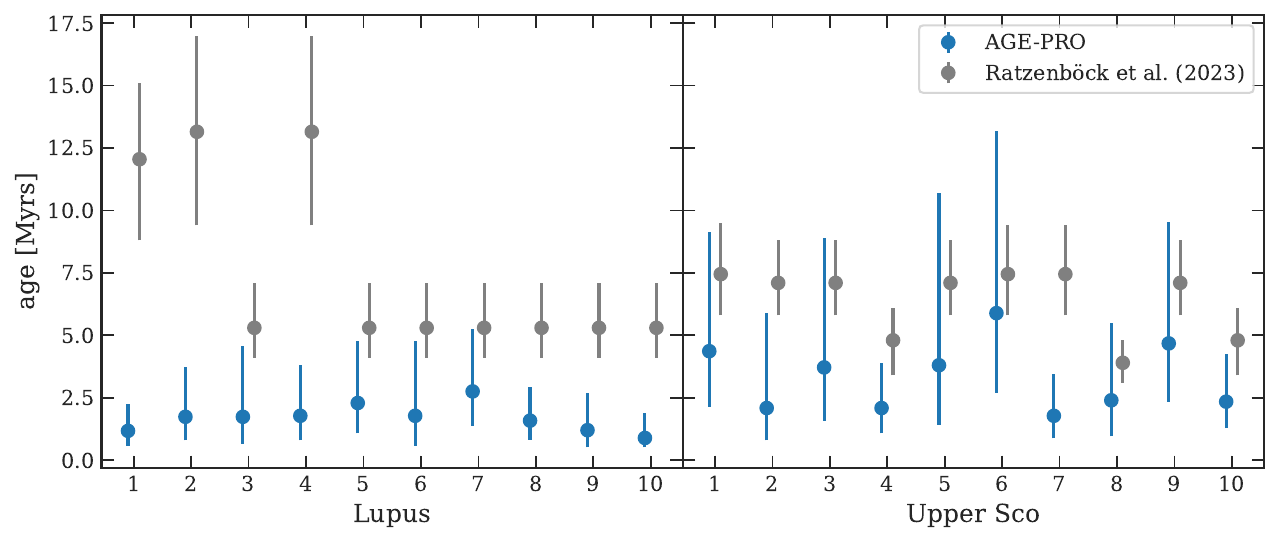}
    \caption{Comparison between the ages derived by the AGE-PRO program \citep{AGEPRO_III_Lupus, AGEPRO_IV_UpperSco} and isochronal ages from CMD fitting of astrometric clusters by \citet{Ratzenbock2023a, Ratzenbock2023b}. Although the ages of \citet{Ratzenbock2023a, Ratzenbock2023b} are older than our results, the median ages of Upper Sco sources are consistently older than that of Lupus sources in both age estimations.}
    \label{fig:age_comparison}
\end{figure*}

In Figure~\ref{fig:age_distribution}, we show the age distributions of the AGE-PRO sample, based on isochronal fitting of individual stars with evolutionary tracks of \citet{Baraffe15,Feiden16}. Figure~\ref{fig:age_comparison} shows the comparison between our age estimations with the estimations from \citet{Ratzenbock2023b}, which were based on fitting a single age to all sources in individual astrometrically identified clusters. Lupus 1, 2, 4 have the largest differences in age estimation. Lupus 1 and 8 are a wide-separation binary system \citep{Miley24}, which are expected to have the same age. But \citet{Ratzenbock2023b} assigned these into two clusters with distinctive ages. It is beyond the scope of this paper to discuss the robustness of cluster membership of individual sources. 

\section{Tests of impact of JvM correction on radial profiles \label{app:JvM}}

\begin{figure*}[htbp]
\centering
\vspace{-0.cm}
\includegraphics[width=0.9\textwidth]{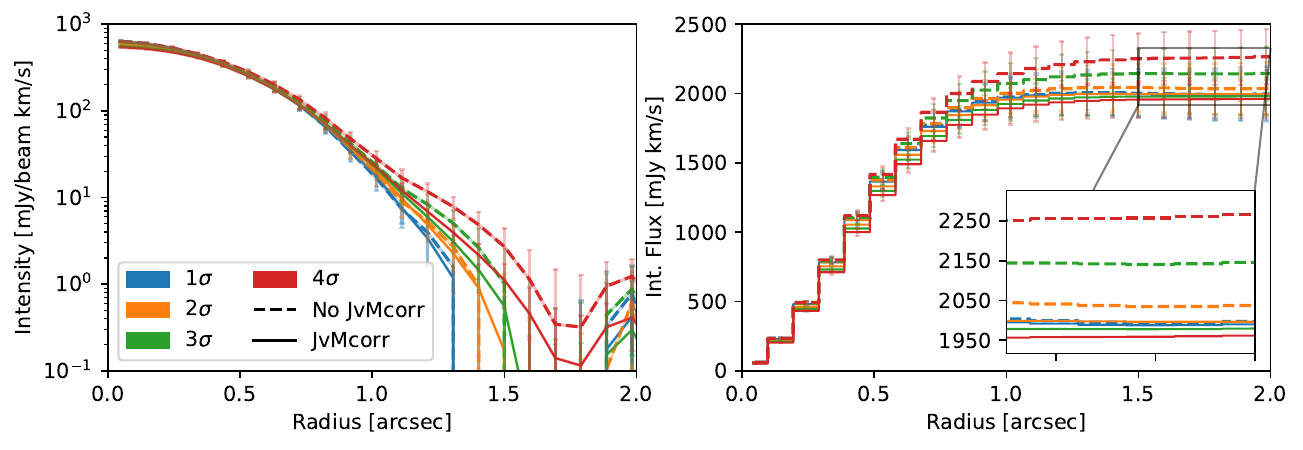}
\vspace{-0.2cm}
\caption{ Left: Radial intensity profile of the $^{12}$CO molecular emission of Upp Sco 8 when cleaning to 1, 2, 3, and 4$\sigma$ (see color code). The solid and dashed lines represent cases where the JvM correction is applied and not applied, respectively. Right:  Curve of growth of the CO integrated flux, obtained from the radial profiles in the left panel. The inset panel shows the convergence of the curve of growth for all cases. Error bars in the inset panel are omitted for clarity. \label{fig:JvM_comparison} 
}
\vspace{-0.cm}
\end{figure*}

For AGE-PRO image products, we aim to have both accurate integrated line fluxes and radial profiles. As discussed in \S\ref{subsec:JvM}, JvM correction improves the accuracy of integrated line fluxes but might cause bias to the radial profiles when the CLEAN is not sufficiently deep, and also make final images appear to have higher signal-to-noise ratio than the real value. To test the effects of JvM correction on AGE-PRO data, we compared line radial profiles and integrated fluxes after cleaning down to 1, 2, 3, and 4\,$\sigma$, with and without the JvM correction. The test results of Upp Sco 8 are shown in Figure~\ref{fig:JvM_comparison}.
The left panel shows the comparison of radial intensity profiles. The results show that when cleaning to 4$\sigma$ level, the intensity profile without JvM is significantly higher than the one with JvM correction. In contrast, when cleaning to 1-2$\sigma$ level, the radial profiles with or without JvM converge and these profiles have lower intensity at larger radii compared with 4$\sigma$ one with JvM. On the right panel, we show the flux difference from the curve of growth method. When cleaning only to 4$\sigma$ level, the non-JvM corrected flux is clearly overestimated and the JvM correction would be necessary. When cleaning down to 1-2$\sigma$ level, the integrated fluxes are similar (and consistent within the error bars) with and without the JvM correction. These results show that when cleaning deeply, to 1-2$\sigma$ level, the JvM correction does not improve radial profiles and flux constraints. Considering that JvM correction also introduces a false boost to the signal-to-noise ratio, we adopted cleaning the AGE-PRO molecular line cubes down to 1\,$\sigma$ without applying the JvM correction.

\section{Median disk properties for four age groups \label{app:age_groups}}

\begin{figure*}[htbp]
\centering
\vspace{-0.cm}
\includegraphics[width=0.9\textwidth]{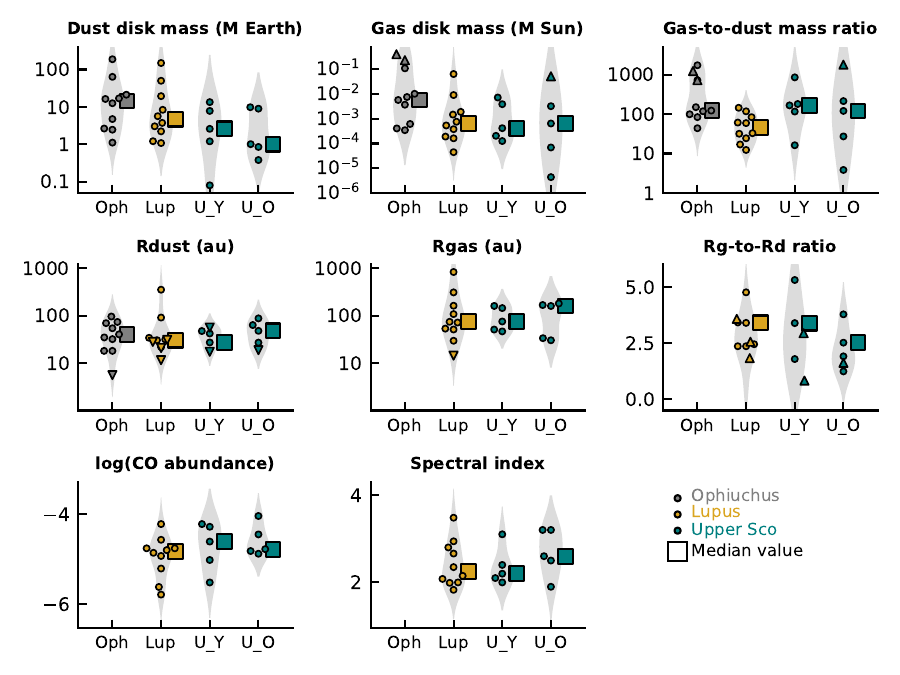}
\vspace{-0.2cm}
\caption{Comparison of disk properties in the Ophiuchus (0.5-1\,Myr), Lupus (1-3\,Myr), young Upper Sco group (2-3\,Myr), and old Upper Sco group (4-6\,Myr). This is the same as Figure~\ref{fig:disk_properties}, but with the Upper Sco sample divided into two groups that show different isochronal ages (see Figure~\ref{fig:age_distribution}). The circles are individual values and the squares are median values, further details in caption of Figure~\ref{fig:disk_properties}.  
\label{fig:4group_disk_properties} 
}
\vspace{-0.cm}
\end{figure*}

As discussed in \S\ref{sec:age_discussion}, the Upper Sco sample seems to be clustered around two different ages, with a younger group (2-3\,Myr) and an older group (4-6\,Myr). In Figure~\ref{fig:4group_disk_properties}, we show the individual disk properties and median disk values when the Upper Sco sample is further divided into two groups. Figure~\ref{fig:4group_disk_properties} shows that the general trends discussed in the main text still hold when four age groups are considered. 

\bibliography{lib2023}{}

\begin{thebibliography}{}
\expandafter\ifx\csname natexlab\endcsname\relax\def\natexlab#1{#1}\fi
\providecommand{\url}[1]{\href{#1}{#1}}
\providecommand{\dodoi}[1]{doi:~\href{http://doi.org/#1}{\nolinkurl{#1}}}
\providecommand{\doeprint}[1]{\href{http://ascl.net/#1}{\nolinkurl{http://ascl.net/#1}}}
\providecommand{\doarXiv}[1]{\href{https://arxiv.org/abs/#1}{\nolinkurl{https://arxiv.org/abs/#1}}}

\bibitem[{{Agurto-Gangas} {et~al.}(2025){Agurto-Gangas}, {P{\'e}rez}, {Sierra},
  {Miley}, {Zhang}, {Pascucci}, {Pinilla}, {Deng}, {Carpenter}, {Trapman},
  {Vioque}, {Rosotti}, {Kurtovic}, {Cieza}, {Anania}, {Tabone}, {Schwarz},
  {Hogerheijde}, {TorresVillanueva}, {Ruiz-Rodriguez}, \&
  {Gonz{\'a}lez-Ruilova}}]{AGEPRO_IV_UpperSco}
{Agurto-Gangas}, C., {P{\'e}rez}, L.~M., {Sierra}, A., {et~al.} 2025, \apj,
  989, 4, \dodoi{10.3847/1538-4357/adc7ab}

\bibitem[{{Aikawa} {et~al.}(2021){Aikawa}, {Cataldi}, {Yamato}, {Zhang},
  {Booth}, {Furuya}, {Andrews}, {Bae}, {Bergin}, {Bergner}, {Bosman},
  {Cleeves}, {Czekala}, {Guzm{\'a}n}, {Huang}, {Ilee}, {Law}, {Le Gal},
  {Loomis}, {M{\'e}nard}, {Nomura}, {{\"O}berg}, {Qi}, {Schwarz}, {Teague},
  {Tsukagoshi}, {Walsh}, \& {Wilner}}]{Aikawa2021_MAPS}
{Aikawa}, Y., {Cataldi}, G., {Yamato}, Y., {et~al.} 2021, \apjs, 257, 13,
  \dodoi{10.3847/1538-4365/ac143c}

\bibitem[{{Alcal{\'a}} {et~al.}(2014){Alcal{\'a}}, {Natta}, {Manara}, {Spezzi},
  {Stelzer}, {Frasca}, {Biazzo}, {Covino}, {Randich}, {Rigliaco}, {Testi},
  {Comer{\'o}n}, {Cupani}, \& {D'Elia}}]{Alcala14_Lupus}
{Alcal{\'a}}, J.~M., {Natta}, A., {Manara}, C.~F., {et~al.} 2014, \aap, 561,
  A2, \dodoi{10.1051/0004-6361/201322254}

\bibitem[{{Alcal{\'a}} {et~al.}(2017){Alcal{\'a}}, {Manara}, {Natta}, {Frasca},
  {Testi}, {Nisini}, {Stelzer}, {Williams}, {Antoniucci}, {Biazzo}, {Covino},
  {Esposito}, {Getman}, \& {Rigliaco}}]{Alcala17_Lupus}
{Alcal{\'a}}, J.~M., {Manara}, C.~F., {Natta}, A., {et~al.} 2017, \aap, 600,
  A20, \dodoi{10.1051/0004-6361/201629929}

\bibitem[{{ALMA Partnership} {et~al.}(2015){ALMA Partnership}, {Brogan},
  {P{\'e}rez}, {Hunter}, {Dent}, {Hales}, {Hills}, {Corder}, {Fomalont},
  {Vlahakis}, {Asaki}, {Barkats}, {Hirota}, {Hodge}, {Impellizzeri}, {Kneissl},
  {Liuzzo}, {Lucas}, {Marcelino}, {Matsushita}, {Nakanishi}, {Phillips},
  {Richards}, {Toledo}, {Aladro}, {Broguiere}, {Cortes}, {Cortes}, {Espada},
  {Galarza}, {Garcia-Appadoo}, {Guzman-Ramirez}, {Humphreys}, {Jung}, {Kameno},
  {Laing}, {Leon}, {Marconi}, {Mignano}, {Nikolic}, {Nyman}, {Radiszcz},
  {Remijan}, {Rod{\'o}n}, {Sawada}, {Takahashi}, {Tilanus}, {Vila Vilaro},
  {Watson}, {Wiklind}, {Akiyama}, {Chapillon}, {de Gregorio-Monsalvo}, {Di
  Francesco}, {Gueth}, {Kawamura}, {Lee}, {Nguyen Luong}, {Mangum}, {Pietu},
  {Sanhueza}, {Saigo}, {Takakuwa}, {Ubach}, {van Kempen}, {Wootten},
  {Castro-Carrizo}, {Francke}, {Gallardo}, {Garcia}, {Gonzalez}, {Hill},
  {Kaminski}, {Kurono}, {Liu}, {Lopez}, {Morales}, {Plarre}, {Schieven},
  {Testi}, {Videla}, {Villard}, {Andreani}, {Hibbard}, \& {Tatematsu}}]{alma15}
{ALMA Partnership}, {Brogan}, C.~L., {P{\'e}rez}, L.~M., {et~al.} 2015, \apjl,
  808, L3, \dodoi{10.1088/2041-8205/808/1/L3}

\bibitem[{{Anania} {et~al.}(2025){Anania}, {Rosotti}, {G{\'a}rate}, {Pinilla},
  {Vioque}, {Trapman}, {Carpenter}, {Zhang}, {Pascucci}, {Cieza}, {Sierra},
  {Kurtovic}, {Miley}, {P{\'e}rez}, {Tabone}, {Hogerheijde}, {Deng},
  {Agurto-Gangas}, {Ruiz-Rodriguez}, {Gonz{\'a}lez-Ruilova}, \&
  {TorresVillanueva}}]{AGEPRO_VIII_ext_photoevap}
{Anania}, R., {Rosotti}, G.~P., {G{\'a}rate}, M., {et~al.} 2025, \apj, 989, 8,
  \dodoi{10.3847/1538-4357/adb587}

\bibitem[{{Anderson} {et~al.}(2019){Anderson}, {Blake}, {Bergin}, {Zhang},
  {Carpenter}, {Schwarz}, {Huang}, \& {{\"O}berg}}]{Anderson19}
{Anderson}, D.~E., {Blake}, G.~A., {Bergin}, E.~A., {et~al.} 2019, \apj, 881,
  127, \dodoi{10.3847/1538-4357/ab2cb5}

\bibitem[{{Anderson} {et~al.}(2022){Anderson}, {Cleeves}, {Blake}, {Bergin},
  {Zhang}, {Carpenter}, \& {Schwarz}}]{Anderson22}
{Anderson}, D.~E., {Cleeves}, L.~I., {Blake}, G.~A., {et~al.} 2022, \apj, 927,
  229, \dodoi{10.3847/1538-4357/ac517e}

\bibitem[{{Andrews}(2020)}]{Andrews20_review}
{Andrews}, S.~M. 2020, \araa, 58, 483,
  \dodoi{10.1146/annurev-astro-031220-010302}

\bibitem[{{Andrews} {et~al.}(2013){Andrews}, {Rosenfeld}, {Kraus}, \&
  {Wilner}}]{Andrews2013}
{Andrews}, S.~M., {Rosenfeld}, K.~A., {Kraus}, A.~L., \& {Wilner}, D.~J. 2013,
  \apj, 771, 129, \dodoi{10.1088/0004-637X/771/2/129}

\bibitem[{{Andrews} {et~al.}(2018{\natexlab{a}}){Andrews}, {Terrell},
  {Tripathi}, {Ansdell}, {Williams}, \& {Wilner}}]{Andrews18a}
{Andrews}, S.~M., {Terrell}, M., {Tripathi}, A., {et~al.} 2018{\natexlab{a}},
  \apj, 865, 157, \dodoi{10.3847/1538-4357/aadd9f}

\bibitem[{{Andrews} {et~al.}(2016){Andrews}, {Wilner}, {Zhu}, {Birnstiel},
  {Carpenter}, {P{\'e}rez}, {Bai}, {{\"O}berg}, {Hughes}, {Isella}, \&
  {Ricci}}]{andrews16}
{Andrews}, S.~M., {Wilner}, D.~J., {Zhu}, Z., {et~al.} 2016, \apjl, 820, L40,
  \dodoi{10.3847/2041-8205/820/2/L40}

\bibitem[{{Andrews} {et~al.}(2018{\natexlab{b}}){Andrews}, {Huang},
  {P{\'e}rez}, {Isella}, {Dullemond}, {Kurtovic}, {Guzm{\'a}n}, {Carpenter},
  {Wilner}, {Zhang}, {Zhu}, {Birnstiel}, {Bai}, {Benisty}, {Hughes},
  {{\"O}berg}, \& {Ricci}}]{Andrews18b}
{Andrews}, S.~M., {Huang}, J., {P{\'e}rez}, L.~M., {et~al.} 2018{\natexlab{b}},
  \apj, 869, L41, \dodoi{10.3847/2041-8213/aaf741}

\bibitem[{{Ansdell} {et~al.}(2017){Ansdell}, {Williams}, {Manara}, {Miotello},
  {Facchini}, {van der Marel}, {Testi}, \& {van Dishoeck}}]{ansdell17}
{Ansdell}, M., {Williams}, J.~P., {Manara}, C.~F., {et~al.} 2017, \aj, 153,
  240, \dodoi{10.3847/1538-3881/aa69c0}

\bibitem[{{Ansdell} {et~al.}(2016){Ansdell}, {Williams}, {van der Marel},
  {Carpenter}, {Guidi}, {Hogerheijde}, {Mathews}, {Manara}, {Miotello},
  {Natta}, {Oliveira}, {Tazzari}, {Testi}, {van Dishoeck}, \& {van
  Terwisga}}]{ansdell16}
{Ansdell}, M., {Williams}, J.~P., {van der Marel}, N., {et~al.} 2016, \apj,
  828, 46, \dodoi{10.3847/0004-637X/828/1/46}

\bibitem[{{Ansdell} {et~al.}(2018){Ansdell}, {Williams}, {Trapman}, {van
  Terwisga}, {Facchini}, {Manara}, {van der Marel}, {Miotello}, {Tazzari},
  {Hogerheijde}, {Guidi}, {Testi}, \& {van Dishoeck}}]{ansdell18}
{Ansdell}, M., {Williams}, J.~P., {Trapman}, L., {et~al.} 2018, \apj, 859, 21,
  \dodoi{10.3847/1538-4357/aab890}

\bibitem[{{Aso} {et~al.}(2017){Aso}, {Ohashi}, {Aikawa}, {Machida}, {Saigo},
  {Saito}, {Takakuwa}, {Tomida}, {Tomisaka}, \& {Yen}}]{Aso2017}
{Aso}, Y., {Ohashi}, N., {Aikawa}, Y., {et~al.} 2017, \apj, 849, 56,
  \dodoi{10.3847/1538-4357/aa8264}

\bibitem[{{Astropy Collaboration} {et~al.}(2018){Astropy Collaboration},
  {Price-Whelan}, {Sip{\H{o}}cz}, {G{\"u}nther}, {Lim}, {Crawford}, {Conseil},
  {Shupe}, {Craig}, {Dencheva}, {Ginsburg}, {VanderPlas}, {Bradley},
  {P{\'e}rez-Su{\'a}rez}, {de Val-Borro}, {Aldcroft}, {Cruz}, {Robitaille},
  {Tollerud}, {Ardelean}, {Babej}, {Bach}, {Bachetti}, {Bakanov}, {Bamford},
  {Barentsen}, {Barmby}, {Baumbach}, {Berry}, {Biscani}, {Boquien}, {Bostroem},
  {Bouma}, {Brammer}, {Bray}, {Breytenbach}, {Buddelmeijer}, {Burke},
  {Calderone}, {Cano Rodr{\'\i}guez}, {Cara}, {Cardoso}, {Cheedella}, {Copin},
  {Corrales}, {Crichton}, {D'Avella}, {Deil}, {Depagne}, {Dietrich}, {Donath},
  {Droettboom}, {Earl}, {Erben}, {Fabbro}, {Ferreira}, {Finethy}, {Fox},
  {Garrison}, {Gibbons}, {Goldstein}, {Gommers}, {Greco}, {Greenfield},
  {Groener}, {Grollier}, {Hagen}, {Hirst}, {Homeier}, {Horton}, {Hosseinzadeh},
  {Hu}, {Hunkeler}, {Ivezi{\'c}}, {Jain}, {Jenness}, {Kanarek}, {Kendrew},
  {Kern}, {Kerzendorf}, {Khvalko}, {King}, {Kirkby}, {Kulkarni}, {Kumar},
  {Lee}, {Lenz}, {Littlefair}, {Ma}, {Macleod}, {Mastropietro}, {McCully},
  {Montagnac}, {Morris}, {Mueller}, {Mumford}, {Muna}, {Murphy}, {Nelson},
  {Nguyen}, {Ninan}, {N{\"o}the}, {Ogaz}, {Oh}, {Parejko}, {Parley}, {Pascual},
  {Patil}, {Patil}, {Plunkett}, {Prochaska}, {Rastogi}, {Reddy Janga},
  {Sabater}, {Sakurikar}, {Seifert}, {Sherbert}, {Sherwood-Taylor}, {Shih},
  {Sick}, {Silbiger}, {Singanamalla}, {Singer}, {Sladen}, {Sooley},
  {Sornarajah}, {Streicher}, {Teuben}, {Thomas}, {Tremblay}, {Turner},
  {Terr{\'o}n}, {van Kerkwijk}, {de la Vega}, {Watkins}, {Weaver}, {Whitmore},
  {Woillez}, {Zabalza}, \& {Astropy Contributors}}]{Astropy18}
{Astropy Collaboration}, {Price-Whelan}, A.~M., {Sip{\H{o}}cz}, B.~M., {et~al.}
  2018, \aj, 156, 123, \dodoi{10.3847/1538-3881/aabc4f}

\bibitem[{{Bai} \& {Stone}(2013)}]{bai13}
{Bai}, X.-N., \& {Stone}, J.~M. 2013, \apj, 769, 76,
  \dodoi{10.1088/0004-637X/769/1/76}

\bibitem[{{Bailer-Jones} {et~al.}(2021){Bailer-Jones}, {Rybizki}, {Fouesneau},
  {Demleitner}, \& {Andrae}}]{BailerJones2021}
{Bailer-Jones}, C.~A.~L., {Rybizki}, J., {Fouesneau}, M., {Demleitner}, M., \&
  {Andrae}, R. 2021, \aj, 161, 147, \dodoi{10.3847/1538-3881/abd806}

\bibitem[{{Baraffe} {et~al.}(2015){Baraffe}, {Homeier}, {Allard}, \&
  {Chabrier}}]{Baraffe15}
{Baraffe}, I., {Homeier}, D., {Allard}, F., \& {Chabrier}, G. 2015, \aap, 577,
  A42, \dodoi{10.1051/0004-6361/201425481}

\bibitem[{{Barenfeld} {et~al.}(2016){Barenfeld}, {Carpenter}, {Ricci}, \&
  {Isella}}]{barenfeld16}
{Barenfeld}, S.~A., {Carpenter}, J.~M., {Ricci}, L., \& {Isella}, A. 2016,
  \apj, 827, 142, \dodoi{10.3847/0004-637X/827/2/142}

\bibitem[{{Benz} {et~al.}(2014){Benz}, {Ida}, {Alibert}, {Lin}, \&
  {Mordasini}}]{benz14}
{Benz}, W., {Ida}, S., {Alibert}, Y., {Lin}, D., \& {Mordasini}, C. 2014,
  Protostars and Planets VI, 691,
  \dodoi{10.2458/azu_uapress_9780816531240-ch030}

\bibitem[{{Bergin} \& {Williams}(2017)}]{bergin17}
{Bergin}, E.~A., \& {Williams}, J.~P. 2017, {The Determination of
  Protoplanetary Disk Masses}, Vol. 445 (Astrophysics and Space Science
  Library), 1, \dodoi{10.1007/978-3-319-60609-5\_1}

\bibitem[{{Bergin} {et~al.}(2013){Bergin}, {Cleeves}, {Gorti}, {Zhang},
  {Blake}, {Green}, {Andrews}, {Evans}, {Henning}, {{\"O}berg}, {Pontoppidan},
  {Qi}, {Salyk}, \& {van Dishoeck}}]{bergin13}
{Bergin}, E.~A., {Cleeves}, L.~I., {Gorti}, U., {et~al.} 2013, \nat, 493, 644,
  \dodoi{10.1038/nature11805}

\bibitem[{{Bergner} {et~al.}(2020){Bergner}, {{\"O}berg}, {Bergin}, {Andrews},
  {Blake}, {Carpenter}, {Cleeves}, {Guzm{\'a}n}, {Huang}, {J{\o}rgensen}, {Qi},
  {Schwarz}, {Williams}, \& {Wilner}}]{Bergner_20evolution}
{Bergner}, J.~B., {{\"O}berg}, K.~I., {Bergin}, E.~A., {et~al.} 2020, \apj,
  898, 97, \dodoi{10.3847/1538-4357/ab9e71}

\bibitem[{{B{\'e}thune} {et~al.}(2017){B{\'e}thune}, {Lesur}, \&
  {Ferreira}}]{Bethune17}
{B{\'e}thune}, W., {Lesur}, G., \& {Ferreira}, J. 2017, \aap, 600, A75,
  \dodoi{10.1051/0004-6361/201630056}

\bibitem[{{Birnstiel}(2023)}]{Birnstiel23_dustreview}
{Birnstiel}, T. 2023, arXiv e-prints, arXiv:2312.13287,
  \dodoi{10.48550/arXiv.2312.13287}

\bibitem[{{Birnstiel} {et~al.}(2010){Birnstiel}, {Dullemond}, \&
  {Brauer}}]{Birnstiel2010}
{Birnstiel}, T., {Dullemond}, C.~P., \& {Brauer}, F. 2010, \aap, 513, A79,
  \dodoi{10.1051/0004-6361/200913731}

\bibitem[{{Blandford} \& {Payne}(1982)}]{Blandford82}
{Blandford}, R.~D., \& {Payne}, D.~G. 1982, \mnras, 199, 883,
  \dodoi{10.1093/mnras/199.4.883}

\bibitem[{{Bohlin} {et~al.}(1978){Bohlin}, {Savage}, \& {Drake}}]{Bohlin78}
{Bohlin}, R.~C., {Savage}, B.~D., \& {Drake}, J.~F. 1978, \apj, 224, 132,
  \dodoi{10.1086/156357}

\bibitem[{{Bosman} {et~al.}(2018){Bosman}, {Walsh}, \& {van
  Dishoeck}}]{Bosman18}
{Bosman}, A.~D., {Walsh}, C., \& {van Dishoeck}, E.~F. 2018, \aap, 618, A182,
  \dodoi{10.1051/0004-6361/201833497}

\bibitem[{{Calahan} {et~al.}(2021){Calahan}, {Bergin}, {Zhang}, {Teague},
  {Cleeves}, {Bergner}, {Blake}, {Cazzoletti}, {Guzm{\'a}n}, {Hogerheijde},
  {Huang}, {Kama}, {Loomis}, {{\"O}berg}, {Qi}, {van Dishoeck}, {Terwisscha van
  Scheltinga}, {Walsh}, \& {Wilner}}]{Calahan21_twhya}
{Calahan}, J.~K., {Bergin}, E., {Zhang}, K., {et~al.} 2021, \apj, 908, 8,
  \dodoi{10.3847/1538-4357/abd255}

\bibitem[{{Carpenter} {et~al.}(2025){Carpenter}, {Esplin}, {Luhman}, {Mamajek},
  \& {Andrews}}]{Carpenter2025}
{Carpenter}, J.~M., {Esplin}, T.~L., {Luhman}, K.~L., {Mamajek}, E.~E., \&
  {Andrews}, S.~M. 2025, \apj, 978, 117, \dodoi{10.3847/1538-4357/ad8ebc}

\bibitem[{{CASA Team} {et~al.}(2022){CASA Team}, {Bean}, {Bhatnagar}, {Castro},
  {Donovan Meyer}, {Emonts}, {Garcia}, {Garwood}, {Golap}, {Gonzalez Villalba},
  {Harris}, {Hayashi}, {Hoskins}, {Hsieh}, {Jagannathan}, {Kawasaki},
  {Keimpema}, {Kettenis}, {Lopez}, {Marvil}, {Masters}, {McNichols},
  {Mehringer}, {Miel}, {Moellenbrock}, {Montesino}, {Nakazato}, {Ott}, {Petry},
  {Pokorny}, {Raba}, {Rau}, {Schiebel}, {Schweighart}, {Sekhar}, {Shimada},
  {Small}, {Steeb}, {Sugimoto}, {Suoranta}, {Tsutsumi}, {van Bemmel},
  {Verkouter}, {Wells}, {Xiong}, {Szomoru}, {Griffith}, {Glendenning}, \&
  {Kern}}]{CASA2022}
{CASA Team}, {Bean}, B., {Bhatnagar}, S., {et~al.} 2022, \pasp, 134, 114501,
  \dodoi{10.1088/1538-3873/ac9642}

\bibitem[{{Casassus} \& {C{\'a}rcamo}(2022)}]{Casassus22}
{Casassus}, S., \& {C{\'a}rcamo}, M. 2022, \mnras, 513, 5790,
  \dodoi{10.1093/mnras/stac1285}

\bibitem[{{Cauley} {et~al.}(2021){Cauley}, {France}, {Herzceg}, \&
  {Johns-Krull}}]{Cauley2021}
{Cauley}, P.~W., {France}, K., {Herzceg}, G.~J., \& {Johns-Krull}, C.~M. 2021,
  \aj, 161, 217, \dodoi{10.3847/1538-3881/abea21}

\bibitem[{{Cieza} {et~al.}(2019){Cieza}, {Ru{\'\i}z-Rodr{\'\i}guez}, {Hales},
  {Casassus}, {P{\'e}rez}, {Gonzalez-Ruilova}, {C{\'a}novas}, {Williams},
  {Zurlo}, {Ansdell}, {Avenhaus}, {Bayo}, {Bertrang}, {Christiaens}, {Dent},
  {Ferrero}, {Gamen}, {Olofsson}, {Orcajo}, {Pe{\~n}a Ram{\'\i}rez},
  {Principe}, {Schreiber}, \& {van der Plas}}]{Cieza19_ODISEA_I}
{Cieza}, L.~A., {Ru{\'\i}z-Rodr{\'\i}guez}, D., {Hales}, A., {et~al.} 2019,
  \mnras, 482, 698, \dodoi{10.1093/mnras/sty2653}

\bibitem[{{Cieza} {et~al.}(2021){Cieza}, {Gonz{\'a}lez-Ruilova}, {Hales},
  {Pinilla}, {Ru{\'\i}z-Rodr{\'\i}guez}, {Zurlo}, {Casassus}, {P{\'e}rez},
  {C{\'a}novas}, {Arce-Tord}, {Flock}, {Kurtovic}, {Marino}, {Nogueira},
  {Perez}, {Price}, {Principe}, \& {Williams}}]{Cieza21_ODISEAIII}
{Cieza}, L.~A., {Gonz{\'a}lez-Ruilova}, C., {Hales}, A.~S., {et~al.} 2021,
  \mnras, 501, 2934, \dodoi{10.1093/mnras/staa3787}

\bibitem[{{Cleeves} {et~al.}(2013){Cleeves}, {Adams}, \& {Bergin}}]{cleeves13a}
{Cleeves}, L.~I., {Adams}, F.~C., \& {Bergin}, E.~A. 2013, \apj, 772, 5,
  \dodoi{10.1088/0004-637X/772/1/5}

\bibitem[{{Cleeves} {et~al.}(2015){Cleeves}, {Bergin}, {Qi}, {Adams}, \&
  {{\"O}berg}}]{cleeves15}
{Cleeves}, L.~I., {Bergin}, E.~A., {Qi}, C., {Adams}, F.~C., \& {{\"O}berg},
  K.~I. 2015, \apj, 799, 204, \dodoi{10.1088/0004-637X/799/2/204}

\bibitem[{{Cleeves} {et~al.}(2018){Cleeves}, {{\"O}berg}, {Wilner}, {Huang},
  {Loomis}, {Andrews}, \& {Guzman}}]{Cleeves18}
{Cleeves}, L.~I., {{\"O}berg}, K.~I., {Wilner}, D.~J., {et~al.} 2018, \apj,
  865, 155, \dodoi{10.3847/1538-4357/aade96}

\bibitem[{{Comer{\'o}n}(2008)}]{Comeron2008}
{Comer{\'o}n}, F. 2008, in Handbook of Star Forming Regions, Volume II, ed.
  B.~{Reipurth}, Vol.~5, 295

\bibitem[{{Cuello} {et~al.}(2023){Cuello}, {M{\'e}nard}, \&
  {Price}}]{Cuello2023}
{Cuello}, N., {M{\'e}nard}, F., \& {Price}, D.~J. 2023, European Physical
  Journal Plus, 138, 11, \dodoi{10.1140/epjp/s13360-022-03602-w}

\bibitem[{{Czekala} {et~al.}(2021){Czekala}, {Loomis}, {Teague}, {Booth},
  {Huang}, {Cataldi}, {Ilee}, {Law}, {Walsh}, {Bosman}, {Guzm{\'a}n}, {Le Gal},
  {{\"O}berg}, {Yamato}, {Aikawa}, {Andrews}, {Bae}, {Bergin}, {Bergner},
  {Cleeves}, {Kurtovic}, {M{\'e}nard}, {Nomura}, {P{\'e}rez}, {Qi}, {Schwarz},
  {Tsukagoshi}, {Waggoner}, {Wilner}, \& {Zhang}}]{Czekala21_MAPS}
{Czekala}, I., {Loomis}, R.~A., {Teague}, R., {et~al.} 2021, \apjs, 257, 2,
  \dodoi{10.3847/1538-4365/ac1430}

\bibitem[{Davidson-Pilon(2024)}]{lifelines_davidson_pilon_2024}
Davidson-Pilon, C. 2024, lifelines, survival analysis in Python, v0.29.0,
  Zenodo, \dodoi{10.5281/zenodo.12549337}

\bibitem[{{de Geus} {et~al.}(1989){de Geus}, {de Zeeuw}, \& {Lub}}]{deGeus1989}
{de Geus}, E.~J., {de Zeeuw}, P.~T., \& {Lub}, J. 1989, \aap, 216, 44

\bibitem[{{de Valon} {et~al.}(2020){de Valon}, {Dougados}, {Cabrit}, {Louvet},
  {Zapata}, \& {Mardones}}]{deValon20}
{de Valon}, A., {Dougados}, C., {Cabrit}, S., {et~al.} 2020, \aap, 634, L12,
  \dodoi{10.1051/0004-6361/201936950}

\bibitem[{{Delage} {et~al.}(2023){Delage}, {G{\'a}rate}, {Okuzumi}, {Yang},
  {Pinilla}, {Flock}, {Stammler}, \& {Birnstiel}}]{Delage23}
{Delage}, T.~N., {G{\'a}rate}, M., {Okuzumi}, S., {et~al.} 2023, \aap, 674,
  A190, \dodoi{10.1051/0004-6361/202244731}

\bibitem[{{Delage} {et~al.}(2022){Delage}, {Okuzumi}, {Flock}, {Pinilla}, \&
  {Dzyurkevich}}]{Delage22}
{Delage}, T.~N., {Okuzumi}, S., {Flock}, M., {Pinilla}, P., \& {Dzyurkevich},
  N. 2022, \aap, 658, A97, \dodoi{10.1051/0004-6361/202141689}

\bibitem[{Deng {et~al.}(2025)Deng, Pascucci, \&
  Fernandes}]{Deng_2025_ysoisochrone}
Deng, D., Pascucci, I., \& Fernandes, R.~B. 2025, Journal of Open Source
  Software, 10, 7493, \dodoi{10.21105/joss.07493}

\bibitem[{{Deng} {et~al.}(2023){Deng}, {Ruaud}, {Gorti}, \&
  {Pascucci}}]{Deng2023}
{Deng}, D., {Ruaud}, M., {Gorti}, U., \& {Pascucci}, I. 2023, \apj, 954, 165,
  \dodoi{10.3847/1538-4357/acdfcc}

\bibitem[{{Deng} {et~al.}(2025){Deng}, {Vioque}, {Pascucci}, {P{\'e}rez},
  {Zhang}, {Kurtovic}, {Trapman}, {TorresVillanueva}, {Agurto-Gangas},
  {Carpenter}, {Pinilla}, {Gorti}, {Tabone}, {Sierra}, {Rosotti}, {Cieza},
  {Anania}, {Gonz{\'a}lez-Ruilova}, {Hogerheijde}, {Miley}, {Ruiz-Rodriguez},
  {Ruaud}, \& {Schwarz}}]{AGEPRO_III_Lupus}
{Deng}, D., {Vioque}, M., {Pascucci}, I., {et~al.} 2025, \apj, 989, 3,
  \dodoi{10.3847/1538-4357/add43a}

\bibitem[{{Drazkowska} {et~al.}(2023){Drazkowska}, {Bitsch}, {Lambrechts},
  {Mulders}, {Harsono}, {Vazan}, {Liu}, {Ormel}, {Kretke}, \&
  {Morbidelli}}]{Drakazkowska_PPVII}
{Drazkowska}, J., {Bitsch}, B., {Lambrechts}, M., {et~al.} 2023, in
  Astronomical Society of the Pacific Conference Series, Vol. 534, Protostars
  and Planets VII, ed. S.~{Inutsuka}, Y.~{Aikawa}, T.~{Muto}, K.~{Tomida}, \&
  M.~{Tamura}, 717, \dodoi{10.48550/arXiv.2203.09759}

\bibitem[{{Ediss} {et~al.}(2004){Ediss}, {Carter}, {Cheng}, {Effland},
  {Grammer}, {Horner}, {Kerr}, {Koller}, {Lauria}, {Morris}, {Pan}, {Reiland},
  \& {Sullivan}}]{ALMA_B6_receiver}
{Ediss}, G.~A., {Carter}, M., {Cheng}, J., {et~al.} 2004, in Fifteenth
  International Symposium on Space Terahertz Technology, ed. G.~{Narayanan},
  181--188

\bibitem[{{Enoch} {et~al.}(2009){Enoch}, {Evans}, {Sargent}, \&
  {Glenn}}]{Enoch09}
{Enoch}, M.~L., {Evans}, Neal~J., I., {Sargent}, A.~I., \& {Glenn}, J. 2009,
  \apj, 692, 973, \dodoi{10.1088/0004-637X/692/2/973}

\bibitem[{{Evans} {et~al.}(2009){Evans}, {Dunham}, {J{\o}rgensen}, {Enoch},
  {Mer{\'\i}n}, {van Dishoeck}, {Alcal{\'a}}, {Myers}, {Stapelfeldt}, {Huard},
  {Allen}, {Harvey}, {van Kempen}, {Blake}, {Koerner}, {Mundy}, {Padgett}, \&
  {Sargent}}]{Evans09}
{Evans}, Neal~J., I., {Dunham}, M.~M., {J{\o}rgensen}, J.~K., {et~al.} 2009,
  \apjs, 181, 321, \dodoi{10.1088/0067-0049/181/2/321}

\bibitem[{{Fang} {et~al.}(2023){Fang}, {Pascucci}, {Edwards}, {Gorti},
  {Hillenbrand}, \& {Carpenter}}]{Fang2023}
{Fang}, M., {Pascucci}, I., {Edwards}, S., {et~al.} 2023, \apj, 945, 112,
  \dodoi{10.3847/1538-4357/acb2c9}

\bibitem[{{Favre} {et~al.}(2013){Favre}, {Cleeves}, {Bergin}, {Qi}, \&
  {Blake}}]{favre13}
{Favre}, C., {Cleeves}, L.~I., {Bergin}, E.~A., {Qi}, C., \& {Blake}, G.~A.
  2013, \apjl, 776, L38, \dodoi{10.1088/2041-8205/776/2/L38}

\bibitem[{{Fedele} {et~al.}(2010){Fedele}, {van den Ancker}, {Henning},
  {Jayawardhana}, \& {Oliveira}}]{Fedele10}
{Fedele}, D., {van den Ancker}, M.~E., {Henning}, T., {Jayawardhana}, R., \&
  {Oliveira}, J.~M. 2010, \aap, 510, A72, \dodoi{10.1051/0004-6361/200912810}

\bibitem[{{Feiden}(2016)}]{Feiden16}
{Feiden}, G.~A. 2016, \aap, 593, A99, \dodoi{10.1051/0004-6361/201527613}

\bibitem[{{Flaherty} {et~al.}(2020){Flaherty}, {Hughes}, {Simon}, {Qi}, {Bai},
  {Bulatek}, {Andrews}, {Wilner}, \& {K{\'o}sp{\'a}l}}]{Flaherty20}
{Flaherty}, K., {Hughes}, A.~M., {Simon}, J.~B., {et~al.} 2020, \apj, 895, 109,
  \dodoi{10.3847/1538-4357/ab8cc5}

\bibitem[{{Flaherty} {et~al.}(2015){Flaherty}, {Hughes}, {Rosenfeld},
  {Andrews}, {Chiang}, {Simon}, {Kerzner}, \& {Wilner}}]{flaherty15}
{Flaherty}, K.~M., {Hughes}, A.~M., {Rosenfeld}, K.~A., {et~al.} 2015, \apj,
  813, 99, \dodoi{10.1088/0004-637X/813/2/99}

\bibitem[{{Flores} {et~al.}(2023){Flores}, {Ohashi}, {Tobin}, {J{\o}rgensen},
  {Takakuwa}, {Li}, {Lin}, {van't Hoff}, {Plunkett}, {Yamato}, {Sai (Insa
  Choi)}, {Koch}, {Yen}, {Aikawa}, {Aso}, {de Gregorio-Monsalvo}, {Kido},
  {Kwon}, {Lee}, {Lee}, {Looney}, {Santamar{\'\i}a-Miranda}, {Sharma},
  {Thieme}, {Williams}, {Han}, {Narayanan}, \& {Lai}}]{Flores2023}
{Flores}, C., {Ohashi}, N., {Tobin}, J.~J., {et~al.} 2023, \apj, 958, 98,
  \dodoi{10.3847/1538-4357/acf7c1}

\bibitem[{{Galli} {et~al.}(2020){Galli}, {Bouy}, {Olivares}, {Miret-Roig},
  {Vieira}, {Sarro}, {Barrado}, {Berihuete}, {Bertout}, {Bertin}, \&
  {Cuillandre}}]{Galli2020}
{Galli}, P.~A.~B., {Bouy}, H., {Olivares}, J., {et~al.} 2020, \aap, 643, A148,
  \dodoi{10.1051/0004-6361/202038717}

\bibitem[{{Gressel} {et~al.}(2015){Gressel}, {Turner}, {Nelson}, \&
  {McNally}}]{Gressel15}
{Gressel}, O., {Turner}, N.~J., {Nelson}, R.~P., \& {McNally}, C.~P. 2015,
  \apj, 801, 84, \dodoi{10.1088/0004-637X/801/2/84}

\bibitem[{{Haisch} {et~al.}(2001){Haisch}, {Lada}, \& {Lada}}]{Haisch01}
{Haisch}, Karl~E., J., {Lada}, E.~A., \& {Lada}, C.~J. 2001, \apjl, 553, L153,
  \dodoi{10.1086/320685}

\bibitem[{{Harrison} {et~al.}(2021){Harrison}, {Looney}, {Stephens}, {Li},
  {Teague}, {Crutcher}, {Yang}, {Cox}, {Fern{\'a}ndez-L{\'o}pez}, \&
  {Shinnaga}}]{Harrison2021}
{Harrison}, R.~E., {Looney}, L.~W., {Stephens}, I.~W., {et~al.} 2021, \apj,
  908, 141, \dodoi{10.3847/1538-4357/abd94e}

\bibitem[{{Harsono} {et~al.}(2018){Harsono}, {Bjerkeli}, {van der Wiel},
  {Ramsey}, {Maud}, {Kristensen}, \& {J{\o}rgensen}}]{Harsono18}
{Harsono}, D., {Bjerkeli}, P., {van der Wiel}, M. H.~D., {et~al.} 2018, Nature
  Astronomy, 2, 646, \dodoi{10.1038/s41550-018-0497-x}

\bibitem[{{Hendler} {et~al.}(2020){Hendler}, {Pascucci}, {Pinilla}, {Tazzari},
  {Carpenter}, {Malhotra}, \& {Testi}}]{Hendler20}
{Hendler}, N., {Pascucci}, I., {Pinilla}, P., {et~al.} 2020, \apj, 895, 126,
  \dodoi{10.3847/1538-4357/ab70ba}

\bibitem[{{Hildebrand}(1983)}]{Hildebrand1983}
{Hildebrand}, R.~H. 1983, \qjras, 24, 267

\bibitem[{{Hsieh} {et~al.}(2024){Hsieh}, {Arce}, {Maureira}, {Pineda},
  {Segura-Cox}, {Mardones}, {Dunham}, \& {Arun}}]{Hsieh_2024_CAMPOS}
{Hsieh}, C.-H., {Arce}, H.~G., {Maureira}, M.~J., {et~al.} 2024, \apj, 973,
  138, \dodoi{10.3847/1538-4357/ad6152}

\bibitem[{{Hughes} {et~al.}(2018){Hughes}, {Duch{\^e}ne}, \&
  {Matthews}}]{Hughes18_debrisdisks}
{Hughes}, A.~M., {Duch{\^e}ne}, G., \& {Matthews}, B.~C. 2018, \araa, 56, 541,
  \dodoi{10.1146/annurev-astro-081817-052035}

\bibitem[{{Hunter}(2007)}]{Hunter2007}
{Hunter}, J.~D. 2007, Computing in Science and Engineering, 9, 90,
  \dodoi{10.1109/MCSE.2007.55}

\bibitem[{{Johansen} {et~al.}(2014){Johansen}, {Blum}, {Tanaka}, {Ormel},
  {Bizzarro}, \& {Rickman}}]{johansen14}
{Johansen}, A., {Blum}, J., {Tanaka}, H., {et~al.} 2014, Protostars and Planets
  VI, 547, \dodoi{10.2458/azu_uapress_9780816531240-ch024}

\bibitem[{{Jorsater} \& {van Moorsel}(1995)}]{JvM95}
{Jorsater}, S., \& {van Moorsel}, G.~A. 1995, \aj, 110, 2037,
  \dodoi{10.1086/117668}

\bibitem[{{Kadam} {et~al.}(2025){Kadam}, {Vorobyov}, {Woitke}, {Basu}, \& {van
  Terwisga}}]{Kadam2025}
{Kadam}, K., {Vorobyov}, E., {Woitke}, P., {Basu}, S., \& {van Terwisga}, S.
  2025, arXiv e-prints, arXiv:2502.00161, \dodoi{10.48550/arXiv.2502.00161}

\bibitem[{{Kama} {et~al.}(2016){Kama}, {Bruderer}, {van Dishoeck},
  {Hogerheijde}, {Folsom}, {Miotello}, {Fedele}, {Belloche}, {G{\"u}sten}, \&
  {Wyrowski}}]{kama16}
{Kama}, M., {Bruderer}, S., {van Dishoeck}, E.~F., {et~al.} 2016, \aap, 592,
  A83, \dodoi{10.1051/0004-6361/201526991}

\bibitem[{{Kama} {et~al.}(2020){Kama}, {Trapman}, {Fedele}, {Bruderer},
  {Hogerheijde}, {Miotello}, {van Dishoeck}, {Clarke}, \& {Bergin}}]{Kama20}
{Kama}, M., {Trapman}, L., {Fedele}, D., {et~al.} 2020, \aap, 634, A88,
  \dodoi{10.1051/0004-6361/201937124}

\bibitem[{{Kennedy} \& {Kenyon}(2008)}]{kennedy08}
{Kennedy}, G.~M., \& {Kenyon}, S.~J. 2008, \apj, 673, 502,
  \dodoi{10.1086/524130}

\bibitem[{{Krijt} {et~al.}(2020){Krijt}, {Bosman}, {Zhang}, {Schwarz},
  {Ciesla}, \& {Bergin}}]{Krijt20}
{Krijt}, S., {Bosman}, A.~D., {Zhang}, K., {et~al.} 2020, \apj, 899, 134,
  \dodoi{10.3847/1538-4357/aba75d}

\bibitem[{{Krijt} {et~al.}(2023){Krijt}, {Kama}, {McClure}, {Teske}, {Bergin},
  {Shorttle}, {Walsh}, \& {Raymond}}]{Krijt_PPVII}
{Krijt}, S., {Kama}, M., {McClure}, M., {et~al.} 2023, in Astronomical Society
  of the Pacific Conference Series, Vol. 534, Protostars and Planets VII, ed.
  S.~{Inutsuka}, Y.~{Aikawa}, T.~{Muto}, K.~{Tomida}, \& M.~{Tamura}, 1031,
  \dodoi{10.48550/arXiv.2203.10056}

\bibitem[{{Krijt} {et~al.}(2018){Krijt}, {Schwarz}, {Bergin}, \&
  {Ciesla}}]{Krijt18}
{Krijt}, S., {Schwarz}, K.~R., {Bergin}, E.~A., \& {Ciesla}, F.~J. 2018, \apj,
  864, 78, \dodoi{10.3847/1538-4357/aad69b}

\bibitem[{{Kurtovic} {et~al.}(2025){Kurtovic}, {G{\'a}rate}, {Pinilla},
  {Zhang}, {Rosotti}, {Anania}, {Pascucci}, {Tabone}, {Trapman}, {Deng},
  {Vioque}, {Carpenter}, {Cieza}, {P{\'e}rez}, {Agurto-Gangas}, {Sierra},
  {Ruiz-Rodriguez}, {Miley}, {Gonz{\'a}lez-Ruilova}, {Torres-Villanueva}, \&
  {Kuznetsova}}]{AGEPRO_VI_dustevolution}
{Kurtovic}, N.~T., {G{\'a}rate}, M., {Pinilla}, P., {et~al.} 2025, \apj, 989,
  6, \dodoi{10.3847/1538-4357/add1d0}

\bibitem[{{Law} {et~al.}(2021){Law}, {Loomis}, {Teague}, {{\"O}berg},
  {Czekala}, {Andrews}, {Huang}, {Aikawa}, {Alarc{\'o}n}, {Bae}, {Bergin},
  {Bergner}, {Boehler}, {Booth}, {Bosman}, {Calahan}, {Cataldi}, {Cleeves},
  {Furuya}, {Guzm{\'a}n}, {Ilee}, {Le Gal}, {Liu}, {Long}, {M{\'e}nard},
  {Nomura}, {Qi}, {Schwarz}, {Sierra}, {Tsukagoshi}, {Yamato}, {van't Hoff},
  {Walsh}, {Wilner}, \& {Zhang}}]{law_maps_radial}
{Law}, C.~J., {Loomis}, R.~A., {Teague}, R., {et~al.} 2021, \apjs, 257, 3,
  \dodoi{10.3847/1538-4365/ac1434}

\bibitem[{{Lee} \& {Chiang}(2016)}]{Lee2016}
{Lee}, E.~J., \& {Chiang}, E. 2016, \apj, 817, 90,
  \dodoi{10.3847/0004-637X/817/2/90}

\bibitem[{{Lesur} {et~al.}(2023){Lesur}, {Flock}, {Ercolano}, {Lin}, {Yang},
  {Barranco}, {Benitez-Llambay}, {Goodman}, {Johansen}, {Klahr}, {Laibe},
  {Lyra}, {Marcus}, {Nelson}, {Squire}, {Simon}, {Turner}, {Umurhan}, \&
  {Youdin}}]{Lesur_PPVII}
{Lesur}, G., {Flock}, M., {Ercolano}, B., {et~al.} 2023, in Astronomical
  Society of the Pacific Conference Series, Vol. 534, Protostars and Planets
  VII, ed. S.~{Inutsuka}, Y.~{Aikawa}, T.~{Muto}, K.~{Tomida}, \& M.~{Tamura},
  465

\bibitem[{{Long} {et~al.}(2018){Long}, {Pinilla}, {Herczeg}, {Harsono},
  {Dipierro}, {Pascucci}, {Hendler}, {Tazzari}, {Ragusa}, {Salyk}, {Edwards},
  {Lodato}, {van de Plas}, {Johnstone}, {Liu}, {Boehler}, {Cabrit}, {Manara},
  {Menard}, {Mulders}, {Nisini}, {Fischer}, {Rigliaco}, {Banzatti}, {Avenhaus},
  \& {Gully-Santiago}}]{long18}
{Long}, F., {Pinilla}, P., {Herczeg}, G.~J., {et~al.} 2018, \apj, 869, 17,
  \dodoi{10.3847/1538-4357/aae8e1}

\bibitem[{{Long} {et~al.}(2022){Long}, {Andrews}, {Rosotti}, {Harsono},
  {Pinilla}, {Wilner}, {{\"O}berg}, {Teague}, {Trapman}, \&
  {Tabone}}]{Long22_size}
{Long}, F., {Andrews}, S.~M., {Rosotti}, G., {et~al.} 2022, \apj, 931, 6,
  \dodoi{10.3847/1538-4357/ac634e}

\bibitem[{{Louvet} {et~al.}(2018){Louvet}, {Dougados}, {Cabrit}, {Mardones},
  {M{\'e}nard}, {Tabone}, {Pinte}, \& {Dent}}]{Louvet18}
{Louvet}, F., {Dougados}, C., {Cabrit}, S., {et~al.} 2018, \aap, 618, A120,
  \dodoi{10.1051/0004-6361/201731733}

\bibitem[{{Luhman}(2022)}]{Luhman2022_uppstars}
{Luhman}, K.~L. 2022, \aj, 163, 24, \dodoi{10.3847/1538-3881/ac35e2}

\bibitem[{{Mahieu} {et~al.}(2012){Mahieu}, {Maier}, {Lazareff}, {Navarrini},
  {Celestin}, {Chalain}, {Geoffroy}, {Laslaz}, \& {Perrin}}]{ALMA_B7_receiver}
{Mahieu}, S., {Maier}, D., {Lazareff}, B., {et~al.} 2012, IEEE Transactions on
  Terahertz Science and Technology, 2, 29, \dodoi{10.1109/TTHZ.2011.2177734}

\bibitem[{{Manara} {et~al.}(2023){Manara}, {Ansdell}, {Rosotti}, {Hughes},
  {Armitage}, {Lodato}, \& {Williams}}]{Manara_PPVII}
{Manara}, C.~F., {Ansdell}, M., {Rosotti}, G.~P., {et~al.} 2023, in
  Astronomical Society of the Pacific Conference Series, Vol. 534, Protostars
  and Planets VII, ed. S.~{Inutsuka}, Y.~{Aikawa}, T.~{Muto}, K.~{Tomida}, \&
  M.~{Tamura}, 539, \dodoi{10.48550/arXiv.2203.09930}

\bibitem[{{Manara} {et~al.}(2020){Manara}, {Natta}, {Rosotti}, {Alcal{\'a}},
  {Nisini}, {Lodato}, {Testi}, {Pascucci}, {Hillenbrand}, {Carpenter},
  {Scholz}, {Fedele}, {Frasca}, {Mulders}, {Rigliaco}, {Scardoni}, \&
  {Zari}}]{Manara2020}
{Manara}, C.~F., {Natta}, A., {Rosotti}, G.~P., {et~al.} 2020, \aap, 639, A58,
  \dodoi{10.1051/0004-6361/202037949}

\bibitem[{{Mauc{\'o}} {et~al.}(2023){Mauc{\'o}}, {Manara}, {Ansdell},
  {Bettoni}, {Claes}, {Alcala}, {Miotello}, {Facchini}, {Haworth}, {Lodato}, \&
  {Williams}}]{Mauco23}
{Mauc{\'o}}, K., {Manara}, C.~F., {Ansdell}, M., {et~al.} 2023, \aap, 679, A82,
  \dodoi{10.1051/0004-6361/202347627}

\bibitem[{{McClure} {et~al.}(2016){McClure}, {Bergin}, {Cleeves}, {van
  Dishoeck}, {Blake}, {Evans}, {Green}, {Henning}, {{\"O}berg}, {Pontoppidan},
  \& {Salyk}}]{mcclure16}
{McClure}, M.~K., {Bergin}, E.~A., {Cleeves}, L.~I., {et~al.} 2016, \apj, 831,
  167, \dodoi{10.3847/0004-637X/831/2/167}

\bibitem[{{Miley} {et~al.}(2024){Miley}, {Carpenter}, {Booth}, {Jennings},
  {Haworth}, {Vioque}, {Andrews}, {Wilner}, {Benisty}, {Huang}, {Perez},
  {Guzman}, {Ricci}, \& {Isella}}]{Miley24}
{Miley}, J.~M., {Carpenter}, J., {Booth}, R., {et~al.} 2024, \aap, 682, A55,
  \dodoi{10.1051/0004-6361/202347135}

\bibitem[{{Miley} {et~al.}(2025){Miley}, {P{\'e}rez}, {Agurto-Gangas},
  {Sierra}, {Trapman}, {Vioque}, {Kurtovic}, {Pinilla}, {Pascucci}, {Zhang},
  {Anania}, {Carpenter}, {Cieza}, {Deng}, {Gonz{\'a}lez-Ruilova}, {Rosotti},
  {Ruiz-Rodriguez}, \& {TorresVillanueva}}]{AGEPRO_XII_mm_var_USco7}
{Miley}, J.~M., {P{\'e}rez}, L.~M., {Agurto-Gangas}, C., {et~al.} 2025, \apj,
  989, 11, \dodoi{10.3847/1538-4357/add25c}

\bibitem[{{Miotello} {et~al.}(2023){Miotello}, {Kamp}, {Birnstiel}, {Cleeves},
  \& {Kataoka}}]{Miotello23_PPVII}
{Miotello}, A., {Kamp}, I., {Birnstiel}, T., {Cleeves}, L.~C., \& {Kataoka}, A.
  2023, in Astronomical Society of the Pacific Conference Series, Vol. 534,
  Protostars and Planets VII, ed. S.~{Inutsuka}, Y.~{Aikawa}, T.~{Muto},
  K.~{Tomida}, \& M.~{Tamura}, 501, \dodoi{10.48550/arXiv.2203.09818}

\bibitem[{{Miotello} {et~al.}(2016){Miotello}, {van Dishoeck}, {Kama}, \&
  {Bruderer}}]{miotello16}
{Miotello}, A., {van Dishoeck}, E.~F., {Kama}, M., \& {Bruderer}, S. 2016,
  \aap, 594, A85, \dodoi{10.1051/0004-6361/201628159}

\bibitem[{{Miotello} {et~al.}(2017){Miotello}, {van Dishoeck}, {Williams},
  {Ansdell}, {Guidi}, {Hogerheijde}, {Manara}, {Tazzari}, {Testi}, {van der
  Marel}, \& {van Terwisga}}]{miotello17}
{Miotello}, A., {van Dishoeck}, E.~F., {Williams}, J.~P., {et~al.} 2017, \aap,
  599, A113, \dodoi{10.1051/0004-6361/201629556}

\bibitem[{{Molyarova} {et~al.}(2017){Molyarova}, {Akimkin}, {Semenov},
  {Henning}, {Vasyunin}, \& {Wiebe}}]{Molyarova17}
{Molyarova}, T., {Akimkin}, V., {Semenov}, D., {et~al.} 2017, \apj, 849, 130,
  \dodoi{10.3847/1538-4357/aa9227}

\bibitem[{{Morbidelli} \& {Raymond}(2016)}]{morbidelli16b_challenges}
{Morbidelli}, A., \& {Raymond}, S.~N. 2016, Journal of Geophysical Research
  (Planets), 121, 1962, \dodoi{10.1002/2016JE005088}

\bibitem[{{Mordasini} \& {Burn}(2024)}]{Mordasini2024RvMG}
{Mordasini}, C., \& {Burn}, R. 2024, Reviews in Mineralogy and Geochemistry,
  90, 55, \dodoi{10.2138/rmg.2024.90.03}

\bibitem[{{Najita} \& {Bergin}(2018)}]{Najita18_evolution}
{Najita}, J.~R., \& {Bergin}, E.~A. 2018, \apj, 864, 168,
  \dodoi{10.3847/1538-4357/aad80c}

\bibitem[{{Natta} {et~al.}(2007){Natta}, {Testi}, {Calvet}, {Henning},
  {Waters}, \& {Wilner}}]{Natta2007}
{Natta}, A., {Testi}, L., {Calvet}, N., {et~al.} 2007, in Protostars and
  Planets V, ed. B.~{Reipurth}, D.~{Jewitt}, \& K.~{Keil}, 767,
  \dodoi{10.48550/arXiv.astro-ph/0602041}

\bibitem[{{{\"O}berg} {et~al.}(2023){{\"O}berg}, {Facchini}, \&
  {Anderson}}]{Oberg2023ARAA}
{{\"O}berg}, K.~I., {Facchini}, S., \& {Anderson}, D.~E. 2023, \araa, 61, 287,
  \dodoi{10.1146/annurev-astro-022823-040820}

\bibitem[{{Ohashi} {et~al.}(2014){Ohashi}, {Saigo}, {Aso}, {Aikawa},
  {Koyamatsu}, {Machida}, {Saito}, {Takahashi}, {Takakuwa}, {Tomida},
  {Tomisaka}, \& {Yen}}]{Ohashi2014}
{Ohashi}, N., {Saigo}, K., {Aso}, Y., {et~al.} 2014, \apj, 796, 131,
  \dodoi{10.1088/0004-637X/796/2/131}

\bibitem[{{Ohashi} {et~al.}(2023){Ohashi}, {Tobin}, {J{\o}rgensen}, {Takakuwa},
  {Sheehan}, {Aikawa}, {Li}, {Looney}, {Williams}, {Aso}, {Sharma}, {Sai},
  {Yamato}, {Lee}, {Tomida}, {Yen}, {Encalada}, {Flores}, {Gavino}, {Kido},
  {Han}, {Lin}, {Narayanan}, {Phuong}, {Santamar{\'\i}a-Miranda}, {Thieme},
  {van't Hoff}, {de Gregorio-Monsalvo}, {Koch}, {Kwon}, {Lai}, {Lee},
  {Plunkett}, {Saigo}, {Hirano}, {Lam}, \& {Mori}}]{Ohashi_eDisK_2023}
{Ohashi}, N., {Tobin}, J.~J., {J{\o}rgensen}, J.~K., {et~al.} 2023, \apj, 951,
  8, \dodoi{10.3847/1538-4357/acd384}

\bibitem[{{Pascucci} {et~al.}(2023{\natexlab{a}}){Pascucci}, {Cabrit},
  {Edwards}, {Gorti}, {Gressel}, \& {Suzuki}}]{Pascucci_PPVII}
{Pascucci}, I., {Cabrit}, S., {Edwards}, S., {et~al.} 2023{\natexlab{a}}, in
  Astronomical Society of the Pacific Conference Series, Vol. 534, Protostars
  and Planets VII, ed. S.~{Inutsuka}, Y.~{Aikawa}, T.~{Muto}, K.~{Tomida}, \&
  M.~{Tamura}, 567, \dodoi{10.48550/arXiv.2203.10068}

\bibitem[{{Pascucci} {et~al.}(2016){Pascucci}, {Testi}, {Herczeg}, {Long},
  {Manara}, {Hendler}, {Mulders}, {Krijt}, {Ciesla}, {Henning}, {Mohanty},
  {Drabek-Maunder}, {Apai}, {Sz{\H{u}}cs}, {Sacco}, \& {Olofsson}}]{Pascucci16}
{Pascucci}, I., {Testi}, L., {Herczeg}, G.~J., {et~al.} 2016, \apj, 831, 125,
  \dodoi{10.3847/0004-637X/831/2/125}

\bibitem[{{Pascucci} {et~al.}(2023{\natexlab{b}}){Pascucci}, {Skinner}, {Deng},
  {Ruaud}, {Gorti}, {Schwarz}, {Chapillon}, {Vioque}, \&
  {Miley}}]{Pascucci2023}
{Pascucci}, I., {Skinner}, B.~N., {Deng}, D., {et~al.} 2023{\natexlab{b}},
  \apj, 953, 183, \dodoi{10.3847/1538-4357/ace4bf}

\bibitem[{{Pascucci} {et~al.}(2025){Pascucci}, {Beck}, {Cabrit}, {Bajaj},
  {Edwards}, {Louvet}, {Najita}, {Skinner}, {Gorti}, {Salyk}, {Brittain},
  {Krijt}, {Muzerolle Page}, {Ruaud}, {Schwarz}, {Semenov}, {Duch{\^e}ne}, \&
  {Villenave}}]{Pascucci2025wind}
{Pascucci}, I., {Beck}, T.~L., {Cabrit}, S., {et~al.} 2025, Nature Astronomy,
  9, 81, \dodoi{10.1038/s41550-024-02385-7}

\bibitem[{{Pecaut} {et~al.}(2012){Pecaut}, {Mamajek}, \& {Bubar}}]{Pecaut2012}
{Pecaut}, M.~J., {Mamajek}, E.~E., \& {Bubar}, E.~J. 2012, \apj, 746, 154,
  \dodoi{10.1088/0004-637X/746/2/154}

\bibitem[{{Pollack} {et~al.}(1996){Pollack}, {Hubickyj}, {Bodenheimer},
  {Lissauer}, {Podolak}, \& {Greenzweig}}]{pollack96}
{Pollack}, J.~B., {Hubickyj}, O., {Bodenheimer}, P., {et~al.} 1996, Icarus,
  124, 62, \dodoi{10.1006/icar.1996.0190}

\bibitem[{{Preibisch} {et~al.}(2002){Preibisch}, {Brown}, {Bridges},
  {Guenther}, \& {Zinnecker}}]{Preibisch2002}
{Preibisch}, T., {Brown}, A. G.~A., {Bridges}, T., {Guenther}, E., \&
  {Zinnecker}, H. 2002, \aj, 124, 404, \dodoi{10.1086/341174}

\bibitem[{{Preibisch} {et~al.}(2001){Preibisch}, {Guenther}, \&
  {Zinnecker}}]{Preibisch2001}
{Preibisch}, T., {Guenther}, E., \& {Zinnecker}, H. 2001, \aj, 121, 1040,
  \dodoi{10.1086/318774}

\bibitem[{{Pringle}(1981)}]{pringle81}
{Pringle}, J.~E. 1981, \araa, 19, 137,
  \dodoi{10.1146/annurev.aa.19.090181.001033}

\bibitem[{{Ratzenb{\"o}ck} {et~al.}(2023{\natexlab{a}}){Ratzenb{\"o}ck},
  {Gro{\ss}schedl}, {M{\"o}ller}, {Alves}, {Bomze}, \&
  {Meingast}}]{Ratzenbock2023a}
{Ratzenb{\"o}ck}, S., {Gro{\ss}schedl}, J.~E., {M{\"o}ller}, T., {et~al.}
  2023{\natexlab{a}}, \aap, 677, A59, \dodoi{10.1051/0004-6361/202243690}

\bibitem[{{Ratzenb{\"o}ck} {et~al.}(2023{\natexlab{b}}){Ratzenb{\"o}ck},
  {Gro{\ss}schedl}, {Alves}, {Miret-Roig}, {Bomze}, {Forbes}, {Goodman},
  {Hacar}, {Lin}, {Meingast}, {M{\"o}ller}, {Piecka}, {Posch}, {Rottensteiner},
  {Swiggum}, \& {Zucker}}]{Ratzenbock2023b}
{Ratzenb{\"o}ck}, S., {Gro{\ss}schedl}, J.~E., {Alves}, J., {et~al.}
  2023{\natexlab{b}}, \aap, 678, A71, \dodoi{10.1051/0004-6361/202346901}

\bibitem[{{Ribas} {et~al.}(2014){Ribas}, {Mer{\'\i}n}, {Bouy}, \&
  {Maud}}]{Ribas_2014_diskfraction}
{Ribas}, {\'A}., {Mer{\'\i}n}, B., {Bouy}, H., \& {Maud}, L.~T. 2014, \aap,
  561, A54, \dodoi{10.1051/0004-6361/201322597}

\bibitem[{{Rosotti}(2023)}]{Rosotti23}
{Rosotti}, G.~P. 2023, \nar, 96, 101674, \dodoi{10.1016/j.newar.2023.101674}

\bibitem[{{Rosotti} {et~al.}(2019){Rosotti}, {Tazzari}, {Booth}, {Testi},
  {Lodato}, \& {Clarke}}]{Rosotti19}
{Rosotti}, G.~P., {Tazzari}, M., {Booth}, R.~A., {et~al.} 2019, \mnras, 486,
  4829, \dodoi{10.1093/mnras/stz1190}

\bibitem[{{Ruaud} \& {Gorti}(2024)}]{Ruaud2024}
{Ruaud}, M., \& {Gorti}, U. 2024, \apj, 971, 66,
  \dodoi{10.3847/1538-4357/ad5547}

\bibitem[{{Ruaud} {et~al.}(2022){Ruaud}, {Gorti}, \& {Hollenbach}}]{Ruaud2022}
{Ruaud}, M., {Gorti}, U., \& {Hollenbach}, D.~J. 2022, \apj, 925, 49,
  \dodoi{10.3847/1538-4357/ac3826}

\bibitem[{{Ruiz-Rodriguez} {et~al.}(2025){Ruiz-Rodriguez},
  {Gonz{\'a}lez-Ruilova}, {Cieza}, {Zhang}, {Trapman}, {Sierra}, {Pinilla},
  {Pascucci}, {P{\'e}rez}, {Deng}, {Agurto-Gangas}, {Carpenter}, {Tabone},
  {Rosotti}, {Anania}, {Miley}, {Schwarz}, {Kuznetsova}, {Vioque}, \&
  {Kurtovic}}]{AGEPRO_II_Ophiuchus}
{Ruiz-Rodriguez}, D.~A., {Gonz{\'a}lez-Ruilova}, C., {Cieza}, L.~A., {et~al.}
  2025, \apj, 989, 2, \dodoi{10.3847/1538-4357/add2ec}

\bibitem[{{Sanchis} {et~al.}(2021){Sanchis}, {Testi}, {Natta}, {Facchini},
  {Manara}, {Miotello}, {Ercolano}, {Henning}, {Preibisch}, {Carpenter}, {de
  Gregorio-Monsalvo}, {Jayawardhana}, {Lopez}, {Mu{\v{z}}i{\'c}}, {Pascucci},
  {Santamar{\'\i}a-Miranda}, {van Terwisga}, \& {Williams}}]{Sanchis21}
{Sanchis}, E., {Testi}, L., {Natta}, A., {et~al.} 2021, \aap, 649, A19,
  \dodoi{10.1051/0004-6361/202039733}

\bibitem[{{Savvidou} \& {Bitsch}(2023)}]{Savvidou23}
{Savvidou}, S., \& {Bitsch}, B. 2023, \aap, 679, A42,
  \dodoi{10.1051/0004-6361/202245793}

\bibitem[{{Schwarz} {et~al.}(2016){Schwarz}, {Bergin}, {Cleeves}, {Blake},
  {Zhang}, {{\"O}berg}, {van Dishoeck}, \& {Qi}}]{schwarz16}
{Schwarz}, K.~R., {Bergin}, E.~A., {Cleeves}, L.~I., {et~al.} 2016, \apj, 823,
  91, \dodoi{10.3847/0004-637X/823/2/91}

\bibitem[{{Schwarz} {et~al.}(2018){Schwarz}, {Bergin}, {Cleeves}, {Zhang},
  {{\"O}berg}, {Blake}, \& {Anderson}}]{schwarz18}
---. 2018, \apj, 856, 85, \dodoi{10.3847/1538-4357/aaae08}

\bibitem[{{Schwarz} {et~al.}(2021){Schwarz}, {Calahan}, {Zhang}, {Alarc{\'o}n},
  {Aikawa}, {Andrews}, {Bae}, {Bergin}, {Booth}, {Bosman}, {Cataldi},
  {Cleeves}, {Czekala}, {Huang}, {Ilee}, {Law}, {Le Gal}, {Liu}, {Long},
  {Loomis}, {Mac{\'\i}as}, {McClure}, {M{\'e}nard}, {{\"O}berg}, {Teague}, {van
  Dishoeck}, {Walsh}, \& {Wilner}}]{Schwarz2021_MAPS}
{Schwarz}, K.~R., {Calahan}, J.~K., {Zhang}, K., {et~al.} 2021, \apjs, 257, 20,
  \dodoi{10.3847/1538-4365/ac143b}

\bibitem[{{Shakura} \& {Sunyaev}(1973)}]{shakura73}
{Shakura}, N.~I., \& {Sunyaev}, R.~A. 1973, \aap, 24, 337

\bibitem[{{Sierra} {et~al.}(2024){Sierra}, {P{\'e}rez}, {Agurto-Gangas},
  {Miley}, {Zhang}, {Pinilla}, {Pascucci}, {Trapman}, {Kurtovic}, {Vioque},
  {Deng}, {Anania}, {Carpenter}, {Cieza}, {Gonz{\'a}lez-Ruilova},
  {Hogerheijde}, {Kuznetsova}, {Rosotti}, {Ruiz-Rodriguez}, {Schwarz},
  {Tabone}, \& {TorresVillanueva}}]{AGEPRO_IX_Upp1}
{Sierra}, A., {P{\'e}rez}, L.~M., {Agurto-Gangas}, C., {et~al.} 2024, \apj,
  974, 102, \dodoi{10.3847/1538-4357/ad6e73}

\bibitem[{{Simon} {et~al.}(2019){Simon}, {Guilloteau}, {Beck}, {Chapillon}, {Di
  Folco}, {Dutrey}, {Feiden}, {Grosso}, {Pi{\'e}tu}, {Prato}, \&
  {Schaefer}}]{Simon19}
{Simon}, M., {Guilloteau}, S., {Beck}, T.~L., {et~al.} 2019, \apj, 884, 42,
  \dodoi{10.3847/1538-4357/ab3e3b}

\bibitem[{{Soderblom} {et~al.}(2014){Soderblom}, {Hillenbrand}, {Jeffries},
  {Mamajek}, \& {Naylor}}]{Soderblom2014_PPVI}
{Soderblom}, D.~R., {Hillenbrand}, L.~A., {Jeffries}, R.~D., {Mamajek}, E.~E.,
  \& {Naylor}, T. 2014, in Protostars and Planets VI, ed. H.~{Beuther}, R.~S.
  {Klessen}, C.~P. {Dullemond}, \& T.~{Henning}, 219--241,
  \dodoi{10.2458/azu_uapress_9780816531240-ch010}

\bibitem[{{Stammler} {et~al.}(2019){Stammler}, {Drazkowska}, {Birnstiel},
  {Klahr}, {Dullemond}, \& {Andrews}}]{Stammler19}
{Stammler}, S.~M., {Drazkowska}, J., {Birnstiel}, T., {et~al.} 2019, \apjl,
  884, L5, \dodoi{10.3847/2041-8213/ab4423}

\bibitem[{{Sturm} {et~al.}(2023){Sturm}, {Booth}, {McClure}, {Leemker}, \& {van
  Dishoeck}}]{Sturm2023}
{Sturm}, J.~A., {Booth}, A.~S., {McClure}, M.~K., {Leemker}, M., \& {van
  Dishoeck}, E.~F. 2023, \aap, 670, A12, \dodoi{10.1051/0004-6361/202244227}

\bibitem[{{Sullivan} \& {Kraus}(2021)}]{Sullivan2021}
{Sullivan}, K., \& {Kraus}, A.~L. 2021, \apj, 912, 137,
  \dodoi{10.3847/1538-4357/abf044}

\bibitem[{{Tabone} {et~al.}(2020){Tabone}, {Cabrit}, {Pineau des For{\^e}ts},
  {Ferreira}, {Gusdorf}, {Podio}, {Bianchi}, {Chapillon}, {Codella}, \&
  {Gueth}}]{Tabone_2020}
{Tabone}, B., {Cabrit}, S., {Pineau des For{\^e}ts}, G., {et~al.} 2020, \aap,
  640, A82, \dodoi{10.1051/0004-6361/201834377}

\bibitem[{{Tabone} {et~al.}(2025){Tabone}, {Rosotti}, {Trapman}, {Pinilla},
  {Pascucci}, {Somigliana}, {Alexander}, {Vioque}, {Anania}, {Kuznetsova},
  {Zhang}, {P{\'e}rez}, {Cieza}, {Carpenter}, {Deng}, {Agurto-Gangas},
  {Ruiz-Rodriguez}, {Sierra}, {Kurtovic}, {Miley}, {Gonz{\'a}lez-Ruilova},
  {TorresVillanueva}, {Hogerheijde}, {Schwarz}, {Toci}, {Testi}, \&
  {Lodato}}]{AGEPRO_VII_population}
{Tabone}, B., {Rosotti}, G.~P., {Trapman}, L., {et~al.} 2025, \apj, 989, 7,
  \dodoi{10.3847/1538-4357/adc7b1}

\bibitem[{{Takeuchi} {et~al.}(2005){Takeuchi}, {Clarke}, \& {Lin}}]{Takeuchi05}
{Takeuchi}, T., {Clarke}, C.~J., \& {Lin}, D.~N.~C. 2005, \apj, 627, 286,
  \dodoi{10.1086/430393}

\bibitem[{{Tazzari} {et~al.}(2021){Tazzari}, {Clarke}, {Testi}, {Williams},
  {Facchini}, {Manara}, {Natta}, \& {Rosotti}}]{Tazzari2021b}
{Tazzari}, M., {Clarke}, C.~J., {Testi}, L., {et~al.} 2021, \mnras, 506, 2804,
  \dodoi{10.1093/mnras/stab1808}

\bibitem[{Teague(2019)}]{GoFish}
Teague, R. 2019, The Journal of Open Source Software, 4, 1632,
  \dodoi{10.21105/joss.01632}

\bibitem[{{Teague} {et~al.}(2018){Teague}, {Bae}, {Bergin}, {Birnstiel}, \&
  {Foreman-Mackey}}]{Teague18a}
{Teague}, R., {Bae}, J., {Bergin}, E.~A., {Birnstiel}, T., \& {Foreman-Mackey},
  D. 2018, \apj, 860, L12, \dodoi{10.3847/2041-8213/aac6d7}

\bibitem[{Testi {et~al.}(2014)Testi, Birnstiel, Ricci, Andrews, Blum,
  Carpenter, Dominik, Isella, Natta, Williams, \& Wilner}]{testi14}
Testi, L., Birnstiel, T., Ricci, L., {et~al.} 2014, {Protostars and Planets VI}

\bibitem[{{Tobin} {et~al.}(2020){Tobin}, {Sheehan}, {Megeath},
  {D{\'\i}az-Rodr{\'\i}guez}, {Offner}, {Murillo}, {van 't Hoff}, {van
  Dishoeck}, {Osorio}, {Anglada}, {Furlan}, {Stutz}, {Reynolds}, {Karnath},
  {Fischer}, {Persson}, {Looney}, {Li}, {Stephens}, {Chandler}, {Cox},
  {Dunham}, {Tychoniec}, {Kama}, {Kratter}, {Kounkel}, {Mazur}, {Maud},
  {Patel}, {Perez}, {Sadavoy}, {Segura-Cox}, {Sharma}, {Stephenson}, {Watson},
  \& {Wyrowski}}]{Tobin20}
{Tobin}, J.~J., {Sheehan}, P.~D., {Megeath}, S.~T., {et~al.} 2020, \apj, 890,
  130, \dodoi{10.3847/1538-4357/ab6f64}

\bibitem[{{Toci} {et~al.}(2021){Toci}, {Rosotti}, {Lodato}, {Testi}, \&
  {Trapman}}]{Toci2021}
{Toci}, C., {Rosotti}, G., {Lodato}, G., {Testi}, L., \& {Trapman}, L. 2021,
  \mnras, 507, 818, \dodoi{10.1093/mnras/stab2112}

\bibitem[{{Trapman} {et~al.}(2020{\natexlab{a}}){Trapman}, {Ansdell},
  {Hogerheijde}, {Facchini}, {Manara}, {Miotello}, {Williams}, \&
  {Bruderer}}]{Trapman20_drift}
{Trapman}, L., {Ansdell}, M., {Hogerheijde}, M.~R., {et~al.}
  2020{\natexlab{a}}, \aap, 638, A38, \dodoi{10.1051/0004-6361/201834537}

\bibitem[{{Trapman} {et~al.}(2019){Trapman}, {Facchini}, {Hogerheijde}, {van
  Dishoeck}, \& {Bruderer}}]{Trapman19}
{Trapman}, L., {Facchini}, S., {Hogerheijde}, M.~R., {van Dishoeck}, E.~F., \&
  {Bruderer}, S. 2019, \aap, 629, A79, \dodoi{10.1051/0004-6361/201834723}

\bibitem[{{Trapman} {et~al.}(2020{\natexlab{b}}){Trapman}, {Rosotti}, {Bosman},
  {Hogerheijde}, \& {van Dishoeck}}]{Trapman20_viscous}
{Trapman}, L., {Rosotti}, G., {Bosman}, A.~D., {Hogerheijde}, M.~R., \& {van
  Dishoeck}, E.~F. 2020{\natexlab{b}}, \aap, 640, A5,
  \dodoi{10.1051/0004-6361/202037673}

\bibitem[{{Trapman} {et~al.}(2022{\natexlab{a}}){Trapman}, {Tabone}, {Rosotti},
  \& {Zhang}}]{Trapman21_wind}
{Trapman}, L., {Tabone}, B., {Rosotti}, G., \& {Zhang}, K. 2022{\natexlab{a}},
  \apj, 926, 61, \dodoi{10.3847/1538-4357/ac3ed5}

\bibitem[{{Trapman} {et~al.}(2022{\natexlab{b}}){Trapman}, {Zhang}, {van't
  Hoff}, {Hogerheijde}, \& {Bergin}}]{Trapman22_mass}
{Trapman}, L., {Zhang}, K., {van't Hoff}, M. L.~R., {Hogerheijde}, M.~R., \&
  {Bergin}, E.~A. 2022{\natexlab{b}}, \apjl, 926, L2,
  \dodoi{10.3847/2041-8213/ac4f47}

\bibitem[{{Trapman} {et~al.}(2025{\natexlab{a}}){Trapman}, {Zhang}, {Rosotti},
  {Pinilla}, {Tabone}, {Pascucci}, {Agurto-Gangas}, {Anania}, {Carpenter},
  {Cieza}, {Deng}, {Gonz{\'a}lez-Ruilova}, {Hogerheijde}, {Kurtovic},
  {Kuznetsova}, {Miley}, {P{\'e}rez}, {Ruiz-Rodriguez}, {Schwarz}, {Sierra},
  {TorresVillanueva}, \& {Vioque}}]{AGEPRO_V_gasmasses}
{Trapman}, L., {Zhang}, K., {Rosotti}, G.~P., {et~al.} 2025{\natexlab{a}},
  \apj, 989, 5, \dodoi{10.3847/1538-4357/adcd6e}

\bibitem[{{Trapman} {et~al.}(2025{\natexlab{b}}){Trapman}, {Vioque},
  {Kurtovic}, {Zhang}, {Rosotti}, {Pinilla}, {Carpenter}, {Cieza}, {Pascucci},
  {Anania}, {Agurto-Gangas}, {Deng}, {Miley}, {P{\'e}rez}, {Sierra}, {Tabone},
  {Ruiz-Rodriguez}, {Gonz{\'a}lez-Ruilova}, \&
  {TorresVillanueva}}]{AGEPRO_XI_gas_disk_sizes}
{Trapman}, L., {Vioque}, M., {Kurtovic}, N.~T., {et~al.} 2025{\natexlab{b}},
  \apj, 989, 10, \dodoi{10.3847/1538-4357/adc7af}

\bibitem[{{Tripathi} {et~al.}(2017){Tripathi}, {Andrews}, {Birnstiel}, \&
  {Wilner}}]{Tripathi2017}
{Tripathi}, A., {Andrews}, S.~M., {Birnstiel}, T., \& {Wilner}, D.~J. 2017,
  \apj, 845, 44, \dodoi{10.3847/1538-4357/aa7c62}

\bibitem[{{Turner} {et~al.}(2014){Turner}, {Fromang}, {Gammie}, {Klahr},
  {Lesur}, {Wardle}, \& {Bai}}]{turner14}
{Turner}, N.~J., {Fromang}, S., {Gammie}, C., {et~al.} 2014, Protostars and
  Planets VI, 411, \dodoi{10.2458/azu\_uapress\_9780816531240-ch018}

\bibitem[{{Tychoniec} {et~al.}(2020){Tychoniec}, {Manara}, {Rosotti}, {van
  Dishoeck}, {Cridland}, {Hsieh}, {Murillo}, {Segura-Cox}, {van Terwisga}, \&
  {Tobin}}]{Tychoniec20}
{Tychoniec}, {\L}., {Manara}, C.~F., {Rosotti}, G.~P., {et~al.} 2020, \aap,
  640, A19, \dodoi{10.1051/0004-6361/202037851}

\bibitem[{{Vioque} {et~al.}(2025){Vioque}, {Kurtovic}, {Trapman}, {Sierra},
  {P{\'e}rez}, {Zhang}, {Curone}, {Rosotti}, {Carpenter}, {Tabone}, {Pinilla},
  {Deng}, {Pascucci}, {Miley}, {Agurto-Gangas}, {Cieza}, {Anania},
  {Ruiz-Rodriguez}, {Gonz{\'a}lez-Ruilova}, {TorresVillanueva}, \&
  {Kuznetsova}}]{AGEPRO_X_dust_disks}
{Vioque}, M., {Kurtovic}, N.~T., {Trapman}, L., {et~al.} 2025, \apj, 989, 9,
  \dodoi{10.3847/1538-4357/adc7b0}

\bibitem[{{Williams} \& {Best}(2014)}]{williams14}
{Williams}, J.~P., \& {Best}, W. M.~J. 2014, \apj, 788, 59,
  \dodoi{10.1088/0004-637X/788/1/59}

\bibitem[{{Williams} {et~al.}(2019){Williams}, {Cieza}, {Hales}, {Ansdell},
  {Ruiz-Rodriguez}, {Casassus}, {Perez}, \& {Zurlo}}]{Williams19}
{Williams}, J.~P., {Cieza}, L., {Hales}, A., {et~al.} 2019, \apjl, 875, L9,
  \dodoi{10.3847/2041-8213/ab1338}

\bibitem[{{Williams} \& {Cieza}(2011)}]{Williams11}
{Williams}, J.~P., \& {Cieza}, L.~A. 2011, Annual Review of Astronomy and
  Astrophysics, 49, 67, \dodoi{10.1146/annurev-astro-081710-102548}

\bibitem[{{Wilson}(1999)}]{wilson99}
{Wilson}, T.~L. 1999, Reports on Progress in Physics, 62, 143,
  \dodoi{10.1088/0034-4885/62/2/002}

\bibitem[{{Winter} \& {Haworth}(2022)}]{Winter2022_external_photoevaporation}
{Winter}, A.~J., \& {Haworth}, T.~J. 2022, European Physical Journal Plus, 137,
  1132, \dodoi{10.1140/epjp/s13360-022-03314-1}

\bibitem[{{Xu} {et~al.}(2017){Xu}, {Bai}, \& {{\"O}berg}}]{xu17}
{Xu}, R., {Bai}, X.-N., \& {{\"O}berg}, K. 2017, \apj, 835, 162,
  \dodoi{10.3847/1538-4357/835/2/162}

\bibitem[{{Zhang} {et~al.}(2016){Zhang}, {Bergin}, {Blake}, {Cleeves},
  {Hogerheijde}, {Salinas}, \& {Schwarz}}]{zhang16}
{Zhang}, K., {Bergin}, E.~A., {Blake}, G.~A., {et~al.} 2016, \apjl, 818, L16,
  \dodoi{10.3847/2041-8205/818/1/L16}

\bibitem[{{Zhang} {et~al.}(2017){Zhang}, {Bergin}, {Blake}, {Cleeves}, \&
  {Schwarz}}]{zhang17}
{Zhang}, K., {Bergin}, E.~A., {Blake}, G.~A., {Cleeves}, L.~I., \& {Schwarz},
  K.~R. 2017, Nature Astronomy, 1, 0130, \dodoi{10.1038/s41550-017-0130}

\bibitem[{{Zhang} {et~al.}(2020){Zhang}, {Schwarz}, \&
  {Bergin}}]{Zhang20_evolution}
{Zhang}, K., {Schwarz}, K.~R., \& {Bergin}, E.~A. 2020, \apjl, 891, L17,
  \dodoi{10.3847/2041-8213/ab7823}

\bibitem[{{Zhang} {et~al.}(2025){Zhang}, {P{\'e}rez}, {Pascucci}, {Pinilla},
  {Cieza}, {Carpenter}, {Trapman}, {Deng}, {Agurto-Gangas}, {Sierra},
  {Kurtovic}, {Ruiz-Rodriguez}, {Vioque}, {Miley}, {Tabone},
  {Gonz{\'a}lez-Ruilova}, {Anania}, {Rosotti}, {TorresVillanueva},
  {Hogerheijde}, {Schwarz}, \& {Kuznetsova}}]{AGEPRO_I_overview}
{Zhang}, K., {P{\'e}rez}, L.~M., {Pascucci}, I., {et~al.} 2025, \apj, 989, 1,
  \dodoi{10.3847/1538-4357/addebe}

\end{thebibliography}
\bibliographystyle{aasjournal}


\end{document}